%% file: Arxiv_comb_cali_surv.tex
\documentclass[lineno]{biometrika}

\usepackage{amsmath}

\usepackage{times}
\usepackage{bm}
\usepackage{natbib}

\usepackage[linesnumbered,ruled,vlined]{algorithm2e}



\RequirePackage[OT1]{fontenc}
\RequirePackage{amsmath,amssymb,caption,subcaption,graphicx,epstopdf,enumerate,lmodern}
\RequirePackage{bigints}
\RequirePackage[colorlinks,citecolor=blue,urlcolor=blue]{hyperref}
\usepackage{bm,color,fancyvrb,xcolor,mathtools}
\usepackage{amscd}
\usepackage{mathrsfs}
\usepackage{bbm}
\usepackage{tikz}
\usetikzlibrary{decorations.pathreplacing,calc, patterns}

\usepackage{tensor}

\newcommand{\bfm}[1]{\ensuremath{\mathbf{#1}}}

\def\bi{\bfm i}

     \def\SS{\mathbb{S}}

\usepackage{multirow, booktabs}
\newcommand{\indep}{\perp \!\!\! \perp}
\input{notation}
\def\v1{{\mathbf{1}}}
\def\b1{{\boldsymbol{1}}}
\newcommand{\sD}{\mathcal{D}}
\newcommand{\wt}{\widetilde}

\newcommand{\Err}{\mathrm{Err}}

\newcommand{\ignore}[1]{}

\DeclareMathOperator{\pr}{pr} 

\def\newpage{\vfill\eject}
\def\wh{\widehat}

\newcommand{\vertiii}[1]{{\left\vert\kern-0.25ex\left\vert\kern-0.25ex\left\vert #1 
		\right\vert\kern-0.25ex\right\vert\kern-0.25ex\right\vert}}

\usepackage{xr}
\RequirePackage[colorlinks,citecolor=blue,urlcolor=blue]{hyperref}

\begin{document}
	
	\title{Multicalibration for Modeling Censored Survival Data with Universal Adaptability}

	
	
	\markboth{Ye and Li}{Multicalibration for Modeling  Survival Data}
	
	
	\author{Hanxuan Ye}
	\affil{Department of Biostatistics, Epidemiology and Informatics, \\ University of Pennsylvania, Philadelphia, Pennsylvania 19104, U.S.A.
		\email{Huanxuan.Ye@pennmedicine.upenn.edu}}

	\author{HONGZHE LI}
	\affil{Department of Biostatistics, Epidemiology and Informatics, \\ University of Pennsylvania, Philadelphia, Pennsylvania 19104, U.S.A.
		\email{hongzhe@upenn.edu}}

	\maketitle

	\begin{abstract} 
		Traditional statistical and machine learning methods typically assume that the training and test data follow the same distribution. However, this assumption is frequently violated in real-world applications, where the training data in the source domain may under-represent specific subpopulations in the test data of the target domain. This paper addresses target-independent learning under covariate shift, focusing on multicalibration for survival probability and restricted mean survival time. A black-box post-processing boosting algorithm specifically designed for censored survival data is introduced. By leveraging pseudo-observations, our method produces a multicalibrated predictor that is competitive with inverse propensity score weighting in predicting the survival outcome in an unlabeled target domain, ensuring not only overall accuracy but also fairness across diverse subpopulations. Our theoretical analysis of pseudo-observations builds upon the functional delta method and the  $p$-variational norm. The algorithm's sample complexity, convergence properties, and multicalibration guarantees for post-processed predictors are provided. Our results establish a fundamental connection between multicalibration and universal adaptability, demonstrating that our calibrated function is comparable to, or outperforms, the inverse propensity score weighting estimator. Extensive numerical simulations and a real-world case study on cardiovascular disease risk prediction using two large prospective cohort studies validate the effectiveness of our approach.

	\end{abstract}
	

	\section{Introduction}
	
	Traditional statistical and machine learning methods typically assume that the training and test data follow the same distribution. This assumption is often violated in real-world applications, particularly in health care applications, where the training data (source domain) may under represent certain subpopulations in the test (target) domain.
	For example, researchers may collect patient data from multiple hospitals and aim to predict survival probability or restricted mean survival time for patients with unlabeled health records in a new hospital based on their characteristics. At a high level, we consider hospitals with labeled data as source domains and new hospitals with unlabeled data as target domains. In this setting, the distribution of samples from the source domains may differ from that of the target domain. As a result, directly applying models trained on the source domains to predict outcomes in the target domain can lead to poor performance and uncalibrated predictions.
	
	Using data from source domains that differ from target domains presents challenges due to various forms of distributional shift. One such shift arises when the supports of the observed variables in the source and target domains differ, a scenario we refer to as heterogeneous domains. Another type of shift involves changes in the joint distribution of the source and target data, which includes concept shift with difference in the conditional distribution of the outcome given the covariates \citep{moreno2012unifying} and covariate shift, where the marginal distribution of covariates differs across domains \citep{bickel2009discriminative}.
	
	A common approach to obtaining valid statistical inferences under covariate shift is inverse propensity score weighting (\textsc{ipsw}) \citep{rubin1974estimating, weisberg2009selection}. The propensity score quantifies the relative likelihood of observing data under the source and target populations, allowing for reweighting of the source samples to approximate random samples from the target population. While \textsc{ipsw} has been successfully applied in various scientific contexts \citep{frangakis2009calibration, cole2010generalizing}, it is sensitive to model misspecification. Additionally, in the settings where predictions are required across multiple target populations, \textsc{ipsw} requires a separate estimation procedure for each target population, leading to increased computational burden and potential instability.

	An alternative approach is to learn a single estimator that can automatically adapt to source-to-target shifts, enabling efficient inference across multiple target populations and facilitating target-independent learning. However, while a model may achieve high overall predictive accuracy, it can still exhibit poor performance for certain subpopulations \citep{buolamwini2018gender}. Systematic biases against subpopulations may arise inadvertently due to their underrepresentation in the training data, where data from certain minority groups may be significantly less available than that from majority populations.
	
	To mitigate such biases and address the imbalance in training data, we investigate multicalibration, a fairness criterion originally introduced in the seminal work of \cite{johnson18a} and later extended by \cite{kim2019multiaccuracy, Kim22}. Multicalibration ensures that predictions remain well-calibrated across diverse subpopulations, providing a robust framework for achieving fairness in predictive models under distributional shifts.
	
	Notably, given black-box access to an initial predictor (hypothesis) and a relatively small labeled validation set, \cite{kim2019multiaccuracy} proposed a boosting algorithm that audits the initial predictor to assess whether it satisfies subgroup fairness, specifically multicalibration. If the predictor fails to meet the multicalibration criterion, the algorithm iteratively updates it until the condition is satisfied. The resulting multicalibrated predictor offers provable improvements in group-wise fairness.
	Further, \cite{Kim22} established a connection between multicalibration and target-independent learning through the concept of universal adaptability. They demonstrated that, under appropriate conditions, training a multicalibrated predictor on source data enables efficient estimation of statistics on unseen target distributions, thereby facilitating robust adaptation to new domains.
	
	A key advantage of the multicalibration boosting algorithm is that it operates independently of how the initial predictor was trained, whether it is well-constructed or adversarially chosen. This property is particularly valuable in practical settings where users may seek to improve predictions across diverse populations without having access to the underlying training methodology of the prediction system.
	For a deeper understanding of the development of multicalibration, we refer readers to the unpublished  comprehensive notes by Aaron Roth and the references therein. Additionally, for completeness, we highlight related but distinct lines of research on calibration in survival analysis, such as \cite{goldstein2020x, kamran2021estimating}.
	
	The multicalibration framework developed by \cite{kim2019multiaccuracy, Kim22} primarily focuses on binary classification. In this paper, we extend this framework to censored survival data, introducing a novel approach to multicalibration in the context of survival analysis.
	Survival analysis, widely applied in biomedical research, investigates the relationship between individual characteristics and survival time. This relationship is typically modeled using survival or hazard functions, which characterize the conditional distribution of event occurrence. However, survival data are often subject to right censoring, meaning that the true event times for some individuals remain unobserved. This censoring introduces significant challenges for estimation and prediction tasks.
	
	Several methods have been developed to address distributional shifts in survival data. \cite{bellot2019boosting} employed a boosting-based approach to select source samples that more closely resemble the target samples while discarding dissimilar ones. \cite{li2023accommodating} proposed a method to account for time-varying heterogeneity in risk estimation under the Cox model, improving estimation in the target domain. \cite{shaker2023multi} further extended survival domain adaptation to settings with multiple source domains. These works highlight the need for robust methodologies to handle distribution shifts in censored survival data, motivating our development of a multicalibration framework tailored for this setting.
	
	The objective of this paper is to develop a robust predictor for downstream analysis using only labeled data from source domains when the outcome is censored survival data. We focus on scenarios where the source and target data exhibit significant differences due to strong covariate shift.
	To address censoring, we leverage pseudo observations \citep{andersen2010pseudo, graw2009pseudo, overgaard2017asymptotic}, constructed using the Jackknife method with consistent Kaplan-Meier estimators. Compared to widely used survival models such as the proportional hazards model \citep{cox1972regression, zhang2022modern} and the accelerated failure time model \citep{wei1992accelerated}, pseudo observations offer greater flexibility for integrating machine learning algorithms. They enable calibration of any initial estimator against pseudo-labeled outcomes, facilitating adaptation to distributional shifts while maintaining interpretability and robustness in censored survival data settings.
	
	We develop an efficient algorithm tailored for modeling censored survival data and conduct a rigorous analysis of its sample complexity and convergence properties. Our theoretical investigation leverages the functional delta method and the $p$-variational norm to characterize the behavior of pseudo-observations.
	Furthermore, we demonstrate that multicalibration inherently accounts for propensity shifts across potential target distributions. This insight enables us to use multicalibration as a powerful tool for obtaining estimates in the target domain that achieve accuracy comparable to—or even surpassing—that of estimates obtained via propensity scoring methods.
	
	\section{Problem Setup}\label{sec:setup}

	In survival analysis, we denote the time-to-event for individual $i$ as $T_i$ and the censoring time as $C_i$. Under right-censoring, the observed failure time is given by 
	$\wt{T}_i = T_i \wedge C_i$ and the censoring indicator is $\Delta_i = \b1\{T_i \le C_i\}$.   We  let $X \in \mathcal{X} \subset \bbR^d$ represent the $d$-dimensional covariates and define $\sD=\{\sS,\sT\}$  to indicate whether a sample is drawn from the source ($\sS$) or the target domain ($\sT$). Assume that  we collect data in the source domain $\sS$ and target domain $\sT$, with 
	$D_i \in \{ \sS,\sT \}$ serving as an indicator of domain membership. The  observed data for each individual is then represented as   $O_i = ( X_i, \Tilde{T}_i, \Delta_i, D_i)$.   With a slight abuse of notation, we use $\sU_\sS$ and $\sU_\sT$ to denote the covariate distribution over unlabeled samples, conditional on domain membership $D$ as 
	$$X | D = \sS \sim \sU_\sS, \; X | D = \sT \sim \sU_\sT.$$ Similarly, we denote  the joint distribution of $(X, T, C)$ conditional on $D$ as 
	$$(X, T, C)|D = \sS \sim \sD_\sS, \; (X,T,C)|D = \sT \sim \sD_\sT.$$ Following \cite{zeng2021propensity}, we consider two key functions of the survival time $T_i$:
	\begin{enumerate}
		\item 
		the at-the-risk function, which indicates whether an individual remains event-free at time 
		$t$
		$$\nu^{(1)}(T_i; t) = \b1\{ T_i \ge t\}. $$
		\item the truncation function,  which truncates the event time at 
		$t$
		$$\nu^{(2)}(T_i; t) = T_i \wedge t.$$
	\end{enumerate}
	These functions play a central role in our analysis, particularly in modeling censored survival data under distributional shifts between source and target domains.
	
	We impose the following conditional independence assumptions on the random variables defined above:
	\begin{assumption}\label{ass:domain}
		Given $X_i$, we have $( T_i, C_i ) \indep D_i | X_i$.  This assumption states that, given the covariates $X_i$, the joint distribution of the time-to-event $T_i$
		and censoring time 
		$C_i$
		is independent of the domain membership 
		$D_i$. In other words, after conditioning on 
		$X_i$, the distribution of survival and censoring times remains the same across the source and target domains.
		
		This assumption is crucial for ensuring that domain shifts only occur due to differences in the marginal distribution of 
		$X$,  as assumed under the covariate shift, rather than differences in the conditional distribution of 
		$(T,C)$ given 
		$X$.
		
	\end{assumption}
	\begin{assumption}\label{ass:indep}
		(a) The joint distribution of the time-to-event, covariates, and domain membership is independent of the censoring time: $(T_i, X_i, D_i) \indep C_i$. 
		This assumption ensures that the censoring mechanism does not introduce bias related to survival times, covariates, or domain differences. (b) A less restrictive condition allows the censoring time to depend on a coarsened version of the covariates,
		\begin{align}\label{eqn:covariate-dep}
			(T_i, X_i)  \indep C_i | \wt{X}_i,
		\end{align}
		where $\wt{X}_i$ is some function of $X_i$, whose support comprising finite values. 
		This assumption permits censoring to be conditionally independent of survival and covariates given a discrete covariate representation 
		$\wt{X}_i$.
	\end{assumption}

	Our goal is to predict or estimate
	$$\tau^{(k)} (t) = E_{\sD_\sT} \{\nu^{(k)}(T_i; t)\}$$ for a target population using the data collected in the source population only. 
	For $k = 1$, $\tau^{(1)}(t)$ lies in $[0, 1]$ and corresponds to the expected survival probability, while for $k = 2$, $\tau^{(2)}(t)$ represents the expected restricted mean survival time. 
	We restrict our analysis on $t \in \Gamma$, a compacted domain with an upper bound, say $\wt{C}$. If we denote by $m^{(k)}(X; t) = E\{ \nu^{(k)}(T_i; t) |X \}$,  then $m^{(1)}(X; t)$ lies in $[0, 1]$ and $m^{(2)}(X; t) \in [0, \wt{C}]$ as $t \in \Gamma$.
	Assuming absolute continuity and under the above assumptions, the density $f( T, C, X, D)$ can be factored as $f(X) f(D | X) f(T, C | X)$. Consequently, we can express $E_{\sD_\sT}\{ \nu^{(k)} (T_i; t)\}$ in terms of $m^{(k)}(X; t)$ and the conditional probability  $X$ given  $D$, which facilitates the derivation of target population estimate from the source domain data:
	\begin{align*}
		& E_{\sD_\sT}\{ \nu^{(k)} (T_i; t) \} = E_{X \sim \sU_\sT}\{ m^{(k)}(X; t)\} = \int m^{(k)}(X; t) f(X | D = \sT ) \mu(dX) \\
		& = \int m^{(k)}(X; t)  \frac{\pr( D = \sT | X) }{\pr(D = \sS | X)}  \frac{\pr(D = \sS)}{\pr(D = \sT)} \frac{f(X) \pr(D = \sS | X)}{\pr(D = \sS)} \mu(dX). 
	\end{align*}
	
	Given the definition of propensity score $\sigma: \mathcal{X} \rightarrow [0, 1] 
	$ that 
	\begin{equation}\label{eqn:propensity}
		\sigma(X) = \pr(D = \sS | X),
	\end{equation}
	the above equation can be written as 
	\begin{equation}~\label{eqn:covariate-shift}
		E_{\sD_\sS}\left\{ \nu^{(k)}(T_i; t) \frac{1 - \sigma(X)}{\sigma(X)} \frac{\pr(D = \sS)}{\pr(D = \sT)} \right\}.
	\end{equation}
	{\color{blue} 
	}

	\section{Methodology}\label{sec:method}
	\subsection{Pseudo-observations of survival data in source population}
	Survival analysis presents the challenge of handling censored data, where the true survival time 
	$T_i$
	remains unobservable for some individuals. A widely used approach to address this issue involves constructing pseudo observations \citep{graw2009pseudo, andersen2010pseudo, jacobsen2016note} using the Jackknife method \citep{miller1974jackknife}:
	\begin{equation}\label{eqn:normal-pseudo}
		\wh{\theta}_i^{(k)}(t) = N \wh{\theta}^{(k)} (t) - (N - 1) \wh{\theta}_{-i}^{(k)} (t), 
	\end{equation}
	given $N$ samples in the source population, where $\wh{\theta}^{(k)} (t)$ is a consistent estimator for $\tau^{(k)}(t)$, and $\wh{\theta}_{-i}^{(k)} (t)$ is derived by excluding the $i$-th individual. Typically, a good choice for  $\wh{\theta}^{(k)} (t)$ is based on the Kaplan-Meier estimator, 
	$$
	\wh{S}(t) = \prod_{\wt{T}_i \le t } \left\{ 1 - \frac{d N (\wt{T}_i )} {Y(\wt{T}_i)} \right\}, 
	$$
	where $Y(s) = \sum_{i=1}^N \b1\{ \wt{T}_i \ge s \}$ is the at the risk process, and $N(s) = \sum_{i=1}^N \b1\{ \wt{T}_i \le s, \Delta_i = 1\}$ is the counting process for the event. 
	These pseudo observations serve as approximate replacements for the unobserved survival times, enabling the application of standard regression and machine learning techniques to censored survival data.
	
	\begin{remark}
		For the covariate-dependent censoring case,
		we consider the  censoring-weighted pseudo-observations~\citep{overgaard2019pseudo}  
		by replacing
		\begin{equation}\label{eqn:weight-pseudo}
			\wh{\theta}^{(k)} (t) = \frac{1}{N} \sum_{i=1}^N  \frac{\nu^{(k)} (\wt{T}_i; t)\b1\{ C_i \ge \wt{T}_i \wedge t \} }{\wh{G}(\wt{T}_i \wedge t | \wt{X}_i )}
		\end{equation} 
		in~\eqref{eqn:normal-pseudo}, where $\wt{X}_i$ corresponds to the one in~\eqref{eqn:covariate-dep} and $\wh{G}(u | \wt{X}_i)$ is some consistent estimator of the censoring survival function. 
		The modified weighted estimator, as shown in~\cite{overgaard2019pseudo}, produces unbiased and asymptotically normal parameter estimates.   
	\end{remark}
	
	To  accommodate covariate-dependent censoring, additional assumptions in line with~\cite{overgaard2019pseudo} are needed.  
	\begin{assumption}\label{ass:censoring}
		Define $H_{\wt{X}}(s) = \pr ( C \ge s, T \ge s | \wt{X} )  $
		as the conditional survival probability of $T$ and $C$, we have 
		$$
		H_{\wt{X}} (t) = \pr ( C \ge t | \wt{X}) \pr ( T \ge t | \wt{X} ) > 0
		$$
		almost surely, for any $t$ under consideration. Moreover, the covariate-dependent censoring either follows the additive hazard model such that $\Lambda_c (s | \wt{X}) = \int_0^s \wt{X}^\top dB(u)$, or the proportional hazard model 
		$
		\Lambda_c (s | \wt{X}) = \int_0^s \exp(\wt{X}^\top \alpha_0 ) d\Lambda_c^0(u), 
		$
		with some vector function of cumulative parameters $B(\cdot)$, coefficient vector $\alpha_0$, and the baseline hazard function $\Lambda_c^0(\cdot)$. In addition, the hazard $\Lambda_c(s | \wt{X})$ is twice differentiable. 
	\end{assumption}
	With Assumption~\ref{ass:censoring},~\cite{overgaard2019pseudo} provides an estimator  $\wh{G}(s | \wt{X} ) = \prod_0^{s} (1 - \wh{\Lambda}_c(\intd u | \wt{X} ))$ in~\eqref{eqn:weight-pseudo}, where $\wh{\Lambda}_c$ is the estimate for the true censoring hazard. A common choice for $\wh{\Lambda}_c$ is the Breslow estimator derived from Cox regression under the proportional hazard censoring assumption. 
	
	\ignore{
		\begin{figure}[ht!]
			\centering
			\includegraphics[width = 0.47\textwidth]{Figure/Pseudo_Surv.pdf}
			\includegraphics[width = 0.47\textwidth]{Figure/Pseudo_mean.pdf}
			\caption{Illustrative examples of empirical curve derived under the assumption of known survival time, and the curve generated by averaging pseudo observations for $1000$ sample points.
				The hazard follows Weibull distribution $\eta \nu t^{\nu - 1} \exp(X_i^\top \alpha)$ where $\eta = 0.0001, \nu = 3, \alpha = (2, 1.5, -1, 1)^\top$, and the censoring times are uniformly distributed ($U([0, 115])$). The covariates $(X_{i1}, X_{i2})$ are jointly normal with mean zero, variance $2$ and covariance $0.5$, while $X_{i3}$ and $X_{i4}$ are binary with probability $0.4$ and $0.5$, respectively. 
			}
			\label{fig:pseudo}
	\end{figure}   }
	A straightforward estimator for $\tau^{(k)}$ in the source population utilizing the pseudo observations is given  as 
	$$
	\wh{\tau}^{(k)}(t) = N^{-1}\sum_{i=1}^N  \wh{\theta}_i^{(k)}(t),
	$$
	for a specific time point $t$.
	However,  individual pseudo observations $\wh{\theta}_i^{(k)}(t)$ may exhibit considerable noise, fluctuating around the mean function $m^{(k)}(X;t)$.   This variability will be further analyzed in later sections, where we discuss its implications for estimation accuracy and model calibration.
	
	\subsection{Inverse propensity score weighting~(\textsc{ipsw}) using pseudo observations} \label{IPSW}
	
	In domain adaptation, the target sample size is typically limited, and survival outcomes are unobserved, making direct estimation of the target parameters challenging. Additionally, when the source and target domains have different marginal distributions 
	$f(X)$, the resulting covariate shift must be addressed to ensure valid inference.
	A primary approach for handling covariate shift is inverse propensity score weighting (\textsc{ipsw}), which reweights source samples to approximate the distribution of the target domain. 
	In light of~\eqref{eqn:covariate-shift}, substituting $m^{(k)}(X; t)$ with its empirical counterpart  $\wh{\theta}_i^{(k)} (t)$  using source outcomes leads to the empirical \textsc{ipsw} estimator
	of $\tau^{(k)}$ in the target population:
	\begin{equation}\label{eqn:IPSW-emp}
		\wh{\tau}^{(k)}(t) =  \frac{\sum_{i=1}^N  \wh{\theta}_i^{(k)}(t) r \frac{1 - \sigma(X_i)}{\sigma (X_i)}  }{\sum_{i=1}^N r  \frac{1- \sigma (X_i)}{\sigma (X_i)} } = \frac{\sum_{i=1}^N  \wh{\theta}_i^{(k)}(t)  \frac{1 - \sigma(X_i)}{\sigma (X_i)}   }{\sum_{i=1}^N \frac{1- \sigma (X_i)}{\sigma (X_i)} }. 
	\end{equation} 
	where $\sigma(X_i)$ is the propensity score defined in~\eqref{eqn:propensity}, and  $r$ is the ratio of marginal probabilities of belonging to each domain:
	$$
	r := \frac{\pr(D = \sS)}{\pr(D = \sT)}.  
	$$ 
	Previous works, such as~\cite{Kim22}, assume equal marginal probabilities for source and target domain membership,  $\pr(D = \sS) = \pr(D = \sT)$, implying $r = 1$. However, it is important to point out that  the ratio $r$ cancels out in the \textsc{ipsw} estimator~\eqref{eqn:IPSW-emp} and thus is not needed.  Nevertheless, to ensure the validity of our methods and theoretical guarantees, we impose the following standard assumption: 
	\begin{assumption}\label{ass:bounded-propensity}
		The propensity scores satisfy $0 < \sigma_{\sS}(X), \sigma_{\sT}(X) < 1$ for all $X \in \mathcal{X}$. 
	\end{assumption} 
	Assumption~\ref{ass:bounded-propensity} is commonly used in the literature on propensity score reweighting (e.g.,~\cite{zeng2021propensity}) and domain adaptation. It ensures that the weighting functions remain bounded, thereby preventing instability in finite-sample estimation. In addition, it implies that both domains are non-degenerate, i.e., $0 < \pr(D = \sS), \pr(D = \sT) < 1$, which is essential for meaningful comparisons and valid domain adaptation.

	In practice, the true propensity scores are unknown and must be estimated. Various techniques, ranging in complexity, can be used to approximate these scores, with logistic regression being the most commonly employed method. Formally, analysts aim to obtain an optimal estimate 
	$\wh{\sigma}_{\sT} $ from a predefined class of scoring functions 
	$\Sigma$ using unlabeled samples from both the source and target domains. 
	Define the propensity score odds function as
	$$
	w(X) := r \frac{1 - \sigma(X) }{\sigma(X)},
	$$
	the  \textsc{ipsw} estimator is then obtained by reweighting of samples from $\sD_\sS$ 
	\begin{equation}\label{eqn:IPSW}
		\tau^{(k), ps}(t; \wh{w}) = E_{\sD_\sS}\left\{ \wh{w}(X) \nu^{(k)}(T_i; t) \right\},     
	\end{equation} 
	where $\wh{w}(X)$ is 
	the empirical propensity odds   
	$$
	\wh{w}(X) = r \frac{1- \wh{\sigma} (X)}{\wh{\sigma} (X)}, 
	$$  
	
	However, it is crucial to acknowledge the potential misspecification in the estimated conditional membership probabilities, as they may deviate from the true propensity scores. This discrepancy can significantly impact the reweighting procedure and subsequent estimation. The discrepancy between any function $h(\cdot)$ and the true propensity odds  $w(X) = r( 1-\sigma(X) )/\sigma(X)$ can be measured via
	\begin{equation}\label{eqn:specific-error}
		d(h, w) = E_{\sD_\sS }\{ | h(X) - w(X) | \}. 
	\end{equation}
	When $h(X)$ takes the form of estimated propensity odds $h(X) = \wh{w}(X)$, this distance quantifies the impact of propensity score estimation errors on the final weighting scheme.  
	In survival analysis, the \textsc{ipsw} estimator in \eqref{eqn:IPSW} is not directly attainable since the true event times $T_i$
	are unobserved for censored data. Instead, we typically use an empirical estimate that incorporates pseudo observations to approximate the target estimand:
	\begin{equation}\label{eqn:emp-IPSW}
		\wh{\tau}^{(k), ps}(t; \wh{w}) = \frac{\sum_{i=1}^N  \wh{\theta}_i^{(k)}(t) \wh{w}(X_i)  }{\sum_{i=1}^N \wh{w}(X_i) } = \frac{\sum_{i=1}^N  \wh{\theta}_i^{(k)}(t)  \frac{1 - \wh{\sigma}(X_i)}{ \wh{\sigma} (X_i)}   }{\sum_{i=1}^N \frac{1- \wh{\sigma} (X_i)}{ \wh{\sigma} (X_i)} }.  
	\end{equation}

	\ignore{
		Our setup can be extended to more general cases with multiple source domains and a single target domain, as well as the setting with a single source domain and multiple target domains.
		In the setting with multiple source domains $\sS_1, \sS_2, \ldots, \sS_M$ and a singe target domain $\sT$, we define the domain-specific propensity score odds:
		$$
		w_m(X) = r_m \sigma_{\sT}(X)/\sigma_{\sS_m}(X), 
		$$
		where $r_m = \pr(D = \sS_m )/\pr(D = \sT )$,   and propensity score for each domain are 
		$$
		\sigma_{\sT}(X) = \pr( D \in \sT|X), \quad \sigma_{\sS_m }(X) = \pr( D \in \sS_m |X). 
		$$ 
		We denote $\vw = (w_1(X), w_2(X), \ldots, w_M(X))$ as the vector of weight functions.
		Given that the source domain distribution follows: 
		$$
		f(X | D ) = \sum_{m=1}^M f(X | D = \sS_m ) \b1\{D = \sS_m \}, 
		$$
		the corresponding \textsc{ipsw} estimator is: 
		\begin{equation*}
			\wh{\tau}^{(k), ps}(t; \vw)  =  \frac{\sum_{i=1}^N \sum_{m = 1}^M \b1\{D_i \in \sS_m \} \wh{\theta}_i^{(k)}(t) w_m(X_i)  }{\sum_{i=1}^N \sum_{m = 1}^M \b1\{D_i \in \sS_m \} w_m(X_i) }. 
		\end{equation*}
		If probability estimates $\pr(D = \sS_i) (i=1,\cdots, M)$ are required, they can be replaced by their empirical proportions in the observed samples.   
		
	}
	
	\subsection{Multicalibration as an alternative to \textsc{ipsw}}
	An alternative approach, often referred to as imputation, involves learning a function that maps individuals’ covariates to their responses (pseudo labels) using only the source data. The goal is to develop a function with strong generalization ability that performs competitively with the propensity score-based estimate, regardless of the target population. By doing so, we can bypass the need for target-specific propensity scores and leverage unlabeled samples from all domains. For concreteness, suppose $\wt{m}^{(k)}(X; t)$  is a learned approximation of $m^{(k)}(X; t)$, trained on source data. We define its prediction error for the target population as the absolute deviation from the true target parameter:
	\begin{equation}\label{eqn:error}
		\Err_\sT\{\wt{m}^{(k)}(X; t)\} = | E_{\sU_\sT} \wt{m}^{(k)}(X; t) - \tau^{(k)}(t) |.
	\end{equation}
	This quantity measures how well the imputed function 
	$\wt{m}^{(k)}(X; t)$
	approximates the true conditional expectation in the target domain, thereby providing a direct assessment of the imputation strategy's accuracy in adapting to domain shifts. In the context of algorithmic fairness~\citep{dwork2012fairness}, the learned function  should not only provide accurate predictions but also ensure fairness across subpopulations.
	
	We introduce the concept of ``Universal Adaptability" to formalize the objective of learning a function that achieves performance comparable to the \textsc{ipsw} estimator, while avoiding reliance on explicit propensity score estimation.  We first introduce a general notation $\nu(T;t)$ to represent the quantity of interest that depends on the survival time 
	$T$ and $t$.  Let $\tau^{ps}(t, w) = E_{\sD_\sT}[ w(X) \nu(T; t)]$ as its \textsc{ipsw} estimator.  Following the definition of \cite{Kim22}, consider a set of conditional membership probability $\wt{\sigma} \in \Sigma$, and define
	\begin{equation}\label{eqn:odds}
		\mathcal{H}(\Sigma) = \{ \wt{w}(\cdot): = r\{ 1- \wt{\sigma}(\cdot)\}/\wt{\sigma}(\cdot); \wt{\sigma} \in \Sigma \}.   
	\end{equation} 
	\begin{definition}[Universal adaptability]
		For a set of propensity scores $\Sigma$ and a source domain distribution  $\sD_{\sS}$, a predictor $\wt{m}(X; t)$ is said to be $(\Sigma, \alpha)$-universally adaptable if 
		$$
		\Err_\sT \{\wt{m}(X; t)\} \le \Err_{\sT} \{\tau^{ ps}(t; \wh{w})\} + \alpha, 
		$$
		where $\wh{w}  = \argmin_{h \in \sH(\Sigma)} d(h, w) $ is the best-fit approximation of $w$ within the class $\mathcal{H}(\Sigma)$, and $d(h, w)$ is defined as \eqref{eqn:specific-error}.
	\end{definition} 
	
	Universal adaptability characterizes a well-learned predictor $\wt{m}$ to be no worse than an \textsc{ipsw} estimator within a bias threshold $\alpha$. This is a strong notion since accurate prediction on the source domain does not guarantee effective generalization on the target in general. 
	
	Next, we focus on the conditions and procedures necessary to derive 
	$\wt{m}$ that achieves universal adaptability.
	To this end, we introduce the concept of multicalibration \citep{johnson18a, kim2019multiaccuracy}, a key property of prediction functions that was originally studied in the context of algorithmic fairness.
	Multicalibration plays a crucial role in achieving universal adaptability, as it enforces consistency between predictions and true outcomes across diverse groups, ensuring robustness under distributional shifts. 
	
	We formally define multicalibration and demonstrate how it enables the construction of universally adaptable predictors for censored survival data.
	The definition of multicalibration involves a collection of real-valued functions $\mathcal{H}$, ensuring unbiased prediction across every weighted subpopulation defined by $h \in \mathcal{H}$. 
	\begin{definition}[Multicalibration]\label{def:multicali}
		For a given distribution $\sD$ and a class $\mathcal{H}$, a hypothesis~(predictor) $\wt{m}(\cdot; t): \mathcal{X} \rightarrow [0, \wt{C}] $ is $(\mathcal{H}, \alpha)$-multicalibrated, if 
		$$
		\left| E_{\sD} \left[ h(X) \left\{\wt{m}(\cdot; t) - \nu(T; t) \right\} \right] \right| \le \alpha, 
		$$
		for all $h(X) \in \mathcal{H}.$
	\end{definition}
	
	The merits of defining multicalibration in this way are two folds: First, it can be viewed as a generalization of the original definition in~\cite{johnson18a} where the subpopulations of interest are defined in terms of a class of Boolean functions. For instance, $h$ can take form  like $h(X) = c(X)\b1\{X \in G\}$ that combines some function $c(X)$ and subpopulation indicator $\b1\{X \in G\}$, or $h(X) = h(X, m(X) )$ that depends on function values, for more general purposes.  Second, it allows for developing efficient black-box algorithms that agnostically identify function $h \in \mathcal{H}$ by auditing via regression, under which the current prediction violates multicalibration. As such, we do not have to examine all group combinations. If $\mathcal{G}$ is  the collection of disjoint subpopulations of interest, it will be inefficient to take $2^{|\mathcal{G}|}$ inspections by permutating all the combinations.
	
	\subsection{Multicalibration for survival data using pseudo observations}
	A boosting-style algorithm, \texttt{MCBoost}, is publicly available and can generate multicalibrated functions using a small portion of labeled data \citep{pfisterer2021}. However, existing implementations of this algorithm, following \cite{johnson18a, kim2019multiaccuracy}, are primarily designed for classification tasks.
	
	Due to  censoring of true survival outcomes, we construct time-varying pseudo observations via the Jackknife method, aiming to use them as surrogate labels or responses. However, pseudo observations introduce additional variability: their behavior differs between event and censored data, and they are not necessarily bounded.
	For example, in survival probability estimation, pseudo observations may exceed 1 or fall below 0 for certain individuals. Additionally, for each individual, the pseudo observation curve as a function of 
	$t$ may exhibit non-monotonic behavior.  For further insights into the behavior of pseudo observation curves, we refer to Figure 3 of \cite{andersen2010pseudo}.
	
	Unlike classification tasks, where multicalibration is typically applied, our setting is better framed as a regression problem. This distinction is crucial for adapting multicalibration techniques to censored survival data, where predictions involve estimating continuous survival-related quantities rather than discrete class labels.
\begin{algorithm}
	\caption{Boosting for calibrating the censored data}
	\label{alg:MCboost} 
	\KwData{Initial estimator: $m^{(k), 0}(X;t). $\\
		Accuracy parameter $\alpha > 0$; stepsize $\eta$. \\
		Auditing algorithm $\mathcal{A}$. \\
		Calibration set $D = \{ ( X_i, \wt{T}_i, \Delta_i ) \}, i = 1, \ldots, N_1$. \\
		Validation set $V = \{ ( X_i, \wt{T}_i, \Delta_i ) \}, i = N_1 +1, \ldots, N_1 + N_2.$
	}
	\DontPrintSemicolon
	\For{ $b = 0, \ldots, B$ }{
		Compute buckets $\{S_l\}_{l= 1, \ldots, L}$, $S_l : = \left\{ x \in \mathcal{X}: m^{(k),b}(X; t) \in \left[\frac{(l-1)\wt{C}}{L} , \frac{l\wt{C}}{L} \right] \right\}.$ \;
		$h_{b, S_l}(X) \leftarrow \mathcal{A}\left(D, \left(m^{(k),b}(X_i; t) - \widehat{\theta}_i^{(k)} (t) \right)_{S_l}  \right).$ \;
		$S^{*}  \leftarrow \argmax_{S_l} \frac{1}{|V|} \left| \sum_{i \in V} h_{b, S_l}(X_i) \cdot \left( m^{(k), b}(X_i; t) - \widehat{\theta}_i^{(k)} (t) \right) \right|. $    
		\;
		$\Delta \leftarrow  \frac{1}{|V|} \left| \sum_{O_i \in V} h_{b, S^* }(X_i) \cdot \left( m^{(k), b}(X_i; t) - \widehat{\theta}_i^{(k)} (t) \right) \right|. $ \;
		\If{$\Delta > \alpha$}{
			$m^{(k), b+1}(X; t) = m^{(k), b}(X; t) - \eta \cdot h_{b, S^* }(X) $. 
		}
		\Else{
			\Return{$\wt{m}^{(k)}(X; t) = m^{(k), b}(X; t)$. }
		}
	}
\end{algorithm}
	
	We now propose a modified multicalibration  Algorithm~\ref{alg:MCboost} tailored for censored survival data, building on gradient boosting \citep{efron2021computer}. Our approach calibrates predictions against pseudo observations, leveraging their range to estimate the unknown upper bound 
	$\wt{C}$.  
	This  multicalibration algorithm is built upon gradient boosting, an ensemble method designed to reduce bias by iteratively refining weak learners into a strong predictor. In regression, gradient boosting operates by fitting a function to the residuals using a specified machine learning method and updating the model iteratively by adding the fitted function at each stage. 
	Our multicalibration boosting extends this idea by incorporating a group-wise boosting mechanism that reduces bias across different subpopulations. Instead of merely fitting residuals globally, our approach applies a specialized learning algorithm 
	$\mathcal{A}$  within subgroups, ensuring that predictions remain well-calibrated across structured populations. 
	
	Specifically, at each iteration, the algorithm partitions the observed data into 
	$m$ buckets based on the values of the previously learned function 
	$m^{(k), b}(\cdot; t)$.  An auditing function 
	$\mathcal{A}$ via  ridge regression or decision trees is then applied to each bucket to check whether the predictions satisfy the multicalibration criterion. 
	Alternatively, the buckets can be fixed based on the initial function 
	$m^{(k),0}(\cdot; t)$, aligning with the multiaccuracy approach proposed in \cite{kim2019multiaccuracy}.
	Specifically, our algorithm iteratively searches a function $h \in \sH$ using $\mathcal{A}$  for some $\sH$, on which the current predictor is mis-calibrated, and then refine the predictor. The search for mis-calibration is then reduced to a regression task over a class $\sH$.

	While specifying the audit algorithms is relatively straightforward, characterizing the function class $\mathcal{H}$ is more challenging. Given a hypothesis class  $\mathcal{H}$, it is impractical to exhaustively search through all functions in 
	$\mathcal{H}$ to verify the multicalibration condition.
	Although we may not have precise knowledge of the richness or exact structure of 
	$\sH$ as captured by a given auditing function $\mathcal{A}$,  the agnostic nature of our approach allows 
	$\mathcal{A}$ to implicitly identify a relevant function class through the algorithm’s stopping criteria. This flexibility ensures that, even if 
	$\mathcal{A}$ cannot fully characterize 
	$\sH$, there is  sufficient overlap to effectively reduce bias across subpopulations, provided that 
	$\mathcal{A}$ is well chosen. Thus, our method strikes a balance between the practical limitations of explicitly characterizing a function class and the objective of achieving efficient and robust multicalibration. By leveraging the adaptability of the auditing procedure, we enable systematic bias correction across subgroups while maintaining computational feasibility.

	\section{Theoretical Analysis}\label{sec:theory}
	\subsection{Property of multicalibration}
	
	We first introduce some notation. We use $a_n = O(b_n)$ to state there exists a constant $C > 0$ and an integer $n_0 > 0$ such that for all $n \ge n_0$, $|a_n| \le C |b_n|$. Similarly, $a_n = \Omega(b_n)$ represents there exists a constant $C' > 0$ and an integer $n_0 > 0$ such that for all $n \ge n_0$, $|a_n| \ge C' |b_n|$. We say $a_n = \Theta(b_n)$ if $a_n = O(b_n)$ and $a_n = \Omega(b_n)$.

	Ensuring that our algorithm satisfies the multicalibration guarantee, as defined in Definition~\ref{def:multicali}, is crucial. This definition involves an auditing-related collection of functions, denoted as 
	$\sH$,  whose richness significantly influences the calibration quality. To facilitate theoretical analysis, we impose the following assumption:
	\begin{assumption}\label{ass: finite}
		The auditing algorithm $\mathcal{A}$ agnostically learns a symmetric class $\sH$ such that $ |h(\cdot)| \le C_{\sH}$ for some positive constant $C_{\sH} > 0 $, for every $h \in \sH$.
	\end{assumption} 
	To be more specific, it states that $\mathcal{A}$ is capable of identifying a function class $\mathcal{H}$ such that all of its functions satisfy multicalibration empirically, i.e.,
	$$
	\frac{1}{|V|} \left| \sum_{ O_i \in V} h(X_i) \left\{ m^{(k), b}(X_i,t) - \wh{\theta}_i^{(k)}(t) \right\} \right|  < \alpha,
	$$
	for every $h(X) \in \mathcal{H}$ in Algorithm~\ref{alg:MCboost}, where $\alpha$ is the accuracy parameter, and $V$ is the validation set. 
	This assumption aligns with findings in the literature, where the choice and richness of 
	$\sH$ are crucial  in achieving effective multicalibration. For instance, \cite{johnson18a} demonstrate that a richer function class allows for more nuanced adjustments, leading to better calibration across diverse subpopulations.

	To derive a uniform error bound, we analyze the pseudo-observations through the lens of functional analysis, using a von Mises expansion~\citep{vaart_1998, graw2009pseudo,overgaard2017asymptotic}, 
	$$
	\wh{\theta}_i^{(k)} (t) = \tau^{(k)} (t) + \dot{\phi}^{(k)}(\delta_{O_i} - F; t) + (N-1)^{-1} \sum_{j \neq i} \ddot{\phi}^{(k)} (\delta_{O_i} - F, \delta_{O_j} - F; t) + R_{N, i}^{(k)}(t),
	$$ 
	where $\tau^{(k)}(t)$ is the estimand, and $R_{N, i}^{(k)}(t)$ denotes a negligible remainder. The $\dot{\phi}^{(k)},  \ddot{\phi}^{(k)} $ are the first and second derivatives respectively, evaluated on triplets $\delta_{O_i} = (Y_i, N_{i,0}, N_{i, 1})$, where $Y_i (s) = \b1\{\wt{T}_i \ge s\}$, $N_{i, 0}(s) = \b1\{\wt{T}_i \le s, \Delta_i = 0\}$ and $N_{i, 1}(s) = \b1\{\wt{T}_i \le s , \Delta_i = 1 \}$. Here, $F$ denotes the limit of the empirical process $F_N = N^{-1}\sum_{i=1}^N \delta_{O_i} $.  Full details are provided in Section~\ref{apxsubsubsec:pseudo-bound} of the Supplementary Material.
	
	To ensure both smoothness of the functionals and control over convergence rates, we work within a function space equipped with the $p$-variation norm~\citep{dudley2011concrete}, which balances regularity and convergence. The functional delta method can then be applied to functions that reside in a product Banach space. Further discussion is included in the Supplementary Material.

	Given $t \in \Gamma$ bounded by $\wt{C}$, we impose a uniform Lipschitz condition on the functional derivatives. 
	\begin{assumption}\label{ass:uniform}
		For all $t \in \Gamma$, the linear continuous first-order functional derivative and the bilinear continuous second-order derivative have a uniform Lipschitz constant in the sense that 
		$$|\dot{\phi}^{(k)}(F; t) | \le K_t \|F\|_{[p]} \quad \mbox{ and } \quad |\ddot{\phi}^{(k)}(F_1, F_2; t) | \le K_t \|F_1\|_{[p]} \|F_2\|_{[p]}, 
		$$ 
		with $K_t \le K$ for some universal constant $K$, where $F, F_1, F_2$ are some distribution functions, and $\|F\|_{[p]}$ is the $p$-variation norm of $F$. 
	\end{assumption}
	This assumption strengthens the usual smoothness conditions by a uniform bound across $t \in \Gamma$. 
	While the continuity of the derivatives aligns with the Proposition 3.1 of~\cite{overgaard2017asymptotic}  and Condition 2 of~\cite{overgaard2019pseudo}, we additionally impose $K_t \le K$ to ensure uniform convergence over $t$, as learning rate constants may vary with $t$ unlike in classical regression. 
	
	The following theorem establishes a uniform bound on the deviation between the empirical estimate and the true value of the multicalibration statistics.
	\begin{theorem}\label{thm: algo-multi}
		Under Assumptions~\ref{ass:domain}, consider Algorithm~\ref{alg:MCboost} that multicalibrates predictor against pseudo observations $\wh{\theta}_i^{(k)}(t)$, which are derived according to ~\eqref{eqn:normal-pseudo} under Assumption~\ref{ass:indep}(a), or according to ~\eqref{eqn:weight-pseudo} under Assumptions~\ref{ass:indep}(b) and~\ref{ass:censoring}. Further, suppose $\mathcal{A}$ agnostically learns a class $\mathcal{H}$ satisfying Assumption~\ref{ass: finite}, with $\varepsilon$-covering number $\mathcal{N}_{\varepsilon} = \mathcal{N}(\varepsilon, \mathcal{H}, \|\cdot\|_{L_1})$.
		Then given $\delta > 0$, fixed time point $t \in \Gamma$, and $N$ random samples, and any function $ \check{m}^{(k)} (\cdot; t) \in [0, \wt{C}]$, it follows with probability at least  $1 - \delta$ that:
		\begin{equation}\label{eqn:uniform}
			\begin{aligned}
				&    \sup_{h \in \mathcal{H}} \left[  E_{\sU_\sS}\left\{ h(X) (\check{m}^{(k)}(X; t) - m^{(k)}(X;t) ) \right\}  -   \frac{\sum_{i=1}^N   h(X_i) (\check{m}^{(k)}(X_i; t) - \wh{\theta}_i^{(k)}(t)) }{ N  } \right] \\
				&   \qquad \le    2\wt{C}C_{\sH}  \left\{\frac{\log(2\sN_{\varepsilon})/\delta}{N} \right\}^{1/2} + \wt{C} \epsilon + C_p \left\{ \frac{\log(1/\delta)}{N^{1/p}} + N^{(1-p)/p} \right\} 
			\end{aligned}
		\end{equation}
		for some $p \in (1, 2)$.  
	\end{theorem} 
	The detailed proof of Theorem~\ref{thm: algo-multi} is deferred to Section~\ref{apxsec:proofs} of the Supplementary Material. The sample size $N$ ought to be large enough so that the empirical average is close enough to the mean.
	From  Theorem~\ref{thm: algo-multi}, we conclude
	\begin{corollary}[Sample Complexity] \label{coro:complexity}
		Under the same conditions outlined in Theorem~\ref{thm: algo-multi}, to 
		guarantee a uniform convergence over $\mathcal{H}$ with error~\eqref{eqn:uniform} at most $O(\alpha)$ with failure probability $\delta$, the sample size is required to be $N = \Omega\left\{ \log(2\mathcal{N}_{\Theta(\alpha)} /\delta)/\alpha^2 + (1/\alpha)^{1/\lambda } \right\}$, where $\lambda \in [1/4, 1/2).$
	\end{corollary}
	In the above Corollary,  $\lambda \in [1/4, 1/2)$ can be obtained if we choose $p \in [4/3, 2)$.  A higher sample complexity than standard classification and regression tasks is obtained due to the additional noise from the pseudo observations that potentially decelerates the convergence rate.
	
	The complexity of the function class $\mathcal{H}$ plays an important role in determining sample complexity. Consider $\sH = \{ h_{\theta}(X), \theta \in \Theta \subset \mathbb{R}^d \}$, a class of functions satisfying Lipschitz condition and characterized by a compact parameter space $\Theta$, which is often the case when we consider linear and logistic functions. In this scenario, the covering number $\sN_{\varepsilon} \lesssim (1 + 2\text{diam}(\Theta) /\varepsilon)^d$. Here, the complexity is dominated by $\alpha^{-1/\lambda}$, rendering it slower than typical regression. On the other hand, if we consider $\sH = \mathcal{C}([a,b]^d, B)$, a class of bounded convex functions defined over a compact domain, we have $\log \mathcal{N}_{\varepsilon} \lesssim  \varepsilon^{-d/2} $, and $\log(2\mathcal{N}_{\Theta(\alpha)} /\delta)/\alpha^2$ becomes the dominating factor (see the Supplementary Material for detailed discussion).  
	
	We further show that our Algorithm~\ref{alg:MCboost}  converges in finite iterations. The proof follows by showing that for an appropriately chosen step size $\eta$, every iteration improves the squared loss. Given the squared loss is lower bounded by $0$ and upper bounded by the initial squared loss, the algorithm shall converge in a bounded number of updates. 
	\begin{theorem}[Convergence Analysis]\label{thm:convergence}
		Suppose $\mathcal{A}$ agnostically learns a class $\mathcal{H}$ satisfying Assumption~\ref{ass: finite}. Let bias $\alpha > 0$ and $\delta > 0$, then with probability at least $1 - \delta$, Algorithm~\ref{alg:MCboost} converges to a $(\mathcal{H}, \alpha')$-multicalibrate predictor $\wt{m}^{(k)}(X; t)$ in $B = O\left( \ell_{\sD_{\sS}}\{\wt{m}^{k,0}(\cdot; t)\}/\alpha^2 \right)$ iterations at time point $t \in \Gamma$, from $N = \Omega\left[ \{\log(2\sN_{\Theta(\alpha)}/\delta ) + \log(B) \}/\alpha^2 + (1/\alpha)^{1/\lambda} \right] $ samples, where $\alpha' = O(\alpha)$, and $\lambda \in [1/4, 1/2)$.
	\end{theorem}
	
	\subsection{Universal adaptability}
	The essence of multicalibration boosting lies in its ability to enhance predictive accuracy across diverse subpopulations by iteratively refining residuals, thereby improving statistical estimates within the target domain. To formally elucidate how multicalibration leads to precise predictions under covariate shift, we relate the prediction error obtained using multicalibration to that of inverse propensity score weighting (\textsc{ipsw}). Specifically, under appropriate conditions, a multicalibrated function $\tilde{m}^{(k)}(\cdot; t)$ exhibits universal adaptability. Using the function class agnostically learned by the algorithm  $\mathcal{H}$  and  its corresponding odds class $\mathcal{H}(\Sigma)$ 
	defined in~\eqref{eqn:odds} and the specification error $d(h, w)$ defined in~\eqref{eqn:specific-error}, we  establish the following  universal adaptability from multicalibration.
	\begin{theorem}\label{thm:multicali}
		Under the same conditions outlined in Theorem~\ref{thm: algo-multi}, 
		suppose $\wt{m}^{(k)}(\cdot;t): \mathcal{X} \rightarrow [0, \wt{C}]$ is $(\mathcal{H}, \alpha')$-multicalibrated over the source distribution $\sD_\sS$, 
		then for any target  distribution $\sD_\sT$, and any $\wt{w} \in \Sigma$,  $\wt{m}^{(k)}(X;t)$ satisfies 
		$$
		\Err_{\sT}\{\wt{m}^{(k)}(X; t)\} \le \Err_{\sT}\{\tau^{(k), ps}(t; \wt{w})\} + \wt{C} \inf_{h\in\sH} \{d(h, w) + d(h, \wt{w})\} + \alpha'.   
		$$
		If $\mathcal{H} = \mathcal{H}(\Sigma) $ as defined in~\eqref{eqn:odds}, $\wt{m}^{(k)}(\cdot;t)$ is $(\Sigma, \wt{C}d( \wh{w}, w) + \alpha')$ universal adaptable. Moreover, if $\wt{m}^{(k)}(\cdot; t)$ is $(\mathcal{H}(\Sigma)\otimes \mathcal{C}, \alpha')$-multicalibrated on the source domain, for some function class $\mathcal{C} \subset \{ \mathcal{X} \rightarrow [0, \wt{C}]\}$, $\wt{m}^{(k)}(\cdot; t)$ is also $(\mathcal{C}, \wt{C}^2\inf_{\wt{w} \in \mathcal{H}(\Sigma)} d(w, \wt{w}) + \alpha')$-multicalibrated on the target domain. 
	\end{theorem} 
	Theorem~\ref{thm:multicali} demonstrates that achieving multicalibration in the source domain not only ensures universal adaptability but also preserves a certain degree of multicalibration in the target domain. This result can be interpreted as providing an adaptive guarantee under varying degrees of distribution shift. 
	If $\wt{m}^{(k)}$ is well-calibrated on the source domain, it retains its multicalibration guarantee on target domains that are close in distribution. A simple case is when 
	$\mathcal{H}(\Sigma) = \{ \mbox{Id}(x)\}$ in a setting with no distribution shift.
	In contrast,  if the shift is substantial and requires leveraging the full expressiveness of 
	$\mathcal{H}(\Sigma)$ to correct for distributional changes, the universal adaptability and calibration guarantees deteriorate -
	unless $\Sigma$ is well specified and the auditing algorithm $\mathcal{A}$ can effectively learn $\mathcal{H}(\Sigma)$ or $\mathcal{H}(\Sigma) \otimes \mathcal{C} $ agnostically. When $\Sigma$ is accurately specified,  $w \in \Sigma$, the discrepancy $d$ diminishes, and the multicalibrated estimator is nearly unbiased.

	Additionally, the function class $\mathcal{H}$ may encompass a broader family of functions, including those that depend on both the covariates and the function values such as $h\{X, m^{(k)}(X; t)\}$. In particular, if the learned class $\mathcal{H}$ is sufficiently expressive to include propensity score odds functions
	$w \in \mathcal{H}(\Sigma)$ and function 
	approximations $p \in \mathcal{P} = \{p: \mathcal{X} \mapsto [0, \wt{C}]\}$, then we can establish a connection between multicalibration and the $L_2$ prediction error $E_{\sU_\sT} \{\wt{m}^{(k)}(X;t) - m^{(k)}(X;t)\}^2$, summarized in the following corollary. 
	
	\begin{corollary}\label{col:l2}
		Suppose $w(\cdot) \in \sH(\Sigma)$ with $ |w| \le \wt{C}$ and our algorithm reaches 
		$$
		\sup_{ \wt{w} \in \sH(\Sigma), p \in \mathcal{P}} \left|E_{\sD_{\sS}} \wt{w}(X)\{\wt{m}^{(k)}(X;t )  - p(X) \} \{\wt{m}^{(k)}(X;t ) - m^{(k)}(X; t) \} \right| < \alpha,
		$$
		within $B = O(1/\alpha^2)$ iterations, where $p \in \mathcal{P}$ approximates the true $m^{(k)}(X; t)$, then 
		$$
		E_{\sU_\sT} \{\wt{m}^{(k)}(X;t) - m^{(k)}(X;t)\}^2 \le \alpha + \wt{C}^2  \inf_{ \wt{w} \in \sH(\Sigma), p \in \mathcal{P}} (\|w - \wt{w}\|_{L_2} +  \|m^{(k)} - p\|_{L_2}).  
		$$
	\end{corollary}

	\ignore{
		Our method extends naturally to multiple source domains, leading to an upper bound for the error of $\wt{m}^{(k)}(X; t)$ under the multi-source setting. 
		\begin{theorem}\label{thm:multi-source}
			Under the same conditions as in the previous theorems, consider a setting with multiple source domains $\sS_m, m = 1, 2, \ldots, M$, and a single target $\sT$. Let $\wt{\vw}(X) = (\wt{w}_1(X), \wt{w}_2(X), \ldots, \wt{w}_M(X))$ be a set of weight functions associated with the propensity scores  $ (\wt{\sigma}_{\sT}, \wt{\sigma}_{\sS_1}, \ldots, \wt{\sigma}_{\sS_M}). $ 
			The function  $\wt{m}^{(k)}(X; t)$, produced by the proposed algorithm, satisfies the ``averaged'' muticalibration condition: 
			\begin{equation*}
				\sup_{h \in \mathcal{H}} \left| \sum_{m=1}^M \pr(D = \sS_m | D \in S) \Expect_{\sU_{\sS_m} }\left\{ h(X)  (\wt{m}^{(k)}(X; t) - m^{(k)}(X;t) ) \right\} \right| < \alpha,
			\end{equation*} 
			Furthermore, the target domain error is bounded by 
			\begin{align*}
				\Err_{\sT}(\wt{m}^{(k)}(X; t)) & \le \Err_{\sT}(\tau^{(k), ps}(t; \wt{\vw})) \\
				& \quad  + \alpha + \wt{C} \inf_{h\in\sH} \left[ \sum_{m=1}^M  \pr(D = \sS_m | D \in S) \{ d_{\sS_m}(w_m, h) + d_{\sS_m}(h, \wt{w}_m ) \} \right],
			\end{align*} 
			where the distance metric $d_{\sS_m}$ is defined as $d_{\sS_m}(h_1, h_2): = \Expect_{\sS_m}[|h_1(X) - h_2(X)|]$. 
		\end{theorem}
		The bound provides useful insights: the performance of $\wt{m}^k $ closely matches that of the pseudo-weighted estimator, up to a weighted average of propensity score discrepancies. This supports the use of multicalibration as an adaptive approach for learning under shift across multiple source populations.  
		
	}

	\section{Simulation valuations}\label{sec:simu}
	\subsection{Data generation and comparisons}We perform extensive simulations to evaluate the proposed methods and algorithms and to compare the results based on \textsc{ipsw}. Specifically, we consider the following  models for generating the data: 
	\begin{enumerate}
		\item A Weibull hazards model $\lambda(t|X_i) = \eta \nu t^{\nu - 1}\exp(X_i^\top \alpha)$ is used the model the  failure time.  A model that the failure time violates the proportional hazards assumption,  $\log(T_i) \sim \mathcal{N}(\mu, \sigma^2 = 0.64)$ with $\mu = 3.5 - X_i^\top \alpha$ is also considered. 
		\item Two censoring mechanisms  are considered, including a complete independent censoring with $C_i \sim \text{Unif}(0, 120)$, and covariate dependent censoring, with $C_i$ following a Weibull model $\lambda^c(t | X_i) = \eta_c \nu_c t^{\nu_c - 1}\exp(X_i^\top \alpha_c). $
	\end{enumerate}
	In our simulations, we  specify $\eta = 0.0001, \nu = 3, \alpha = (0, 2, 1, -1.2, 0.8)^\top$, and $\eta_c = 0.0001, \nu_c = 2.7, \alpha_c = (1, 0.5, -0.5, -0.5)^\top$. 
	We generate 5-dimensional covariates $X_i = (1, X_{i1}, X_{i2}, X_{i3}, X_{i4})^\top $, where $(X_{i1}, X_{i2})^\top$ are drawn from a mean zero bivariate normal distribution with correlation $1/4$ and variance $2$, $X_{i3} \sim  \textrm{Binomial}(0.4)$, and $X_{i4} \sim \textrm{Binomial}(0.1X_{i3} + 0.2)$. The unbalanced categorical variables, mimicking practical variables like gender and race, anticipate a 40\% female population, with minority proportions of 20\% among females and 30\% among males.

	To simulate adaptation from source to target domain, we initially consider a single source and target domain, setting the true membership probability odd as 
	$$\log\{\sigma (X_i)/(1 - \sigma(X_i)) \} = X_i^\top \omega, $$
	where $\omega = (0, 0.5, 0.45, -0.9, -0.7)^\top$. For scenario with two source domains, this is extended to 
	$$
	\log \left\{ \frac{\pr(D =\sS_j | X_i)}{\pr(D =\sT | X_i)}  \right\} =  X_i^\top \omega_j, 
	$$
	where $\omega_1 = (0, 0.5, 0.45, -0.9, -0.7)^\top$ and $\omega_2 = (0, 0.4, 0.9, -0.5, -0.9)^\top$. We further consider $$\frac{ \pr^{(q)}(D = \sS | X ) }{\pr^{(q)}(D = \sT | X ) }  = \left\{ \frac{ \sigma (X) }{ 1- \sigma(X) } \right\}^q $$ for $q =  1, 2, 3$, where $q$ determines the degree of covariate shift with $q = 3$ signifying  a large covariate shift. 
	
	In our simulations, we generate a total of 
	1000 samples, with 
	600 drawn from the source domain(s) and 
	400 from the target domain. Under single source domain setting, all 
	600 source samples are split into 
	400 for training, 
	100 for calibration, and 
	100 for validation. Under the setting of two source domains, each has
	300 samples, while the target domain still has 
	400 samples.

	Several methods with details in Table~\ref{method.comp} are compared. The pseudo observation mean in the target domain, serving as a benchmark for comparing all methods in terms of absolute bias $\Err_{\sT} = |E_{\sU_{\sT}} \wt{m}^{(k)}(X; t) - \tau^{(k)} (t) |$ or relative bias $\Err_{\sT}/|\tau^{(k)} (t)|.$
	\begin{table}
		\caption{Methods and their descriptions used in  simulation comparisons }\label{method.comp}
		\begin{tabular}{ll}
			\hline
			naive& Unweighted mean of pseudo-observations from the source domain.\\   
			\textsc{\textsc{ipsw}} & \textsc{ipsw}  weighted mean of pseudo observations from
			the source domain.\\
			&The propensity scores are learned using logistic regression. \\
			\textsc{ipsw-sub} & Subgroup-specific \textsc{ipsw} estimates, each subgroup's propensity score \\
			&learned via logistic regression.  \\
			\textsc{lm} & Naive linear model without calibration fitted for the source domain, \\
			&with the target domain estimate derived from the average prediction. \\
			\textsc{mclm-ridge} & The linear model is post-processed on the auditing set of the source \\
			&using ridge regression; estimates are from the average of \\
			&ridge-multi-calibrated linear predictions. \\
			\textsc{mclm-tree} & Similar post-processing of the linear model \\
			&using a decision tree; estimates are from the average of\\
			&tree-multi-calibrated linear predictions.  \\
			\textsc{rf}& Random forest~(RF) trained on the source domain, with estimates 
			\\
			&from averaged RF predictions over unlabeled target domain samples.\\
			\textsc{mcrf-ridge}& Post-processing of the RF on the source's auditing set using ridge \\
			&regression. The average of ridge-multi-calibrated RF predictions \\
			&gives the estimate. \\
			\textsc{mcrf-tree} &Similar post-processing of RF using a decision tree. The average of \\
			&tree-multi-calibrated RF predictions give the estimate. \\
			\hline
		\end{tabular}
	\end{table}
	
	\begin{figure}[htbp] 
		\centering
		\includegraphics[width=0.9\textwidth, height=0.8\textheight]{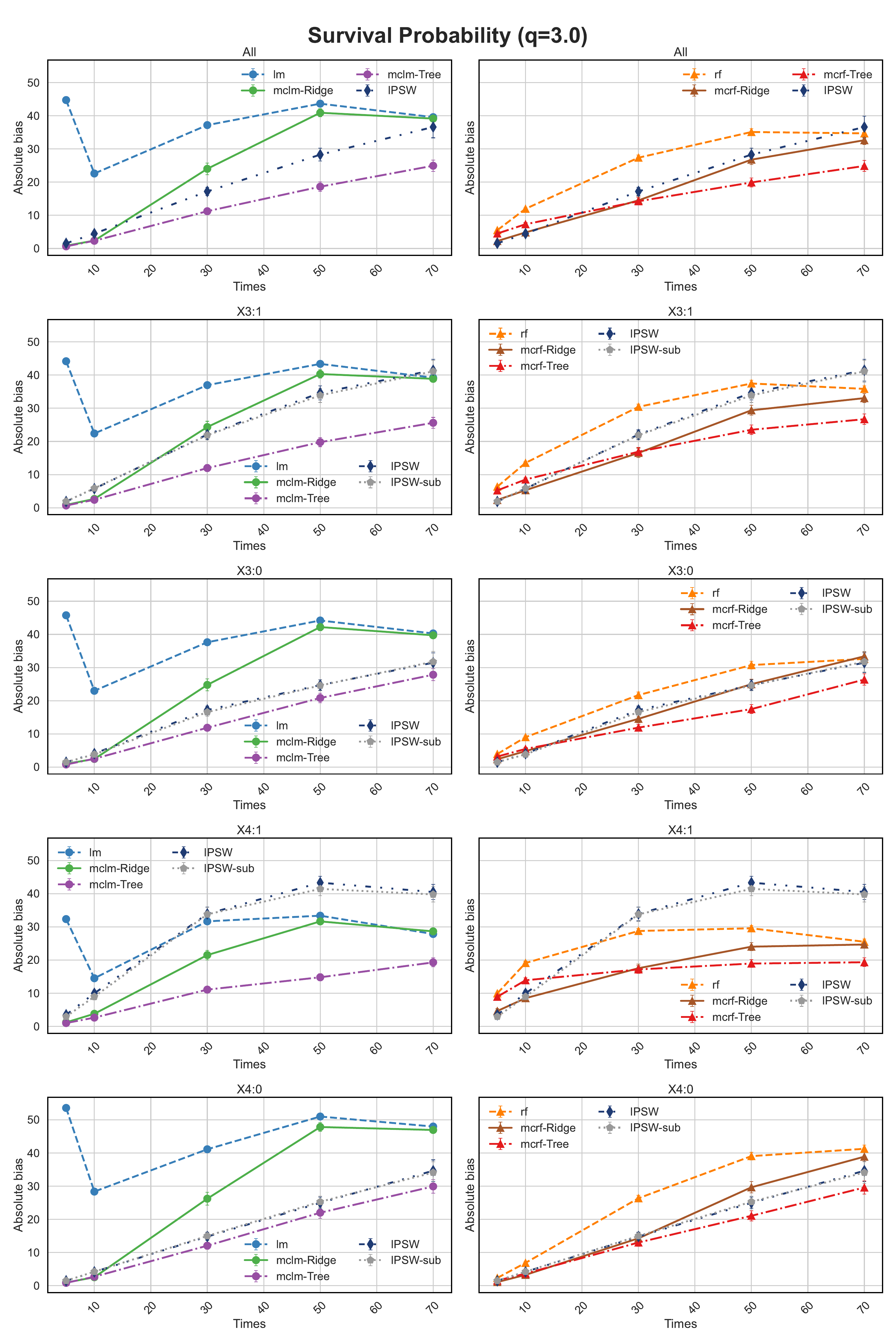}
		\caption{The absolute bias $(\times 10^2)$ for different  methods in estimating survival probability
			under strong covariate shift ($q = 3$) with a total sample size of $N = 1000$ (600 source and 400 target samples). The survival time is generated based on  the proportional hazard model, and the censoring is covariate-independent. Results for all subpopulations are presented based on $100$ simulation replications. Left panel: linear model is used to estimate the survival probability using pseudo observations. Right panel: random forest is used to estimate the survival probability using pseudo observations.}
		\label{fig:SP-PH-indep-3}
	\end{figure}
	
	\subsection{Estimation bias from covariate shift}
	Due to covariate shift, naive estimates of survival probabilities or restricted mean survival time, derived directly from the source population's mean, exhibit significant bias when compared to the target population's mean. This bias arises because unweighted samples from the source do not accurately represent the target domain under the covariate shift, as illustrated in Figures~\ref{fig:IPSW-SP-PH-indep} and~\ref{fig:IPSW-RM-PH-indep} in the Supplementary Material for estimating the survival probability and restricted mean survival time, respectively. The \textsc{ipsw} method substantially reduces this bias, achieving an overall low bias. Estimating propensity scores within each subpopulation, referred to as \textsc{IPSW-sub}, can effectively reduce biases compared to standard \textsc{IPSW}, especially under mild covariate shifts. This improvement results from the method's focused attention on subpopulations, leading to better covariate balance within each group. The enhancement is particularly notable when subpopulations exhibit distinct shift patterns.

	Under a mild or moderate covariate shift with 
	$q=1$ and $q=2$,   the proposed post-processed multicalibrated estimator of survival probability demonstrate comparable or better overall performance than the \textsc{ipsw} method (Figure~\ref{fig:SP-PH-indep-1} and Figure~\ref{fig:SP-PH-indep-2}). 
	Our calibrated estimators demonstrate substantial improvements over the standard  methods, such as linear models and random forests, by mitigating group-wise differences and achieving competitive or lower biases across most time points. 
	
	Under a strong covariate shift with $q=3$,  the absolute biases of survival probability tend to increase at later time points due to censoring (Figure~\ref{fig:SP-PH-indep-3}). Under such conditions, \textsc{ipsw} methods and standard machine learning methods fail to provide accurate estimates in the target domain, whereas the multicalibrated approach performs well and maintains fairness across subpopulations.
	While we observe some slight differences between calibration using ridge regression and decision trees, no consistent trend is observed. Post-processing with ridge regression typically requires a smaller step size and fewer iterations to converge, whereas decision trees require a larger step size and more iterations.
	For simplicity, we use a fixed step size of 
	$\eta=0.3$ for all methods in Figure~\ref{fig:SP-PH-indep-3}.
	Additionally, we observe that carefully tuning the algorithm’s hyperparameters can further improve prediction accuracy.
	
	We observe similar results for the prediction of restricted mean survival time in the target domain, as summarized in Figures~\ref{fig:RM-PH-indep-1}-\ref{fig:RM-PH-indep-3}. Due to a significant increase in absolute error as 
	$t$ increases, we present the relative error for clarity. The multicalibrated approaches learn and calibrate bounded functions, resulting in low prediction errors across all subpopulations and outperforming other methods.
	
	Consistent with previous findings, under moderate covariate shift ($q=2$), the multicalibrated approach continues to outperform standard methods in estimating both survival probability and restricted mean survival time in the target domain. This is evidenced by Figures~\ref{fig:SP-PH-indep-2} and~\ref{fig:RM-PH-indep-2}, which demonstrate the multicalibrated estimator's superior accuracy and fairness across subpopulations.
	
	We also compare the estimation results assuming an accelerated failure time model (Tables~\ref{simu:aft-1}-\ref{simu:aft-3}) and  assuming the  covariate-dependent Weibull censoring (Tables~\ref{simu:prop-dep-sp-1}-\ref{simu:prop-dep-sp-3}) under mild, moderate and strong covariate shift, respectively.  
	In all these settings, our method remains powerful, accommodating  non-proportional hazards and dependent censoring. Due to space limitations, we refer readers to Section~\ref{apxsec:simu} in the Supplementary Material for more details.
	
	\begin{figure}[ht!]
		\centering
		\begin{tabular}{c}
			
			\includegraphics[width = 0.9\textwidth, height=0.38\textheight]{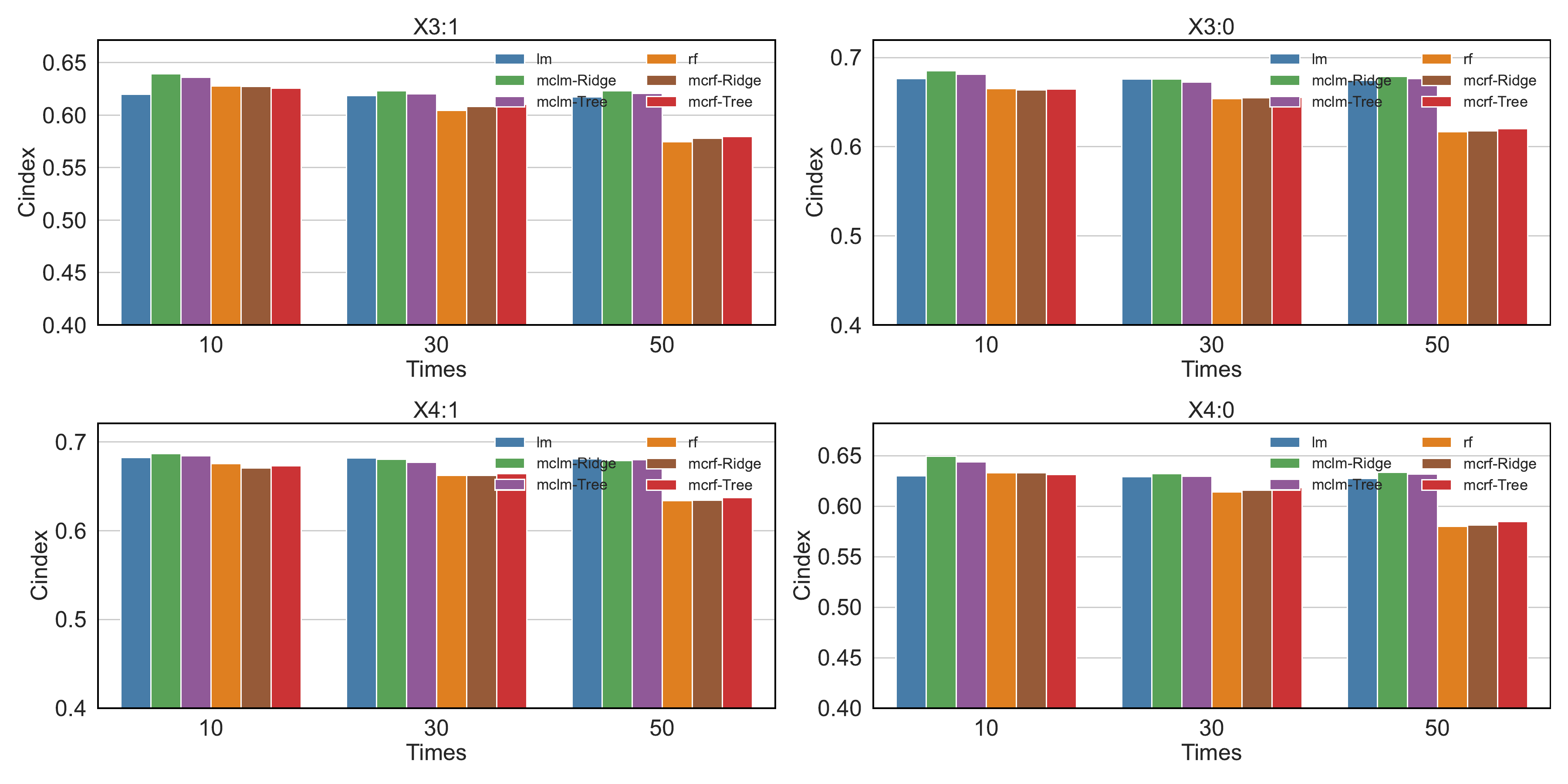}\\
			(a) Mild covariate-shift ($q=1$)\\
			\\	\includegraphics[width = 0.95\textwidth, height=0.38\textheight]{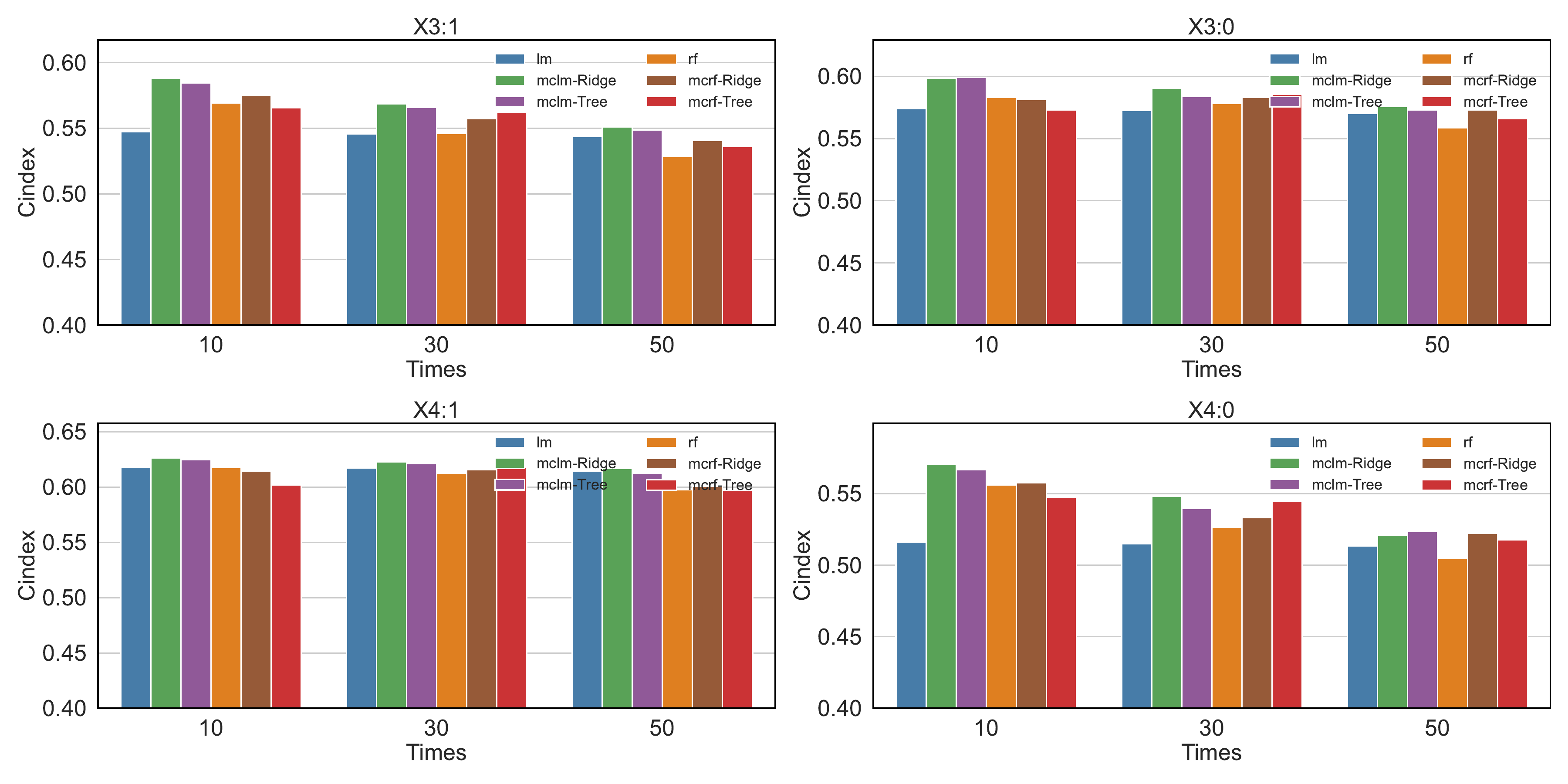}\\
			(b) strong covariate shift ($q = 3$)
		\end{tabular}
		\caption{\textsc{C-index} of different methods under (a) mild covariate shift ($q = 1$) and  (b) strong covariate shift  ($q = 3$) for various subpopulations. The proportional hazard model with independent uniform censoring  is considered. Both linear model and random forest are used to estimate the survival probability using pseudo observations.  Results are based on an average of $100$ simulation replications. 
		}
		\label{fig:Cindex-prop-indep-single-1}
	\end{figure}
	
	\subsection{The concordance index}
	The concordance index~(\textsc{C-index})~\citep{harrell1982evaluating, harrell1984regression} computes the agreement between the predicted survival ranking and the actual survival time, serving as an important metric for the discriminative power of the predictive model, especially involving censored observations. Specifically, it considers instance pairs and checks if the model's prediction ranks the two instances in accordance with their true orders. The \textsc{C-index} takes the form
	\begin{equation}\label{eqn:Cindex}
		C(t) :=  \frac{\sum_{\Delta_i = 1} \sum_{\wt{T}_j > \wt{T}_i } \b1\{ \wh{S}(t | X_j ) > \wh{S}(t | X_i) \}  }{\sum_{\Delta_i = 1} \sum_{\wt{T}_j > \wt{T}_i } \b1\{\wt{T}_j > \wt{T}_i \} },    
	\end{equation}
	where $\wh{S}$ is any prediction for survival probability. Each non-censored instance is compared against all instances that outlive it~(having a larger event or censoring time). Each correct ranking is counted, and the final score is normalized over the total number of pairs.
	
	We observe that multicalibration  leads to increased  \textsc{C-index}~as shown  in Figure~\ref{fig:Cindex-prop-indep-single-1} and Figure~\ref{fig:Cindex-prop-indep-single-2} under strong covariate shift, with similar performance under the weak covariate shift.  Compared to random forests, the linear models result in higher  \textsc{C-index} scores under mild covariate shift, probably attributed to the effective ranking it renders. However,  it generally has higher estimation bias   as supported by our simulations.
	
	\subsection{Extended  simulations to mimic real data}
	\begin{figure}
		\centering
		\includegraphics[width=0.7\linewidth, height=0.8\textheight]{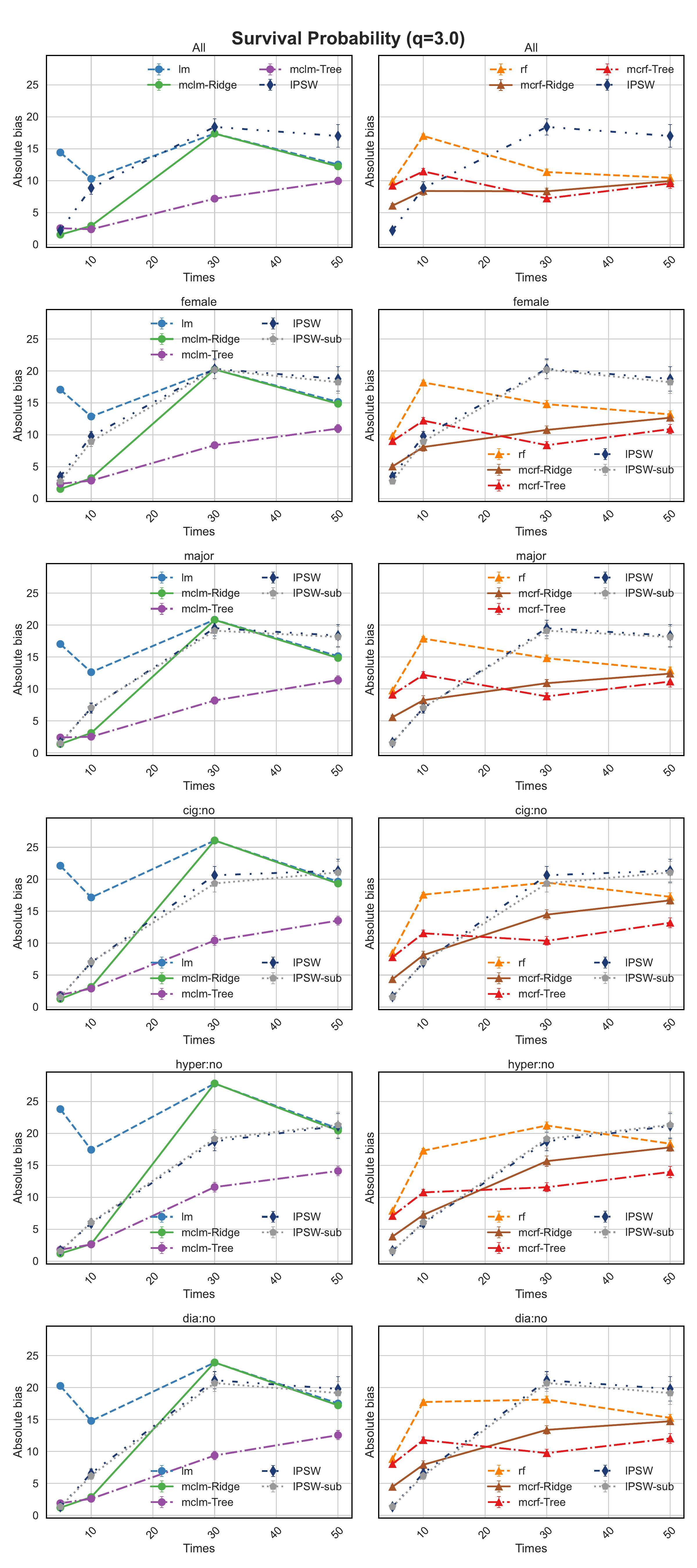}
		\caption{Results for all subpopulations in the setting with $13$ covariates based on $100$ replications. The absolute bias $(\times 10^2)$ for all methods in estimating survival probability are reported under strong covariate shift ($q = 3$) with $1500$ source and $300$ target samples.
			Survival time adheres to the proportional hazard model, and censoring is covariate-independent.   }
		\label{fig:SP-highdim-2}
	\end{figure}
	We conduct additional simulations   with different parameter configurations that mimic our real data sets to gain more insights.  Specifically, we explore the setting with $1500$ source and $300$ target samples and vary $\omega_0 \in \{0.5, 1, 2\}$, the first parameter of $\omega$.   We include  a total of $5$ covariates, including $2$ categorical variables that mimic gender and race, as is done in previous simulations.
	
	The results agree with what we observe in previous simulations, as shown in Figures~\ref{fig:SP-PH-indep-1500-300-1-comb} and~\ref{fig:SP-PH-indep-1500-300-2-comb}.
	Our methods demonstrate better overall accuracy. 
	While traditional machine learning and \textsc{ipsw} benefit from larger sample sizes, their performance deteriorates for smaller subpopulations with significant source-target imbalances. Although imputation and \textsc{ipsw} perform relatively well under mild shifts, our methods consistently match or outperform them. Varying the weight for intercept $\omega_0$, similar to changing the parameter $q$, adjusts the degree of covariate shift, impacting all methods. Nonetheless, our multicalibrated predictor maintains their advantages. 
	
	
	To further simulate real-world complexity, we extend the simulation  setting to include $13$ covariates with $8$ continuous and $5$ categorical variables, which model clinical variables such as sex, race, smoking status, hypertension, and diabetes. Covariates are generated from specific distributions and correlations are introduced to reflect realistic interactions, influencing survival time and domain membership simultaneously. We evaluate all methods at various time points and change the parameter $q$ to reflect the degree of covariate shift. Figures~\ref{fig:SP-highdim-2}  and~\ref{fig:SP-highdim-1}  show bias increases in traditional methods as covariate shift grows, whereas our methods show notable improvement. These findings are consistent with previous simulations, evidencing their efficiency in both simple and complex settings. The details are included in the Supplementary Material.

\section{Application to cardiovascular disease risk prediction}\label{sec:real}
We apply the proposed multicalibration method to data from two prospective cohorts: the Multi-Ethnic Study of Atherosclerosis (\textsc{mesa}) \citep{MESA} and the Chronic Renal Insufficiency Cohort (\textsc{cric}) \citep{CRIC}. \textsc{mesa} is a population-based study investigating sub-clinical cardiovascular disease and its risk factors for progression to clinically overt cardiovascular disease. It includes a sample of 6,814 asymptomatic men and women aged 45–84. In contrast, \textsc{cric} is a longitudinal cohort study examining risk factors for the progression of chronic renal insufficiency (\textsc{cri}) and cardiovascular disease (\textsc{cvd}) among \textsc{cri} patients. The study enrolled adults aged 21–74 with varying degrees of renal disease severity, half of whom were diagnosed with diabetes mellitus.

Both studies aim to predict the 10-year risk of \textsc{cvd}, a composite outcome that includes myocardial infarction, cardiac arrest, confirmed angina requiring revascularization, and coronary heart disease  death. The time to \textsc{cvd} incidence is censored under an unknown mechanism. However, substantial differences exist between the two cohorts (see Figure~\ref{fig:hist_CRIC_MESA} in Supplementary Material) in clinical predictors, indicating a strong covariate shift among subpopulations.
The \textsc{cric} cohort contains more individuals with hypertension and diabetes, suggesting a higher \textsc{cvd} risk. This is further supported by a greater number of observed \textsc{cvd} events in the \textsc{cric} cohort.

In our analysis, we use the \textsc{cric} cohort as the source data to develop a prediction model for 10-year \textsc{cvd} risk. For the testing set, since all \textsc{cric} participants had chronic kidney disease (\textsc{ckd}) with an estimated glomerular filtration rate (\textsc{egfr}) below 60 based on the \textsc{ckd-epi} equations, we restrict our analysis to \textsc{mesa} participants with \textsc{ckd} to estimate their 10-year \textsc{cvd} risk.
After excluding samples with missing values and focusing on Caucasian and Black participants, we obtain a dataset with 1,514 source samples from \textsc{cric} and 383 target samples from \textsc{mesa}. In total, we consider 13 variables: five categorical (e.g., gender, race, hypertension status, diabetic status) and eight continuous (e.g., age, body mass index, blood pressure) factors.

Our objective is to predict the survival probability of \textsc{cvd}  risk for \textsc{mesa} participants with chronic kidney disease, using the \textsc{cric} cohort as the source data. Preliminary analyses indicate that, while survival times differ between the two datasets, their restricted mean survival times (\textsc{rmst}) are remarkably similar. This similarity in \textsc{rmst}  may be attributed to the low incidence of \textsc{cvd}  events observed in both cohorts.

For each subgroup, we first establish a baseline by calculating the mean of the pseudo-observations for the samples within that group. We then assess how our \textsc{ipsw} method and a variety of prediction models, including naive linear models, random forests and multicalibrated predictions using  ridge regression or tree model as the auditing algorithm, perform in estimating survival probabilities in \textsc{MESA} based on \textsc{CRIC} data.  Additionally, we assess prediction performance using the \textsc{C-index}, which serves as a metric for evaluating the models' discriminative ability.

\begin{figure}[htb!]
	\centering
	\includegraphics[width=0.9\textwidth, height=0.85\textheight]{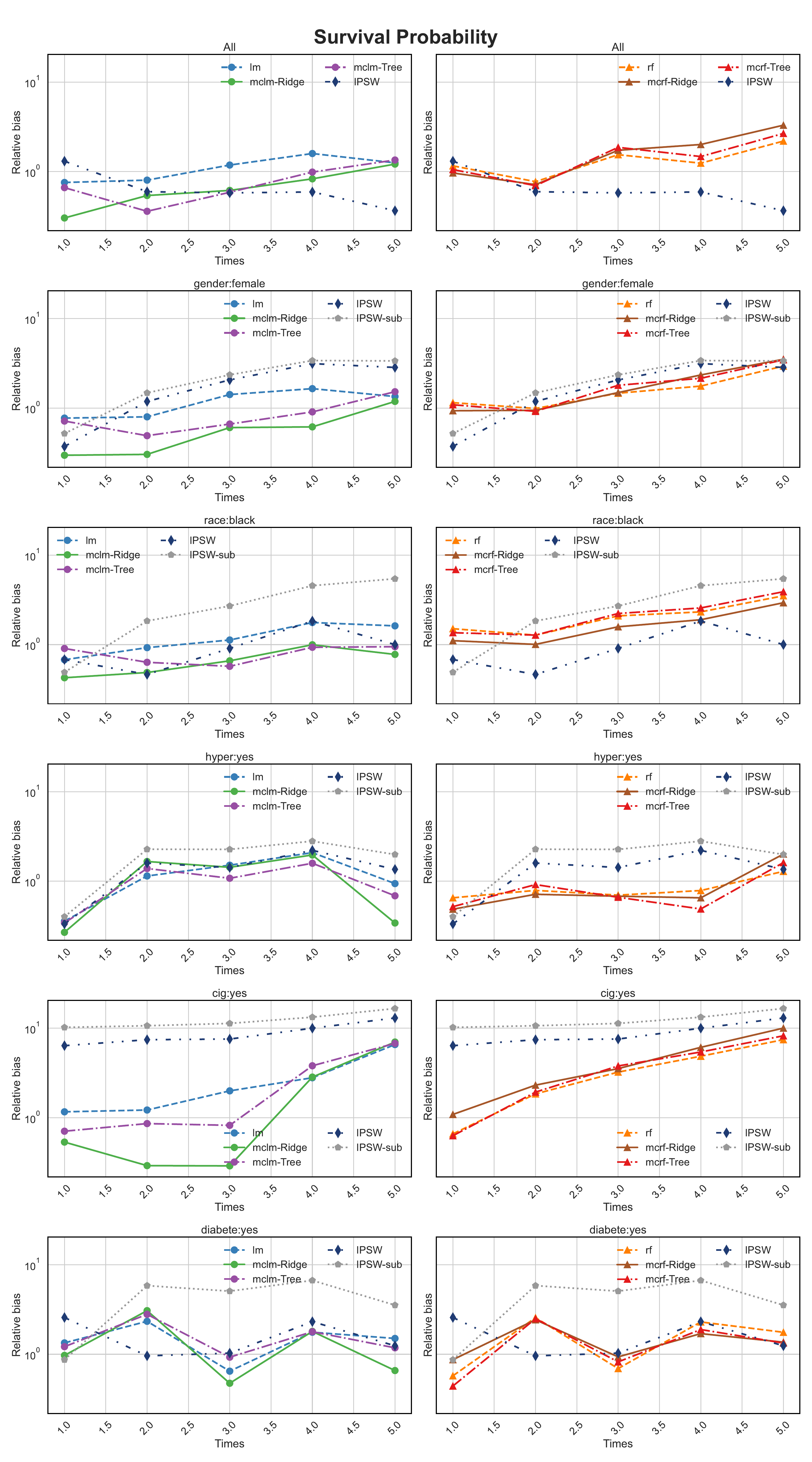}
	\caption{Relative bias  of different  methods in predicting the survival probability for all individuals in the  \textsc{mesa} cohort and various subpopulations at different follow-up time points.  }
	\label{fig:MESA_CRIC_cal_kidInd} 
\end{figure}

We report the prediction error for each method in Figure~\ref{fig:MESA_CRIC_cal_kidInd}. The results indicate that naive estimation using source data alone leads to substantial prediction errors, likely due to significant differences in covariate distributions between the source (\textsc{cric}) and target (\textsc{mesa}) cohorts. In contrast, methods incorporating multicalibration consistently demonstrate lower prediction errors than their uncalibrated counterparts and outperform the \textsc{ipsw} approach. 

As a sensitivity analysis, we also perform the analysis treating none-\textsc{cvd} death as competing risk by estimating the cumulative incidence using the Aalen-Johansen estimatior \citep{AalenJohansen1978} and observe almost the same results on relative biases in estimating the survival probabilities  (see Figure \ref{fig:Competing-real}).

Given the low incidence of \textsc{cvd} in both cohorts, the \textsc{C-index} is generally modest. However, as shown in Figure~\ref{fig:Cindex_MESA_CIRC}, the \textsc{C-index} across various subpopulations indicates that multicalibrated methods achieve higher or comparable \textsc{C-index} values than uncalibrated methods.

\section{Discussion}

\ignore{We have developed methods for multicalibration of models for censored survival data using the idea of pseudo-observations, including methods for predicting survival probability and restricted mean survival time. We compute pseudo-observations at each predetermined time point, based on which we develop a multicalibration method for censored survival data.
	Our proposed multicalibration algorithm, as detailed in Algorithm~\ref{alg:MCboost}, enables the construction of a prediction function that performs nearly as well as the \textsc{ipsw} estimator when applied to unseen target domains, despite relying solely on source domain data. This approach leverages the concept of universal adaptability, ensuring robust performance across diverse populations. These methods readily generalize to multiple domains (see Section \ref{multiple_domains}).}

Our simulation and real data analysis results  highlight the improvement  from a single multicalibrated prediction function trained on source data in generalizing the results to unknown target populations.  Such performances are competitive with shift-specific methods as well as traditional machine learning algorithms. The strong performance across diverse subpopulations illuminates the connection between universal adaptability and the concept of multicalibration: a multicalibrated predictor must robustly model outcome variations, not merely on an aggregate level but across multiple subpopulations. This highlights the potential of multicalibration as a tool to achieve similar  predictive performance in diverse and shifting populations.

The selection of the regression algorithm for auditing $\mathcal{A}$ is a nuanced decision that impacts the effectiveness of multicalibration. In our simulations and examples, we use ridge regression and decision trees as auditors.
The work of \cite{Kim22} demonstrated that linear functions of the form  $\{ 1 - \langle X, \omega \rangle : \omega \in \bbR^d \}$ is an effective approximation to the logistic propensity score odds $\exp(-\langle X, \omega \rangle)$, achieving performance comparable to logistic regression-based propensity scoring. Based on this insight, they employed ridge regression auditing to capture shifts modeled by logistic propensity scores.
Beyond ridge regression and decision trees, other machine learning algorithms such as neural networks and support vector machines  can also be used for auditing. Different auditing algorithms can agnostically learn different function classes 
$\mathcal{H}$ leading to variations in  performance. In general, a richer function class 
$\mathcal{H}$ leads to smaller residual bias, improving calibration across subpopulations. 
However, to mitigate overfitting, several practical considerations should be taken into account, including 
fitting shallow trees, introducing regularization by using a smaller number of buckets $m$ and a smaller step size $\eta$. 

The proposed  methods readily generalize to multiple domains (see Sections \ref{multiple_domains} and \ref{simu.multi} in Supplementary Material). A promising future direction is to adapt our framework for competing risks scenarios, which are common in time-to-event analysis, to estimate cause-specific cumulative incidence functions or the expected cause-specific lost time in a target domain using data pooled from source domains.

\vspace{-10pt}
\section*{Acknowledgment}
This project is supported by grants from the  National Institutes of Health. 

\vspace{-5pt}
\section*{Supplementary Material}
Supplementary Material available at Biometrika online includes extension to multiple source domains, proofs of all theorems, technical assumptions, and additional numerical simulation and real data analysis results.

\bibliographystyle{biometrika}
\bibliography{Reference}

\renewcommand{\theequation}{S.\arabic{equation}} 
\renewcommand{\thesection}{S.\arabic{section}}
\renewcommand{\thesubsection}{S.\arabic{section}.\arabic{subsection}}
\setcounter{equation}{0}
\setcounter{section}{0}
\renewcommand{\thelemma}{{S.\arabic{lemma}}} 
\renewcommand{\thetheorem}{{S.\arabic{theorem}}}
\renewcommand{\thetable}{S.\arabic{table}}
\renewcommand{\thefigure}{S.\arabic{figure}}

\begin{center}
	\Large		Supplementary Material for ``Multicalibration for Modeling Censored Survival Data with Universal Adaptability
\end{center}  




\markboth{Ye and Li}{Multicalibration for Modeling Censored Survival Data}


\author{Hanxuan Ye}
\affil{Department of Biostatistics, Epidemiology and Informatics, \\ University of Pennsylvania, Philadelphia, Pennsylvania 19104, U.S.A.
	\email{Huanxuan.Ye@pennmedicine.upenn.edu}}

\author{HONGZHE LI}
\affil{Department of Biostatistics, Epidemiology and Informatics, \\ University of Pennsylvania, Philadelphia, Pennsylvania 19104, U.S.A.
	\email{hongzhe@upenn.edu}}

\maketitle	 
\section{Extension to multiple source domains}\label{multiple_domains}

Our setup can be extended to more general cases with multiple source domains and a single target domain, as well as the setting with a single source domain and multiple target domains.
In the setting with multiple source domains $\sS_1, \sS_2, \ldots, \sS_M$ and a singe target domain $\sT$, we define the domain-specific propensity score odds:
$$
w_m(X) = r_m \sigma_{\sT}(X)/\sigma_{\sS_m}(X), 
$$
where $r_m = \pr(D = \sS_m )/\pr(D = \sT )$,   and propensity score for each domain are 
$$
\sigma_{\sT}(X) = \pr( D \in \sT|X), \quad \sigma_{\sS_m }(X) = \pr( D \in \sS_m |X). 
$$ 
We denote $\vw = (w_1(X), w_2(X), \ldots, w_M(X))$ as the vector of weight functions.
Given that the source domain distribution follows: 
$$
f(X | D ) = \sum_{m=1}^M f(X | D = \sS_m ) \b1\{D = \sS_m \}, 
$$
the corresponding \textsc{ipsw} estimator is: 
\begin{equation*}
	\wh{\tau}^{(k), ps}(t; \vw)  =  \frac{\sum_{i=1}^N \sum_{m = 1}^M \b1\{D_i \in \sS_m \} \wh{\theta}_i^{(k)}(t) w_m(X_i)  }{\sum_{i=1}^N \sum_{m = 1}^M \b1\{D_i \in \sS_m \} w_m(X_i) }. 
\end{equation*}
If probability estimates $\pr(D = \sS_i) (i=1,\cdots, M)$ are required, they can be replaced by their empirical proportions in the observed samples.   

We have the following  upper bound for the error of $\wt{m}^{(k)}(X; t)$ under the multi-source setting. 
\begin{theorem}\label{thm:multi-source}
	Under the same conditions as in the previous theorems, consider a setting with multiple source domains $\sS_m, m = 1, 2, \ldots, M$, and a single target $\sT$. Let $\wt{\vw}(X) = (\wt{w}_1(X), \wt{w}_2(X), \ldots, \wt{w}_M(X))$ be a set of weight functions associated with the propensity scores  $ (\wt{\sigma}_{\sT}, \wt{\sigma}_{\sS_1}, \ldots, \wt{\sigma}_{\sS_M}). $ 
	The function  $\wt{m}^{(k)}(X; t)$, produced by the proposed algorithm, satisfies the ``averaged'' muticalibration condition: 
	\begin{equation*}
		\sup_{h \in \mathcal{H}} \left| \sum_{m=1}^M \pr(D = \sS_m | D \in S) \Expect_{\sU_{\sS_m} }\left\{ h(X)  (\wt{m}^{(k)}(X; t) - m^{(k)}(X;t) ) \right\} \right| < \alpha,
	\end{equation*} 
	Furthermore, the target domain error is bounded by 
	\begin{align*}
		\Err_{\sT}(\wt{m}^{(k)}(X; t)) & \le \Err_{\sT}(\tau^{(k), ps}(t; \wt{\vw})) \\
		& \quad  + \alpha + \wt{C} \inf_{h\in\sH} \left[ \sum_{m=1}^M  \pr(D = \sS_m | D \in S) \{ d_{\sS_m}(w_m, h) + d_{\sS_m}(h, \wt{w}_m ) \} \right],
	\end{align*} 
	where the distance metric $d_{\sS_m}$ is defined as $d_{\sS_m}(h_1, h_2): = \Expect_{\sS_m}[|h_1(X) - h_2(X)|]$. 
\end{theorem}
The bound provides useful insights: the performance of $\wt{m}^k $ closely matches that of the pseudo-weighted estimator, up to a weighted average of propensity score discrepancies. This supports the use of multicalibration as an adaptive approach for learning under shift across multiple source populations.

\section{Technical Proofs}\label{apxsec:proofs}
Recall that for two sequences of positive number $\{a_n \}$ and $\{b_n\}$, we write $a_n \lesssim b_n$ if $|a_n| \le c |b_n|$ for some universal constant $c > 0$, and $a_n \gtrsim b_n$ if $|a_n| \ge c' |b_n|$ for some constant $c' > 0$. We say $a_n \asymp b_n $ if  $a_n \lesssim b_n$ and $a_n \gtrsim b_n$. Moreover, we use $a_n = O(b_n)$ to represent $|a_n| \le C |b_n|$, and $a_n = \Omega(b_n)$ to represent $|a_n| \ge C' |b_n|$, for some $C, C' > 0$. We say $a_n = \Theta(b_n)$ if $a_n = O(b_n)$ and $a_n = \Omega(b_n)$. 

In contrast to binary classification problems, where indicators are available, our algorithm calibrates with respect to pseudo-observations: 
$$
\frac{1}{|V|} \left| \sum_{i} h(X_i) \left(\wt{m}^{(k)} (X_i; t) - \wh{\theta}_i^{(k)} (t) \right) \right|. 
$$
Question arises that whether this approach ensure multicalibration, in the sense of Definition~\ref{def:multicali}, remains unknown. Since the size of validation data set is usually a negligible portion of total sample size $N$, we can, for simplicity, assume $|V| = N$. 

Notice that 
\begin{equation}\label{eqn: empirical}
	\begin{aligned}
		& \sup_{h \in \mathcal{H}} \left|  E_{\sU_\sS}\left\{ h(X) (\wt{m}^{(k)}(X; t) - m^{(k)}(X;t) ) \right\} \right| \\
		& \le \sup_{h \in \mathcal{H}}\left| \frac{\sum_{i=1}^N   h(X_i) (\wt{m}^{(k)}(X_i; t) - \wh{\theta}_i^{(k)}(t)) }{ N  } \right| \\
		& \quad + \sup_{h \in \mathcal{H}}\left| \frac{\sum_{i=1}^N   h(X_i) (\wt{m}^{(k)}(X_i; t) - \wh{\theta}_i^{(k)}(t)) }{ N  }  - E_{\sU_\sS}\left\{ h(X) (\wt{m}^{(k)}(X; t) - m^{(k)}(X;t) ) \right\} \right|  \\
		& \le \alpha +  \sup_{h \in \mathcal{H}}\left| E_{\sU_\sS}  h(X) \wt{m}^{(k)}(X; t) -   \frac{\sum_{i=1}^N  \wt{m}^{(k)}(X_i; t) h(X_i) }{ N  }  \right| \\
		& \qquad + \sup_{h \in \mathcal{H}} \left|  \frac{\sum_{i=1}^N  \wh{\theta}_i^{(k)}(t) h(X_i)  }{N }   -    E_{\sU_\sS}\left\{ h(X)m^{(k)} (X; t) \right\}
		\right|. 
	\end{aligned}
\end{equation} 
If we are able to establish $o_p(1)$ bound for the last two terms in the preceding inequality, we can then assert multicalibration can be attained with bias $\alpha' = \alpha + o_p(1)$.

\subsection{Proof for Theorem~\ref{thm: algo-multi}}

To start with, we prove the uniform convergence using a standard metric entropy argument for the term
$$\sup_{h \in \mathcal{H}}\left| E_{\sU_\sS}  h(X) \wt{m}^{(k)}(X; t) -   \frac{\sum_{i=1}^N  \wt{m}^{(k)}(X_i; t) h(X_i) }{ N  }  \right|.$$  
\begin{lemma}\label{lem: Hoeffding}
	Suppose $\mathcal{A}$ agnostically learns a class $\mathcal{H}$ which has $\varepsilon$-covering number $\sN_{\varepsilon} = \sN(\varepsilon, \sH, \|\cdot\|_{L_1})$, then with probability at least $1 - \delta$, then for any function $\check{m}^{(k)}(X; t)$
	$$
	\sup_{h \in \sN_{\epsilon} }\left| E_{\sU_\sS} h(X) \check{m}^{(k)}(X; t) -   \frac{\sum_{i=1}^N  \check{m}^{(k)}(X_i; t) h(X_i)  }{N  }
	\right| \le  \wt{C} C_{\sH} \sqrt{ \frac{ \log (2\mathcal{N}_{\varepsilon}/\delta)  }{N} } 
	$$
\end{lemma} 
\begin{proof}
	Suppose the class $\mathcal{H}$ has $\varepsilon$-covering number $\sN_{\varepsilon} = \sN(\varepsilon, \mathcal{H}, \|\cdot\|_{L_1})$.
	Using the McDiamid's inequality or general Hoeffding inequality~\citep{wainwright_2019, vershynin_2018},
	we have 
	$$
	\mathbb{P}\left( \left| E_{\sU_\sS} h(X) \check{m}^{(k)}(X; t) -   \frac{\sum_{i=1}^N  \check{m}^{(k)}(X_i; t) h(X_i)  }{N  }  \right| \ge t \right) \le 2\exp\left( - \frac{ Nt^2}{\wt{C}^2 C_{\sH}^2 }\right). 
	$$
	Then 
	$$
	\mathbb{P}\left( \sup_{h \in \mathcal{N}_{\varepsilon}}\left| E_{\sU_\sS} h(X) \check{m}^{(k)}(X; t) -   \frac{\sum_{i=1}^N  \check{m}^{(k)}(X_i; t) h(X_i)  }{N  }  \right| \ge t \right) \le 2 \sN_{\varepsilon} \exp\left( - \frac{ Nt^2}{\wt{C}^2 C_{\sH}^2 }\right). 
	$$
	Moreover, for any $h \in \sH$, there exists some $\wt{h} \in \mathcal{N}_{\varepsilon}$ such that $E[|h - \wt{h}|] \le \epsilon $. Therefore, 
	$$
	\sup_{h \in \sH }\left| E_{\sU_\sS} h(X) \check{m}^{(k)}(X; t) -   \frac{\sum_{i=1}^N  \check{m}^{(k)}(X_i; t) h(X_i)  }{N  }  \right| \le \wt{C} C_{\sH} \sqrt{ \frac{ \log (2\mathcal{N}_{\varepsilon}/\delta)  }{N} } + \wt{C} \varepsilon, 
	$$
	for any $\epsilon > 0$ with large probability. 
\end{proof} 
The calibration property is assured by limiting the supreme terms in the last inequality of~\eqref{eqn: empirical} to a negligibly small $o_p(1)$ term. In fact, the third term follows the result of~\cite{graw2009pseudo, jacobsen2016note} that  
\begin{align*}
	\frac{\sum_{i=1}^N  \wh{\theta}_i^{(k)}(t) h(X_i) }{ N  } & = E_{\sD_s} \left\{ h(X) \wh{\theta}_i^{(k)}(t) \right\} + o_p(1) 
	= E_{\sU_s} \left\{ h(X_i) E[ \nu^{(k)}(T_i; t) | X_i \} \right] + o_p(1)  \\
	& = E_{\sU_s} \left\{ h(X_i) m^{(k)}(X_i ; t) \right\} + o_p(1),
\end{align*} 
based on which we provide a supremum bound for all $h \in \sH$. Hence, if $\wt{m}^{(k)}(X; t)$ is well multi-calibrated via the algorithm, i.e., 
$$
\sup_{h \in \mathcal{H}}\left| \frac{\sum_{i=1}^N   h(X_i) (\wt{m}^{(k)}(X_i; t) - \wh{\theta}_i^{(k)}(t)) }{ N  } \right| \le \alpha,
$$
then $\alpha' = \sup_{h \in \mathcal{H}}\left| E_{\sU_\sS}\left\{ h(X) (\wt{m}^{(k)}(X; t) - m^{(k)}(X;t) ) \right\} \right| \le \alpha + o_p(1)$.
In the remainder, we establish the specific bound for the error incurred from pseudo observations.

{\noindent \bf Bound for pseudo-observations. \label{apxsubsubsec:pseudo-bound} }
In time-to-event analysis, an observation is a tuple $O_i = (\wt{T}_i, \Delta_i, X_i)$\footnote{ We omit the domain $D_i$ since only source data is used for multicalibration}, and there is a correspondence between $O_i$ and $\delta_{O_i} = (Y_i, N_{i,0}, N_{i, 1})$, a gathering of three counting processes, where $Y_i (s) = \b1\{\wt{T}_i \ge s\}$, $N_{i, 0}(s) = \b1\{\wt{T}_i \le s, \Delta_i = 0\}$ and $N_{i, 1}(s) = \b1\{\wt{T}_i \le s , \Delta_i = 1 \}$.  A limit function $F = (H, H_0, H_1)$, as a correspondence to the average of counting processes  $F_N = N^{-1}\sum_{i=1}^N (Y_i, N_{i,0}, N_{i,1})$, is a triplets where $H(s) = \pr(\wt{T}_i \ge s)$, $H_0 (s) = \pr(\wt{T}_i \le s, \Delta_i = 0) $, and $H_1 (s) = \pr(\wt{T}_i \le s, \Delta_i = 1) $.

Considered as a continuous differentiable functional, the pseudo observation can be von Mises expanded~\citep{vaart_1998} similarly to a smooth function that can be ``Taylor expanded". To be specific, we introduce the Nelson-Aalen functional $\psi(F ; t) = \int_0^t \frac{\b1\{H(s) > 0 \} }{H(s)} \intd H_1 (s) = \Lambda_1 (t)$. The Kaplan-Meier estimator can then be represented as $\wh{S}(t) = \phi^1 (F_N; t)$ where  $\phi^1 (F; t)$ is defined as 
$$
\phi^1 (F; t)= \prod_0^{t} (1 - \psi(F; du)).
$$
The restricted mean survival time is the integral of the Kaplan-Meier functional $\phi^2 (F; t) = \int_0^t \phi^1 (F; u) du$. 
Given $\phi^{(k)}, k = 1,2$ and $\wh{\theta}_i^{(k)}(t) = N \phi^{(k)} (F_N; t) - (N-1) \phi^{(k)} (F_N^{(i)}; t) $,
it allows for a decomposition of a pseudo-observation into an essential and a remainder part~\citep{overgaard2017asymptotic},  
\begin{align*}
	\wh{\theta}_i^{k}(t) & =  \wh{\theta}_i^{k*}(t) + R_{n,k}(t) = \tau^{(k)}(t) + \dot{\phi}^{(k)}(O_i; t) + (N-1)^{-1} \sum_{j \neq i}\ddot{\phi}^{(k)}(O_i, O_j; t)  + R_{n,k}(t),
\end{align*}
where $\dot{\phi}^{(k)}(O_i; t)$ and $\ddot{\phi}^{(k)}(O_i, O_j; t) $ are abbreviation for the first functional derivative $\dot{\phi}_{F}^{(k)}(\delta_{O_i} - F; t) $and the second funcitonal derivative $\ddot{\phi}_{F}^{(k)}(\delta_{O_i} - F, \delta_{O_j} - F; t) $, respectively. They are bounded, continuous, and linear~(bilinear). For survival probability and restricted mean derived from the Kaplan-Meier estimator, the functional derivatives have explicit forms, 
\begin{align}
	&   \dot{\phi}^{1}(O_i; t) = -S(t) \dot{\psi}(O_i; t),  \label{eqn:first-diff}\\
	& \ddot{\phi}^{1}(O_i, O_j; t) = -S(t)\{ \ddot{\psi}(O_i, O_j; t) - \dot{\psi}(O_i; t)\dot{\psi}(O_j; t) + \b1\{ i = j \} \int_0^t \frac{1}{H^2(s)} dN_{i,1}(s) \}, \\
	&   \dot{\phi}^{2}(O_i; t) = \int_0^t \dot{\phi}^{1}(O_i; s) ds , \qquad \ddot{\phi}^{2}(O_i, O_j; t) = \int_0^t \ddot{\phi}^{1}(O_i, O_j; s) ds, 
\end{align}
where the first and second-order derivatives of $\psi$ is given by~\cite{james1997study},
\begin{align*}
	&  \dot{\psi}(O_i; t) = \int_0^t \frac{1}{H(s)} d M_{i,1}(s), \\
	& \ddot{\psi}(O_i, O_j; t) = \int_0^t  \frac{H(s) - Y_j(s)}{H(s)^2} dM_{i,1} (s) - \int_0^t \frac{H(s) - Y_i(s)}{H(s)^2} d M_{j,1}(s),
\end{align*}
where $M_{i,1}(s) = N_{i,1}(s) - \int_0^s Y_i(u) d \Lambda_1 (u)$. Therefore, 
\begin{align*}
	& \sup_{h \in \mathcal{H}} \left| \sum_{i=1}^N  \frac{h(X_i)\wh{\theta}_i^{(k)}(t) }{N}  - E\left[ h(X) m^{(k)}(X; t)\right] \right| \\
	& = \sup_{h \in \mathcal{H}} | \sum_{i=1}^N \frac{h(X_i)(\tau^{(k)}(t)  + \dot{\phi}^{(k)}(O_i; t) )}{N} - E\left[ h(X) m^{(k)}(X; t)\right] \\
	& \quad  + \sum_{i=1}^N \frac{h(X_i)\ddot{\phi}_{F}^{(k)}(\delta_{O_i} - F, F_{N}^{(i)} - F; t) }{N} 
	+ \sum_{i=1}^N\frac{h(X_i)R_{N,i}(t)}{N}  | \\
	& \le \sup_{h \in \mathcal{H}} \left| \sum_{i=1}^N \frac{h(X_i)(\tau^{(k)}(t) + \dot{\phi}^{(k)}(O_i; t) )}{N} - E\left[ h(X) m^{(k)}(X; t)\right] \right| \\
	& \qquad + \sup_{h \in \mathcal{H}}  \left| \sum_{i=1}^N \frac{h(X_i)\ddot{\phi}_{F}^{(k)}(\delta_{O_i} - F, F_{N}^{(i)} - F; t) }{N} \right|  +  \sup_{h \in \mathcal{H}}  \left|  \sum_{i=1}^N\frac{h(X_i)R_{N,i}(t)}{N} \right|. 
\end{align*}
To bound this, we invoke the following Lemma, which is adapted from Proposition 1 of~\cite{jacobsen2016note}, (4.7) of~\cite{overgaard2017asymptotic} and Condition 3 of~\cite{overgaard2019pseudo}.
\begin{lemma}\label{lem:first-order}
	In modeling the survival probability~(SP) and the restricted mean~(RE) under the scenario
	of completely independent censoring as outlined in Assumption~\ref{ass:indep}(a), the following relation holds:  
	\begin{equation}
		\tau^{(k)}(t) + E[ \dot{\phi}^{(k)}(O_i; t) | X_i ] =  m^{(k)}(X_i; t).
	\end{equation} 
\end{lemma}
Then according to Proposition 3.1 of~\cite{overgaard2017asymptotic}, we have 
\begin{align}
	& \sup_{h \in \mathcal{H}} \left| \sum_{i=1}^N  \frac{h(X_i)\wh{\theta}_i^{(k)}(t) }{N}  - E\left[ h(X) m^{(k)}(X; t)\right] \right| \notag \\
	& \le \sup_{h \in \mathcal{H}} \left| \sum_{i=1}^N \frac{h(X_i) (\tau^{(k)}(t) + \dot{\phi}^{(k)} (O_i; t) )}{N} - E\left[ h(X) m^{(k)}(X; t)\right] \right| \notag \\
	& \quad + \sup_{h \in \mathcal{H}}  | \sum_{i=1}^N \frac{h(X_i)\ddot{\phi}_{F}^{(k)}(\delta_{O_i} - F, F_{N}^{(i)} - F; t) }{N} | + o_p(N^{-2\lambda}), \notag \\
	& = \sup_{h \in \mathcal{H}} \left| \sum_{i=1}^N \frac{h(X_i) (\tau^{(k)}(t) + \dot{\phi}^{(k)} (O_i; t) )}{N} - E\left[ h(X) (\tau^{(k)}(t) + \dot{\phi}^{(k)} (O_i; t) ) \right] \right| \notag \\
	& \quad + \sup_{h \in \mathcal{H}}  | \sum_{i=1}^N \frac{h(X_i)\ddot{\phi}_{F}^{(k)}(\delta_{O_i} - F, F_{N}^{(i)} - F; t) }{N} | + o_p(N^{-2\lambda}), \label{eqn:pseudo_error}
\end{align}
where $\lambda \in [1/4, 1/2)$.

The paper of~\cite{overgaard2017asymptotic} mentioned that $\dot{\phi}^{(k)}$ and $\ddot{\phi}^{(k)}$ are bound linear and bilinear operator, respectively. We have made uniform boundness of Lipschitz constant for all $t \in \mathbf{T}$ where $\mathbf{T}$ with upperbound $\wt{C}$,as stated in Assumption~\ref{ass:uniform}. 
\begin{remark}
	We argue that Assumption~\ref{ass:uniform} is reasonable if we investigate the form of functional derivatives as shown in Equation~\eqref{eqn:first-diff}, 
	\begin{align*}
		\dot{\phi}^{1}(O_i; t) = -S(t) \dot{\psi}(O_i; t) & = -S(t) \int_0^t \frac{1}{H(s)} d M_{i,1}(s) \\
		& = -S(t) \int_0^t \frac{1}{H(s)} d (N_{i,1}(s) - \int_0^s Y_i(u) d\Lambda(u)) \\
		& = -S(t) \left( \int_0^t \frac{1}{H(s)} d\b1\{ \wt{T}_i \le s, \Delta_i = 1\} - \int_0^t \frac{Y_i(s)}{H(s)} \lambda(s) ds  \right) \\
		& = -S(t) \left( \int_0^t \frac{1}{H(\wt{T}_i)} \b1\{ \wt{T}_i \le t, \Delta_i = 1\} - \int_0^{t \wedge \wt{T}_i } \frac{1}{H(s)} \lambda(s) ds  \right).
	\end{align*}
	Suppose $\lambda(\cdot)$ is a continuous, and 
	\begin{align*}
		H(s) = \pr(T \ge s ) \pr(C \ge s ) > c^2 > 0, 
	\end{align*} 
	for some constant $c$ for $t \in \mathbf{T} \subset [0, \wt{C}]$ $(T \le \wt{C})$, under independent censoring assumption, then 
	$ | \dot{\phi}^1 | $ is bounded.  A similar rationale applies to the second derivatives. Given the bounded nature of these functionals, the uniform boundedness is reasonable.  
\end{remark}
Under Assumption~\ref{ass:uniform}, 
$\dot{\phi}^{(k)}$ is a linear bounded operator such that  
$$
\dot{\phi}^{(k)}(O_i; t) \le K \|\delta_{O_i} - F\|_{[p]} \le K (\|\delta_{O_i} \|_{[p]} + \| F\|_{[p]} ), 
$$ 
for some Lipschitz constant $K$.  Recall the norm we consider is the $p$-variation norm that $\|f\|_{[p]} = v_p (f; [a, b])^{1/p} + \|f\|_{\infty}$, where $\|\cdot\|_{\infty}$ is the supremum norm. The functional $v_p (f; [a, b]) = \sup \sum_{i=1}^m |f(x_{i-1}) - f(x_i)|^p$, where the supremum is over $m \in \mathbb{N}$ and points $x_0 < x_1 < \ldots < x_m$ in interval $[a, b]$. Note that $\|\delta_{O_i}\|_{[p]}$ is bounded for any $x$ and $p$,  given that $\delta_{O_i}$ is empirical distribution containing 1 left-continuous entry and 2 right-continuous entries~\citep{overgaard2017asymptotic}.  The first term of~\eqref{eqn:pseudo_error} can be bounded in the same way as brought up earlier, using McDiarmid's inequality.

It then reduces to bound the second-order term $\sup_{h \in \mathcal{H}}  | N^{-1} \sum_{i=1}^N h(X_i)\ddot{\phi}_{F}^{(k)}(\delta_{O_i} - F, F_{N}^{(i)} - F)  |$. As for the second derivative term, note that $F_N^{(i)} - F = (F_N - F) + (N-1)^{-1} (F_N - \delta_{O_i})$. 
The Lipschitz continuity of the second-order derivative indicates that there exists a constant $K > 0$ such that 
\begin{equation}\label{eqn:pseudo-bound}
	\begin{aligned}
		& \sup_{h \in \mathcal{H}} \left| \ddot{\phi}_{F}^{(k)} \left( \sum_{i=1}^N \frac{h(X_i)(\delta_{O_i} - F) }{N},  F_N - F; t \right)   \right| + \sup_{h \in \mathcal{H}}  \left| \sum_{i=1}^N \frac{ h(X_i)\ddot{\phi}_{F}^{(k)} (\delta_{O_i} - F, F_N - \delta_{O_i}; t) }{N(N-1)} \right| \\
		& \le  K \sup_{h} \left \| \sum_{i=1}^N \frac{h(X_i)(\delta_{O_i} - F) }{N}  \right \|_{[p]} \| F_N - F \|_{[p]} \\
		& \qquad +  C_{\mathcal{H}} (N-1)^{-1} K (\max_i \|\delta_{O_i}\|_{[p]} + \|F\|_{[p]})( \max_i \|\delta_{O_i}\|_{[p]} + \|F\|_{[p]} + \|F - F_N \|_{[p]}) \\
		& = O_p(N^{\frac{1-p}{p}}(\log \log N)^{1/2} )  + O_p(N^{-1}) = O_p(N^{-\lambda}).
	\end{aligned}
\end{equation}
The last equality is because $\|F_N - F \|_{[p]} = O_p(N^{\frac{1-p}{p}}(\log \log N)^{1/2} ) $ for $1 \le p < 2$ \citep{dudley1999differentiability}, and $\|\delta_{O_i}\|_{[p]} $ is bounded for any $x$ and $p$. If we take $p \in (4/3, 2)$, we can achieve the rate $N^{-\lambda}, \lambda \in [1/4, 1/2)$. Combining Lemma~\ref{lem: Hoeffding} and~\eqref{eqn:pseudo-bound}
together, we can get the desired results. 

A refined bound elaborated in Section~\ref{apx:sec:sample} indicates that 
\begin{align*}
	& \sup_{h \in \mathcal{H}}\left| E_{\sU_\sS}  h(X) \check{m}^{(k)}(X; t) -   \frac{\sum_{i=1}^N  \check{m}^{(k)}(X_i; t) h(X_i) }{ N  }  \right| \\
	& \quad \quad  + \sup_{h \in \mathcal{H}} \left|  \frac{\sum_{i=1}^N  \wh{\theta}_i^{(k)}(t) h(X_i)  }{N }   -    E\left[ h(X)m^{(k)} (X; t) \right]
	\right|  \\
	& \le  2\wt{C}C_{\sH} \sqrt{
		\frac{\log(2\sN_{\varepsilon})/\delta}{N}
	} + \frac{\log(1/\delta)}{N^{1/p}} + C_p N^{\frac{1-p}{p}}.     
\end{align*}
Thus, a well-calibrated post-processed predictor $\wt{m}^{(k)}$ is able to achieve bias 
$$\alpha' \le \alpha + 2\wt{C}C_{\sH} \sqrt{
	\frac{\log(2\sN_{\varepsilon})/\delta}{N}
} + \frac{\log(1/\delta)}{N^{1/p}} + C_p N^{\frac{1-p}{p}} .$$

\begin{remark}
	
	Under covariate-dependent censoring, we define $G$ as a functional by $$G(F; s|\wt{X}) = \prod_{0}^{s^- } ( 1 - \Lambda_c (F; du| \wt{X}) ), F \in \mathbb{D},
	$$
	such that 
	$\wh{G}(s | \wt{X}) = G(F_N; s | \wt{X})$, where $\Lambda_c$ is the Nelson-Aalen functional for the cumulative hazard of censoring time $C$. The pseudo observations can be expressed as 
	$$
	\wh{\theta}_i^{(k)} (t) = N \wh{\theta}^{(k)} (t) - (N-1) \wh{\theta}_{-i}^{(k)} (t),
	$$ 
	where 
	\begin{align*}
		&  \wh{\theta}^{(k)} (t) = N^{-1} \sum_{i=1}^N \frac{\nu^{(k)}(\wt{T}_i; t) \b1\{ C_i \ge \wt{T}_i \wedge t \}  }{G(F_N; \wt{T}_i \wedge t | \wt{X}_i)}, \\
		&  \wh{\theta}_{-i}^{(k)} (t) = (N-1)^{-1} \sum_{j \neq i } \frac{\nu^{(k)}(\wt{T}_i; t) \b1\{ C_i \ge \wt{T}_i \wedge t \}  }{G(F_N^{(i)}; \wt{T}_i \wedge t | \wt{X}_i)},
	\end{align*} 
	as per~\eqref{eqn:weight-pseudo}. Approaching from the functional perspective, we can view the pseudo observation as functional mapping from distribution space $\mathbb{D}$ to $\bbR$, which is given by 
	$$
	\phi^{(k)} (F) = \int \frac{\nu^{(k)}(F; t)}{ G(F; \wt{T}\wedge t | \wt{X} )} dF,
	$$
	where we define the functional $\nu^{(k)}(F; t) = \nu^{(k)}(\wt{T}; t) \b1\{C \ge \wt{T} \wedge t\}$ for $F \in \mathbb{D}$ Functional $\phi^{(k)}$ is measurable, $2$-times continuously differentiable with Lipschitz continuous second order derivative in the neighbor of $F$. The first functional derivative will be 
	$$
	\dot{\phi}^{(k)} (O_i) = \int \frac{\nu^{(k)}(\wt{T}; t) \b1\{C \ge \wt{T} \wedge t\} }{ G(F; \wt{T}\wedge t | \wt{X} )} d \delta_{O_i} - \int \frac{\nu^{(k)}(\wt{T}; t) \b1\{C \ge \wt{T} \wedge t\} }{ G(F; \wt{T}\wedge t | \wt{X} )^2 } G_F'(\delta_{O_i}; \wt{T} \wedge t | \wt{X}) d F, 
	$$
	where $G_F'(g; s | \wt{X})$ is the derivative of functional $G$ at $F$ along direction $g$. Assuming the censoring survival function $G$ is consistently estimated, say, by a Cox model, the result of Lemma~\ref{lem:first-order} still holds for the condition where the censoring distribution is covariate-dependent, following the additive or proportional hazard model (Assumption~\ref{ass:indep}(b) and Assumption~\ref{ass:censoring}). 
	
	The detailed derivation of the second derivative and its attributes, the reader is directed to Proposition 2 and Appendix of~\cite{overgaard2019pseudo}. Hence, similar results can be concluded by following the exact procedure in the proof of Theorem~\ref{thm: algo-multi}.

\end{remark}
\subsection{Proof of Corollary~\ref{coro:complexity}}\label{apx:sec:sample}
In this part, we delve into the aspect of 
the sample complexity of Algorithm~\ref{alg:MCboost}. As illuminated by Lemma~\ref{lem: Hoeffding}, 
$$
\sup_{h \in \sN_{\epsilon} }\left| E_{\sU_\sS} h(X) \check{m}^{(k)}(X; t) -   \frac{\sum_{i=1}^N  \check{m}^{(k)}(X_i; t) h(X_i)  }{N  }
\right| \le  \wt{C} C_{\sH} \sqrt{ \frac{ \log (2\mathcal{N}_{\varepsilon}/\delta)  }{N} } 
$$
holds uniformly for $h \in \sN_{\epsilon} $. Given an $h \in \sH$, there exists some $\wt{h} \in \mathcal{N}_{\varepsilon}$ such that $E[|h - \wt{h}|] \le \epsilon $, leading to
\begin{equation}{\label{eqn:MicDiamid}}
	\sup_{h \in \sH }\left| E_{\sU_\sS} h(X) \check{m}^{(k)}(X; t) -   \frac{\sum_{i=1}^N  \check{m}^{(k)}(X_i; t) h(X_i)  }{N  }  \right| \le \wt{C} C_{\sH} \sqrt{ \frac{ \log (2\mathcal{N}_{\varepsilon}/\delta)  }{N} } + \wt{C} \varepsilon, 
\end{equation}
for any $\varepsilon > 0$ with probability $1 - \delta$. To achieve the multicalibration, the empirical estimate for each $h \in \mathcal{H}$ must be within $t = O(\alpha)$ with probability at least $1 - \delta$, for some small enough $ \delta$. We set $\epsilon = \Theta(\alpha)$, and the sample size will be $N = O \left( \frac{ \log(2\sN_{\Theta(\alpha)} /\delta) }{ \alpha^2 } \right)$. It depends on $\delta, \alpha$ and the metric entropy that characterizes the complexity of the function class. Illustrative instances of covering numbers include:
\begin{example}\label{ex:covering}
	(I) (Lipschitz function) Let $\sH = \{ h_{\theta}(X), \theta \in \Theta \subset \mathbb{R}^d \}$ denote a class of Lipschitz functions characterized by a compact parameter space $\Theta$, satisfying Lipschitz condition $ | h_{\theta_1}(X) -  h_{\theta_2}(X)| \le L(X) \|\theta_1 - \theta_2 \|$. Under the assumption that $E[L(X)] < \infty$, we establish the following relationship: 
	$$
	\sN(\varepsilon, \sH, \|\cdot\|_{L_1})  \le  \sN\left( \frac{\varepsilon}{E[L(X)]}, \Theta, \|\cdot \| \right).
	$$
	Then the covering number $\sN_{\varepsilon} \lesssim (1 + \frac{2\text{diam}(\Theta) }{\varepsilon})^d$. Notable instances falling under this paradigm include linear regression and logistic regression. (II) (Convex functions)~\citep{guntuboyina2012l1} The covering number of the set of bounded multivariate convex function $\log \mathcal{N}(\varepsilon, \sC([a, b]^d, B),  \|\cdot\|_{L_1}) \lesssim  \varepsilon^{-d/2} $.  
\end{example}
For (I), the bound in~\eqref{eqn:MicDiamid} satisfy 
\begin{align*}
	&\sup_{h \in \sH }\left| E_{\sU_\sS} h(X) \check{m}^{(k)}(X; t) -   \frac{\sum_{i=1}^N  \check{m}^{(k)}(X_i; t) h(X_i)  }{N  }  \right| \\
	& \quad \lesssim \sqrt{\frac{\log(2N) + d\log(1 + \frac{\text{diam}(\Theta)}{\alpha})}{N}} + \alpha,    
\end{align*}
taking $\varepsilon = \Theta(\alpha), \delta = 1/N$. For (II), this bound $\lesssim N^{-1/2} \sqrt{ \log(2N) + \alpha^{-d/2} } + \alpha $. 

As for error incurred by pseudo observations, the rate $O_p(N^{-\lambda})$ in~\eqref{eqn:pseudo-bound} is not enough. Obtaining the finite sample bound for our estimator is urgent to understand the complexity of our algorithm. The difficulty lies in that the pseudo observations are neither bounded nor independent, rendering the inviability of directly applying the bounded difference inequality. Suppose our data is ordered with observed time $t_1 < t_2 < \ldots < t_N$, then $\wh{\theta}_i^{(k)} (t) = N \wh{S}(t) - (N-1) \wh{S}(t) = N \prod_{t_i \le t } ( \frac{N - i}{N - i + 1}) - (N-1) \prod_{t_i \le t }  (\frac{N - i - 1}{N - i + 1})  = 1 $, where $\wh{S}(t) = \prod_{t_i \le t } ( \frac{N - i}{N - i + 1})^{\Delta_i } $, when there is no censoring. Notably, this is not always the case, as in some extreme case where $\Delta_i = 0 $ for $ i = 1, \ldots, N-3 $ is censored and $\Delta_{N-2} = 1$, we have $ \wh{\theta}_{N-2}^{(k)} (t_{N-2}) = \frac{2}{3} N - \frac{N}{2} = \frac{N}{6} $, which is not bounded when $N$ getting large. Typically, $N \cdot \frac{N - i }{ N - s} = (N-1) \cdot \frac{N - i - 1}{ N - s - 1 } \ge 1$ and this gap increases as $s$ gets larger. Fortunately, a stronger results for bounding the empirical process $\sqrt{N}(F_N - F)$ can be found. 
\begin{lemma}[Restatement of Theorem 4.2 of~\cite{qian1998p}]
	For $1 < p < 2$, if $F$ is any continuous distribution function, then there exists some constant $C_p$ that depends on $p$ only such that 
	$$
	1 \le \inf_{N \ge 1} \frac{\|F - F_{N} \|_{[p]}}{N^{\frac{1-p}{p} }} \le \limsup_{n \rightarrow \infty} \frac{\|F - F_{N} \|_{[p]} }{N^{\frac{1-p}{p} } } \le C_p. 
	$$
\end{lemma}
The probabilistic bound can also be established from Lemma 2.1 of~\cite{kuelbs1977kolmogorov} and Lemma 2.6 of~\cite{dudley1983invariance}. More specifically, we have 
\begin{align*}
	\pr \{ \| F - F_N\|_{[p]} \ge t N^{\frac{1-p}{p} } \} \le \exp(- (t - C_p ) N^{\frac{2}{p} - 1}) 
\end{align*} 
for any $t > C_p$. In light of this, 
\begin{align*}
	&  K \sup_{h} \left \| \sum_{i=1}^N \frac{h(X_i)(\delta_{O_i} - F) }{N}  \right \|_{[p]} \| F_N - F \|_{[p]} \\
	& \qquad +  C_{\mathcal{H}} (N-1)^{-1} K (\max_i \|\delta_{O_i}\|_{[p]} + \|F\|_{[p]})( \max_i \|\delta_{O_i}\|_{[p]} + \|F\|_{[p]} + \|F - F_N \|_{[p]})  \\
	& \lesssim   (\max_i \|\delta_{O_i}\|_{[p]} + \|F\|_{[p]}) \|F - F_N \|_{[p]} \\
	& \qquad + (N-1)^{-1} (\max_i \|\delta_{O_i}\|_{[p]} + \|F\|_{[p]})( \max_i \|\delta_{O_i}\|_{[p]} + \|F\|_{[p]} + \|F - F_N \|_{[p]}).
\end{align*}
Meanwhile, the remainder~\citep{overgaard2017asymptotic} 
\begin{align*}
	& \max_i |R_{N, i}| \lesssim  (1 + \max_i \|\delta_{O_i}\|_{[p]} ) \| F_N - F \|_{[p]}^2 \\
	& \quad + N^{-1} \max_i \|\delta_{O_i}\|_{[p]}^2 ( 1 + \| F_N - F \|_{[p]} ) + N^{-2} \max_i \|\delta_{O_i}\|_{[p]}^3. 
\end{align*}

Given bounded $\|\delta_{O_i}\|_{[p]}$, with probability $\delta$, 
\begin{equation}\label{eqn:pseudo-bound2}
	\sup_{h \in \mathcal{H}} \left| \sum_{i=1}^N  \frac{h(X_i)\wh{\theta}_i^{(k)}(t) }{N}  - E\left[ h(X) m^{(k)}(X; t)\right] \right| \lesssim  \frac{\log(1/\delta)}{N^{1/p}} + C_p N^{(1-p)/p}. 
\end{equation}
As per equation~\eqref{eqn:MicDiamid} and~\eqref{eqn:pseudo-bound2}, we have
\begin{equation}\label{eqn:finite-bound}
	\begin{aligned}
		& \sup_{h \in \sH }\left| E_{\sU_\sS} h(X) \check{m}^{(k)}(X; t) -   \frac{\sum_{i=1}^N  \check{m}^{(k)}(X_i; t) h(X_i)  }{N  }  \right| \\
		& \quad + \sup_{h \in \mathcal{H}} \left| \sum_{i=1}^N  \frac{h(X_i)\wh{\theta}_i^{(k)}(t) }{N}  - E\left[ h(X) m^{(k)}(X; t)\right] \right|  \\
		& \lesssim  \sqrt{ \frac{ \log (2\mathcal{N}_{\alpha}/\delta)  }{N} } + \alpha  + \frac{\log(1/\delta)}{N^{1/p}} + C_p N^{(1-p)/p}. 
	\end{aligned}    
\end{equation}
Firstly, $1 - \frac{1}{p} < \frac{1}{p}$ when $ 1 < p < 2$. The last term $C_p N^{(1-p)/p}$ dominates the third term~\eqref{eqn:finite-bound}: they have the same order if and only if $\delta^{-1} \asymp \exp(N^{\frac{2}{p} - 1}) $. However, $\sqrt{\log(2/\delta)/N}$ diverges when $\delta$ is chosen such small. $N^{(1-p)/p}$ is slower for fixed $\delta$ and large $N$.
Secondly, the first term is related to the covering number. For (I) in Example~\ref{ex:covering}, the first term is dominated by the last term as $ ( \frac{ d\log(1/\alpha)}{\alpha} )^2 < ( \frac{1}{\alpha})^{\frac{p}{p-1}}$, e.g., $p = 4/3$; 
For bounded convex function (II) in Example~\ref{ex:covering},  $\sqrt{\alpha^{-d/2}/N} = O(\alpha)$ only if $N \ge O( \alpha^{-(\frac{d}{2}+2) } )$, which dominates $(\frac{1}{\alpha})^\frac{p}{p-1}$ if $\frac{d}{2} + 2 > \frac{p}{p-1}$.
In summary, to achieve a $O(\alpha)$ bound for the above upper bound, we require sample size $N = O \left( \frac{ \log(2\sN_{\Theta(\alpha)} /\delta) }{ \alpha^2 } + (\frac{1}{\alpha})^\frac{p}{p-1} \right) $.

\subsection{Proof of Theorem~\ref{thm:convergence}}
We have reached a stage for analyzing the convergence of the algorithm. Rather than deploying the conventional multiplicative updates that are mainly apt for classification tasks, we employs additive updates $\wt{m}^{k, b + 1}(X, t) = \wt{m}^{k, b}(X, t) - \eta h(X)$, as of the case in~\cite{roth2022uncertain}. We will track the progress of Algorithm~\ref{alg:MCboost} by tracking the squared error $\ell_{\sD_\sS}(\wt{m}^{k, b}(X, t)) = E_{\sU_\sS} (\wt{m}^{k, b}(X, t) - m^{(k)}(X;t))^2 .$ 
\begin{lemma}\label{lem: iteration}
	The algorithm~\ref{alg:MCboost} converges to a $(\mathcal{H}, \alpha')$-multicalibrated function $\wt{m}^{(k)}(X;t): = m^{(k), B}(X;t) $ in $B = O( \frac{\ell_{\sD_\sS}(\wt{m}^{k, 0}\left(\cdot, t) \right) }{\alpha^2})$ iterations, if we choose step size $\eta = \alpha/(2C_{\mathcal{H}}^2)$.
\end{lemma}
\begin{proof}
	Under Assumption~\ref{ass: finite}, $|h| \le C_{\sH}, \forall h \in \mathcal{H} $, the squared error 
	\begin{align*}
		& \ell_{\sD_\sS}(\wt{m}^{k, b}(X, t)) - \ell_{\sD_\sS}(\wt{m}^{k, b+1}(X, t))\\
		& = E_{\sU_\sS} \left[ (\wt{m}^{k, b}(X, t) - m^{(k)}(X;t))^2 - (\wt{m}^{k, b+1}(X, t) - m^{(k)}(X;t))^2 \right] \\
		& = E_{\sU_\sS} \left[ 2(\wt{m}^{k, b}(X, t) - m^{(k)}(X;t)) \eta h(X) - \eta^2 h^2(X)\right] \\
		& \ge  \eta \alpha - \eta^2 E[h^2(X)] \\
		& \ge  \frac{\alpha^2}{2C_{\sH}^2} - \frac{\alpha^2 E[h^2(X)]}{4 C_{\sH}^4}
		\ge \frac{\alpha^2}{4 C_{\sH}^2},
	\end{align*}
	if we choose the step size $\eta = \alpha/(2C_{\sH}^2 )$. 
	The second to last inequality is due to~\eqref{eqn:finite-bound}
	\begin{align*}
		& \sup_{h \in \sH} E_{\sU_\sS} [h(X) (\wt{m}^{(k), b}(X, t) - m^{(k)}(X;t)) ] \\
		& \ge \sup_{h \in \sH} \left| N^{-1} \sum_{i=1}^N \left[ h(X_i)   (\wt{m}^{(k), b}(X_i, t) - \wh{\theta}^{(k)}_i (t) ) \right]  \right| \\
		& \quad - \sup_{h \in \mathcal{H} } \left| E_{\sU_\sS} h(X) \wt{m}^{(k), b}(X; t)  -   \frac{\sum_{i=1}^N  \wt{m}^{(k), b}(X_i; t) h(X_i)  }{N  }  \right| \\
		& \quad - \sup_{h \in \mathcal{H}} \left| \sum_{i=1}^N  \frac{h(X_i)\wh{\theta}_i^{(k)}(t) }{N}  - E\left[ h(X) m^{(k)}(X; t)\right] \right|  \\
		& \ge \sup_{h \in \sH} \left| N^{-1} \sum_{i=1}^N \left[ h(X_i)   (\wt{m}^{(k), b}(X_i, t) - \wh{\theta}^{(k)}_i (t) ) \right]  \right| - o_p(1) \\
		& > \frac{\alpha}{2},    
	\end{align*}
	for sufficiently large $N$ before the algorithm halts. Since we have $\ell_{\sD_\sS}(\wt{m}^{(k), 0}(X, t)) \le \wt{C}^2 $, we run a circuits to obtain a multi-calibrated function in $B = O( \frac{\ell_{\sD_\sS}(\wt{m}^{(k), 0}\left(\cdot, t) \right) }{\alpha^2}) $ iterations.
\end{proof}
Finally, to achieve multicalibration with a high probability ($1- \delta$) with a finite number of iterations, we need to require 
\begin{align*}
	& \sup_{h \in \mathcal{H} } \left| E_{\sU_\sS} h(X) \wt{m}^{(k), b}(X; t) -   \frac{\sum_{i=1}^N  \wt{m}^{(k), b}(X_i; t) h(X_i)  }{N  }  \right| \\
	& \quad + \sup_{h \in \mathcal{H}} \left| \sum_{i=1}^N  \frac{h(X_i)\wh{\theta}_i^{(k)}(t) }{N}  - E_{\sU_\sS} \left[ h(X) m^{(k)}(X; t)\right] \right|  \\
	& \quad <  \alpha/2  
\end{align*}
occurs with probability at least $1 - \delta/B$ in each iteration, 
and the squared loss error consistently decreases in each iteration before the algorithm halts. As such, $m^{(k), B}(\cdot, t)$ will be multicalibrated from $N \ge \Omega\left( \frac{\log(2\sN_{\Theta(\alpha)}/\delta ) + \log(B) }{\alpha^2} + (\frac{1}{\alpha})^{\frac{p}{p-1}} \right) $ samples with probability at least $1 - \delta$. 

\subsection{Proof for Theorem~\ref{thm:multicali}}

From Theorem~\ref{thm: algo-multi}, we have 
\begin{align*}
	& \Err_\sT(\wt{\tau}^{(k)}(t))  = | E_{\sU_\sT} \wt{m}^{(k)}(X; t) - \tau^{(k)}(t) | \\
	& \le | E_{\sU_\sT} \wt{m}^{(k)}(X; t) - \tau^{(k), ps}(t; \wt{w}) | + | \tau^{(k), ps}(t;\wt{w})- \tau^{(k)}(t) | \\
	& \le \inf_{h \in \sH} \left\{ \left|E_{\sU_\sS} \left[ \left( w(X) - h(X) \right) \wt{m}^{(k)}(X; t) \right]  + E_{\sU_\sS}\left[ h(X)  (\wt{m}^{(k)}(X; t) - m^{(k)}(X;t) ) \right] \right| \right. \\ 
	& \left. \qquad + \left| E_{\sU_s}[ ( h(X) - \wt{w}(X) ) m^{(k)}(X; t) ] \right| \right\} + \Err_{\sT}(\tau^{(k), ps}(t; \wt{w})) \\
	& \le \wt{C} \inf_{h\in\sH} (d(h, w) + d(h, \wt{w})) + \alpha'  + \Err_{\sT}(\tau^{(k), ps}(t; \wt{w}))\\
	& =  \wt{C} \inf_{h\in\sH} (d(h, w) + d(h, \wt{w}))  + \alpha  + \Err_{\sT}(\tau^{(k), ps}(t; \wt{w})) + o_p(1), 
\end{align*}
for any $\wt{w} \in \mathcal{H}(\Sigma)$.
Suppose then we choose $\wt{w} = \wh{w}$, which is the best-fit approximation as mentioned in Definition~\ref{def:multicali}. In that case, if $\wt{C} \inf_{h\in\sH} (d(h, w) + d(h, \wh{w}))$ is small enough, the above arguments indicate that $\wt{\tau}^{(k)}(t)$ is universally adaptable, whose estimation error is only a small range worse than the propensity score estimator using estimated conditional membership probability. 
Particularly, we obtain the bound $\wt{C} d(\wt{w}, w)   + \alpha  + \Err_{\sT}(\tau^{(k), ps}(t; \wt{w})) + o_p(1)$ if $\wt{w} \in \mathcal{H}$.

Furthermore, if $\wt{m}^{(k)}$ is $(\mathcal{H}(\Sigma) \otimes \mathcal{C}, \alpha')$-multicalibrated, where $\mathcal{C}$ is a function class that belongs to $\{ \mathcal{X} \rightarrow [0, \wt{C}]\}$; then for any $c(\cdot) \in \mathcal{C}$ and any $\wt{w} \in \mathcal{H}(\Sigma)$, we have 
\begin{align*}
	& | E_{\sU_\sT} [ c(X) (\wt{m}^{(k)}(X; t) - m^{(k)}(X;t) )  ] |  = | E_{\sU_\sS} [ w(X)c(X)(\wt{m}^{(k)}(X; t) - m^{(k)}(X;t) ) ] |  \\
	&  \le | E_{\sU_\sS} [ \wt{w}(X)c(X)(\wt{m}^{(k)}(X; t) - m^{(k)}(X;t) ) ] | \\
	& \quad + | E_{\sU_\sS} [ ( w(X) -\wt{w}(X) ) c(X)(\wt{m}^{(k)}(X; t) - m^{(k)}(X;t) ) ] | \\
	& \le \alpha' + \wt{C}^2 d( w, \wt{w}). 
\end{align*}
Hence, $\wt{m}^{(k)}$ is $(\mathcal{C}, \alpha')$-multicalibrated on the target domain if it is $( \mathcal{H}(\Sigma) \otimes \mathcal{C}, \alpha' + \wt{C}^2\inf_{\wt{w} \in \mathcal{H}(\Sigma)} d(w, \wt{w}) )$-multicalibrated. 

\subsection{Proof of Corollary~\ref{col:l2}}

We may consider the universal adaptability with regards to the $L_2$ error, i.e., ensuring $E_{\sU_\tau}( \wt{m}^{(k)}(X; t) - m^{(k)}(X;t) )^2 $ small under covariate shift. Recall that 
\begin{align*}
	& E_{\sU_\sT}( \wt{m}^{(k)}(X; t) - m^{(k)}(X;t) )^2 =E_{\sD_\sT}( \wt{m}^{(k)}(X; t) - m^{(k)}(X;t) )(\wt{m}^{(k)}(X; t) - \nu^{(k)}(T; t))  \\
	&  = E_{\sU_\sS} w(X) ( \wt{m}^{(k)}(X; t) - m^{(k)}(X;t) )(\wt{m}^{(k)}(X; t) - \nu^{(k)}(T; t)).
\end{align*}
Therefore, suppose $h(X) =  w(X) ( \wt{m}^{(k)}(X; t) - m^{(k)}(X;t) ) $ and  
$
| E_{\sU_\sS} h(X) (\wt{m}^{(k)}(X; t) - m^{(k)}(X; t)) | < \alpha, 
$
we have $ E_{\sU_\sT }( \wt{m}^{(k)}(X; t) - m^{(k)}(X;t) )^2 < \alpha$. 

However, we have no clue about $m^{(k)}(X; t)$, $\nu^{(k)}(T;t)$, and the true propensity odd $w(X)$.  Suppose the auditing algorithm agnostically learns the class comprising $\{ \wt{w}(X) (m^{(k)}(X; t) - p(X) )\}$ for $\wt{w} \in \mathcal{H}(\Sigma)$ and function $p(X) \in \mathcal{P}$, where $p(X)$ is an approximation to true conditional mean $m^{(k)}(X;t )$, we can again follow the same procedure for calibrating against pseudo observations and obtain 
$$
\sup_{\wt{w} \in \sH(\Sigma), p \in \mathcal{P} } | E_{U_{\sS} } \wt{w}(X) (\wt{m}^{(k)}(X; t) - p(X) ) ( \wt{m}^{(k)}(X; t) - m^{(k)}(X;t) )    | < \alpha' 
$$
where $\alpha' = O(\alpha)$, 
under the assumption that $m^{(k)}(\cdot), p(\cdot), \wt{w} $ are bounded by constant $\wt{C}$. Then 
\begin{align*}
	& E_{\sU_\sT}( \wt{m}^{(k)}(X; t) - m^{(k)}(X;t) )^2  = E_{\sU_\sS} w(X) ( \wt{m}^{(k)}(X; t) - m^{(k)}(X;t) )(\wt{m}^{(k)}(X; t) - m^{(k)}(X; t))) \\ 
	& \le \alpha' + E_{\sU_\sS} \{  w(X) ( \wt{m}^{(k)}(X; t) - m^{(k)}(X;t) )  -  \wt{w}(X) ( \wt{m}^{(k)}(X; t) - p(X) ) \} (\wt{m}^{(k)}(X; t) - m^{(k)}(X; t)))  \\
	& =   E_{\sU_\sS} \{ ( w(X) - \wt{w}(X) ) ( \wt{m}^{(k)}(X; t) - m^{(k)}(X;t) )  -  \wt{w}(X) ( m^{(k)}(X; t) - p(X) ) \} (\wt{m}^{(k)}(X; t) - m^{(k)}(X; t))) \\
	& \quad + \alpha'  \\
	& \le \alpha' + \wt{C}^2 \{ \|w - \wt{w}\|_{L_2} +  \|m^{(k)} - p\|_{L_2}  \} 
\end{align*}
for any $p \in \mathcal{P}, \wt{w} \in \sH(\Sigma)$, where $ \|f - g\|_{L_2} = \sqrt{ E_{\sU_\sS} (f(x) - g(x))^2 } $ is the $L_2$ distance between two functions. Hence, 
$E_{\sU_\sT}( \wt{m}^{(k)}(X; t) - m^{(k)}(X;t) )^2 \le \alpha' + \beta$ if we denote $\beta := \inf_{p, \wt{w} } \|w - \wt{w}\|_{L_2} +  \|m^{(k)} - p\|_{L_2} $.

\subsection{Proof of Theorem~\ref{thm:multi-source} }
Our method extends naturally to multiple source domains, denoted as $\sS_1, \sS_2, \ldots, \sS_M$, and a single target domain $\sT$. Our goal is to establish an upper bound for the estimation error of $\wt{m}^{(k)}(X; t)$ under this multi-source setting.  

{\noindent \bf Step 1. } First, we verify that $\wt{m}^{(k)}(X; t)$ satisfies the following ``averaged'' multicalibration condition,
\begin{lemma}~\label{lem:average-multi}
	The output function $\wt{m}^{(k)}(X; t)$ of our algorithm applied to multiple source data satisifies the averaged multicalibration condition:  
	\begin{align*}
		\left| \sum_{m=1}^M \pr(D = \sS_m | D \in S) \Expect_{\sU_{\sS_m} }\left\{ h(X)  (\wt{m}^{(k)}(X; t) - m^{(k)}(X;t) ) \right\} \right| < \alpha.
	\end{align*}
\end{lemma} 
\begin{proof}[Sketch of proof]
	We have 
	$$
	f(X | D \in \sS) = \sum_{m=1}^M f(X | D = \sS_m ) \pr(D = \sS_m | D \in S).  
	$$
	We express the expectation of the function over the target domain in terms of the source domains 
	\begin{align*}
		& \Expect_{\sD_\sT }[m^{(k)}(X; t)] = \int f(X|\sT) m^{(k)}(X; t) \mu(dX) \\
		& = \sum_{m=1}^M \pr(D = \sS_m | D \in S) \int \frac{f(X|\sT)}{f(X|\sS_m )} m^{(k)}(X; t) f(X|\sS_m ) \mu(dX)     \\
		& =  \sum_{m = 1}^M \pr(D = \sS_m | D \in S)  \Expect_{\sU_{\sS_m} } \left\{ r_m \frac{\sigma_{\sT }(X)}{\sigma_{\sS_m }(X) } m^{(k)}(X; t) \right\}, 
	\end{align*}
	where the weight function for domain $\sS_m$ is defined as 
	$$
	w_m(X): = r_m  \frac{\sigma_{\sT }(X)}{\sigma_{\sS_m }(X) },  
	$$
	where $r_m = \pr(D = \sS_m)/\pr(D = \sT)$ and $ \sigma_{\sS_m}(X) = \pr(D = \sS_m | X)$. 
	Additionally, we define $p_m = \pr(D = \sS_m | D \in \sS) $ and $\wh{p}_m = N_m/N$, where $N_m$ is the sample size for source domain $\sS_m$. We aim to bound 
	\begin{align*}
		& \sup_{h \in \sH} \left| \sum_{m=1}^M p_m \Expect_{\sU_{\sS_m}}(\wt{m}^{(k)}(X;t) - m^{(k)}(X; t) ) \right| \\
		& \le \sup_{h \in H} \left| \frac{\sum_{i=1}^N h(X_i) \{ \wt{m}^{(k)}(X;t) - \wh{\theta}_i^{(k)}(t) \} }{N} \right| \\
		& \quad + \sup_{h \in \sH} \left| \frac{\sum_{i=1}^N h(X_i) \{ \wt{m}^{(k)}(X;t) - \wh{\theta}_i^{(k)}(t) \} }{N} - \sum_{m=1}^M p_m \Expect_{\sU_{\sS_m}}\{ \wt{m}^{(k)}(X;t) - m^{(k)}(X; t) \} \right| \\
		& < \alpha + \sup_{h \in \sH} \left| \frac{\sum_{i=1}^N h(X_i) \{ \wt{m}^{(k)}(X;t) - \wh{\theta}_i^{(k)}(t) \} }{N} - \sum_{m=1}^M p_m \Expect_{\sU_{\sS_m}}\{ \wt{m}^{(k)}(X;t) - m^{(k)}(X; t) \} \right|. 
	\end{align*}
	The second term in the last line of above inequality can be decomposed into 
	\begin{align*}
		& \sup_{h \in \sH} \left| \frac{\sum_{i=1}^N h(X_i) \wt{m}^{(k)}(X;t)  }{N} - \sum_{m=1}^M p_m \Expect_{\sU_{\sS_m}} \wt{m}^{(k)}(X;t) \right| \\
		& + \sup_{h \in \sH} \left|\frac{\sum_{i=1}^N h(X_i)  \wh{\theta}_i^{(k)}(t) }{N} -  \sum_{m=1}^M p_m \Expect_{\sU_{\sS_m}}m^{(k)}(X; t) \right|. 
	\end{align*}
	We then have 
	\begin{equation}\label{eqn:multisource-1}
		\begin{aligned}
			& \sup_{h \in \sH} \left| \frac{\sum_{i=1}^N h(X_i) \wt{m}^{(k)}(X_i;t)  }{N} - \sum_{m=1}^M p_m \Expect_{\sU_{\sS_m}} \wt{m}^{(k)}(X;t) \right| \\
			& = \sup_{h \in \sH} \left| \sum_{m=1}^M (\wh{p}_m - p_m)\Expect_{\sU_{\sS_m}} \wt{m}^{(k)}(X;t) \right. \\ 
			& \quad \left. + \sum_{m=1}^M \wh{p}_m \{  \frac{\sum_{D_i = \sS_m} h(X_i) \wt{m}^{(k)}(X_i;t)  }{N_m} -  \Expect_{\sU_{\sS_m}} \wt{m}^{(k)}(X;t)
			\} \right|,  
		\end{aligned}
	\end{equation}
	and
	\begin{equation}\label{eqn:multisource-2}
		\begin{aligned}
			& \sup_{h \in \sH} \left| \frac{\sum_{i=1}^N h(X_i) \wh{\theta}_i^{(k)}(t)  }{N} - \sum_{m=1}^M p_m \Expect_{\sU_{\sS_m}} m^{(k)}(X;t) \right| \\
			& = \sup_{h \in \sH} \left| \sum_{m=1}^M (\wh{p}_m - p_m)\Expect_{\sU_{\sS_m}} m^{(k)}(X;t) + \sum_{m=1}^M \wh{p}_m \{  \frac{\sum_{D_i = \sS_m} h(X_i) \wh{\theta}^{(k)}_i(t)  }{N_m} -  \Expect_{\sU_{\sS_m}} m^{(k)}(X;t)
			\} \right|,  
		\end{aligned}
	\end{equation}
	Following the arguments of Theorem~~\ref{thm: algo-multi} and Corollary~\ref{coro:complexity}, we can show~\eqref{eqn:multisource-1} has order $O_p(1/\sqrt{N}) + O_p(\sum_{m=1}^M \sqrt{N_m}/N )$ and~\eqref{eqn:multisource-2} has order 
	$O_p(\sum_{m=1}^M N_m^{1-\lambda}/N )$. As sample size gets larger, 
	$$ \sup_{h \in \sH} \left| \sum_{m=1}^M p_m \Expect_{\sU_{\sS_m}}(\wt{m}^{(k)}(X;t) - m^{(k)}(X; t) ) \right| < \alpha 
	$$ 
	for $\alpha > 0.$ 
\end{proof} 

{\noindent \bf Step 2. } For convenience, we define the vector of weight functions
$$
\vw(X) = (w_1(X), w_2(X), \ldots, w_M(X)).
$$  
Now, using Lemma~\ref{lem:average-multi}, we establish the bound for the error. We begin by decomposing the target error. 
\begin{align*}
	& \Err_{\sT}(\wt{m}^{(k)}(X; t)) \\
	& \le \inf_{h \in \sH} \left[  \left| \sum_{m=1}^M \pr(D = \sS_m | D \in S) E_{\sU_{\sS_m} } \left\{ \left(  w_m(X) - h(X) \right) \wt{m}^{(k)}(X; t) \right\} \right| \right. \\
	& + \left| \sum_{m=1}^M \pr(D = \sS_m | D \in S) \Expect_{\sU_{\sS_m} }\left\{ h(X)  (\wt{m}^{(k)}(X; t) - m^{(k)}(X;t) ) \right\} \right|  \\ 
	& \left. \qquad + \left| \sum_{m=1}^M \pr(D = \sS_m | D \in S) \Expect_{\sU_{\sS_m} }\{ ( h(X) -  \wt{w}_m(X) ) m^{(k)}(X; t) \} \right| \right] + \Err_{\sT}(\tau^{(k), ps}(t;  \wt{\vw})) \\
	&\le \Err_{\sT}(\tau^{(k), ps}(t; \wt{\vw})) + \sum_{m=1}^M \pr(D = \sS_m | D \in S) \alpha \\
	& \quad + \wt{C} \inf_{h\in\sH} \sum_{m=1}^M  \pr(D = \sS_m | D \in S) \{ d_{\sS_m}(w_m, h) + d_{\sS_m}(h, \wt{w}_m ) \} \\
	& = \Err_{\sT}(\tau^{(k), ps}(t; \wt{\vw})) + \alpha + \wt{C} \inf_{h\in\sH} \sum_{m=1}^M  \pr(D = \sS_m | D \in S) \{ d_{\sS_m}(w_m, h) + d_{\sS_m}(h, \wt{w}_m ) \}. 
\end{align*} 
Thus, if $\wt{w}_m \in \mathcal{H}$, the last term in the final line simplifies to $$\wt{C} \sum_{m=1}^M  \pr(D = \sS_m | D \in S) d_{\sS_m}(w_m, \wt{w}_m ).  $$ Alternatively, if there is sufficient overlap, this discrepancy term remains small, ensuring the robust adaptation.

\subsection{The formulation of IPSW estimator}

To estimate the target domain statistics, we express the expectation as
\begin{align*}
	E_{\sD_\sT}[\nu^{(k)} (T_i; t)] 
	=  \frac{ \int_{\mathcal{X}} m^{(k)}(X; t) \frac{\pr( D = \sT | X) }{\pr(D = \sS | X)}  \frac{\pr(D = \sS)}{\pr(D = \sT)} f(X| D = \sS)\mu(dX)  }{ \int_{\mathcal{X}}  \frac{\pr( D = \sT | X) }{\pr(D = \sS | X)}  \frac{\pr(D = \sS)}{\pr(D = \sT)} f(X | D = \sS) \mu(dX) }, 
\end{align*}
This motivates the empirical IPSW estimator 
$$
\frac{\sum_{i=1}^N w(X_i) \wh{\theta}_i^{(k)} (t)}{\sum_{i=1}^N w(X_i) },
$$  
and the constant $r = \pr(D = \sS)/ \pr(D = \sT) $ cancels out, resulting in 
$$
\frac{\sum_{i=1}^N \frac{1 - \sigma(X_i)}{ \sigma(X_i)} \wh{\theta}_i^{(k)} (t)}{\sum_{i=1}^N \frac{1 - \sigma(X_i)}{ \sigma(X_i)} }. 
$$
In a multi-source domain setting, where we have multiple source distribution contributing to the estimation, the expectation under the target domain is expressed as  
\begin{align*}
	& \Expect_{\sD_\sT }[m^{(k)}(X; t)] = \int f(X|\sT) m^{(k)}(X; t) \mu(dX) \\
	& = \sum_{m=1}^M \pr(D = \sS_m | D \in S) \int \frac{f(X|\sT)}{f(X|\sS_m )} m^{(k)}(X; t) f(X|\sS_m ) \mu(dX)     \\
	& =  \sum_{m = 1}^M \pr(D = \sS_m | D \in S)  \Expect_{\sU_{\sS_m} } \left\{ w_m(X) m^{(k)}(X; t) \right\}.  
\end{align*} 
This can be empirically estimated as  
$$
\frac{\sum_{m=1}^M \sum_{i=1}^N \b1\{D_i = \sS_m\} w_m(X_i) \wh{\theta}_i^{(k)}(t) }{ \sum_{m=1}^M\sum_{i=1}^N \b1\{ D_i = \sS_m\} w_m(X_i)}. 
$$

\section{Additional Simulations}\label{apxsec:simu}
We have carried out additional simulations that are complementary to those reported in the main paper. These simulations are designed to explore different scenarios with varying parameter configurations $(t \in \{5, 10, 30, 50, 70\}$, $q = 1, 2, 3)$, thus providing a comprehensive understanding of our model under different conditions. 

\subsection{IPSM estimator improves over the naive estimator}\label{sup_ipsm_est}
\begin{figure}[ht!]
	\centering
	\includegraphics[width=0.95\textwidth]{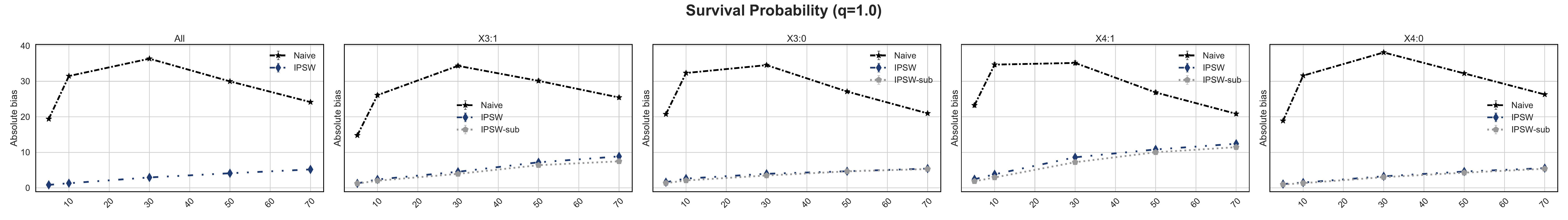}
	\includegraphics[width=0.95\textwidth]{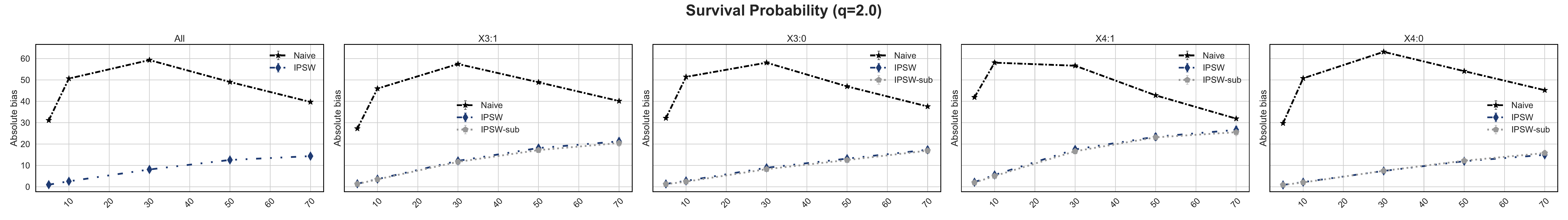}
	\includegraphics[width=0.95\textwidth]{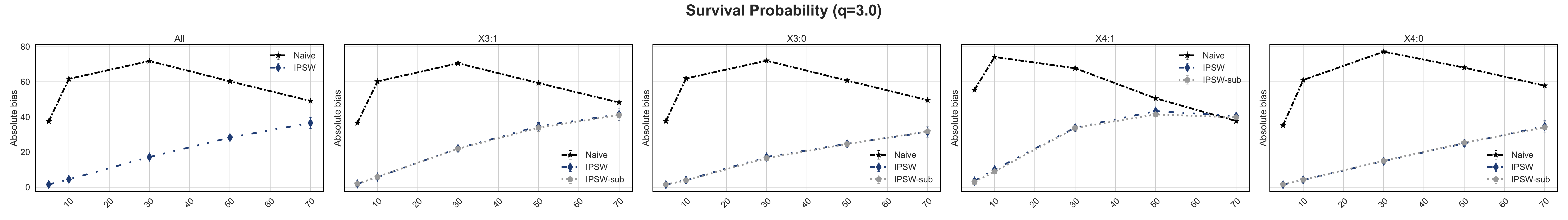}
	\caption{ The absolute bias $(\times 10^2)$ for `Naive,' `IPSW,' and `IPSW-sub' in estimating survival probability under varying degrees of covariate shift ($q = 1, 2, 3$), with a total sample size of $N = 1000$.  The survival time adheres to the proportional hazard model, and the censoring is covariate-independent. Results are aggregated from $100$ simulation replications for all subpopulations.
	}
	\label{fig:IPSW-SP-PH-indep}
\end{figure}

\begin{figure}[ht!]
	\centering
	\includegraphics[width=0.95\textwidth]{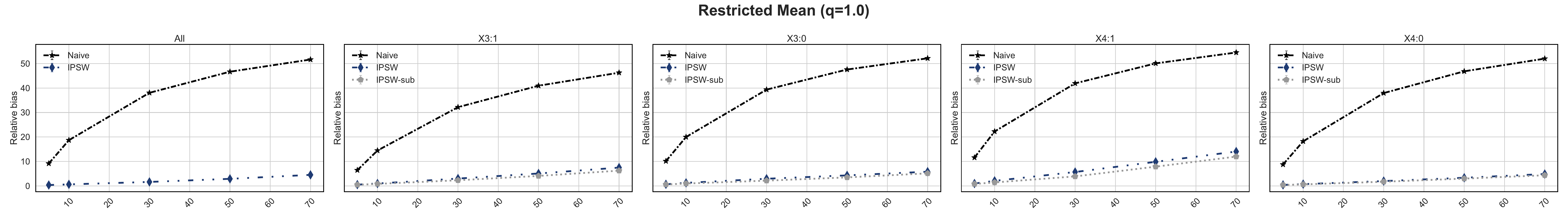}
	\includegraphics[width=0.95\textwidth]{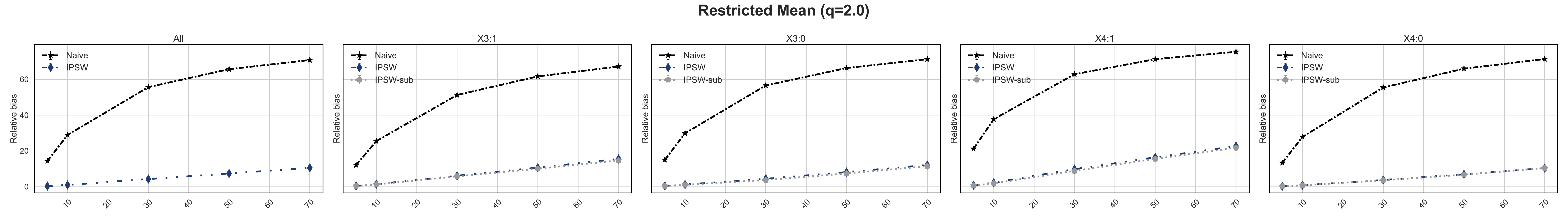}
	\includegraphics[width=0.95\textwidth]{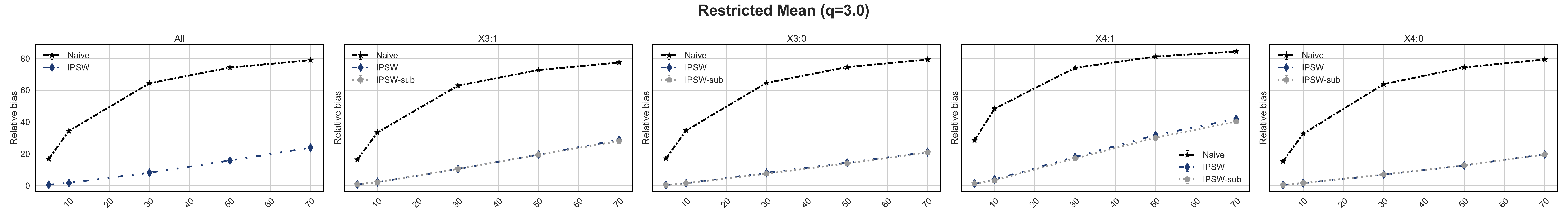}
	\caption{The relative bias $(\times 10^2)$ for `Naive,' `IPSW,' and `IPSW-sub' in estimating restricted mean under varying degrees of covariate shift ($q = 1, 2, 3$), with a total sample size of $N = 1000$.  The survival time adheres to the proportional hazard model, and the censoring is covariate-independent. Results are aggregated from $100$ simulation replications for all subpopulations.  }
	\label{fig:IPSW-RM-PH-indep}
\end{figure}

Figures~\ref{fig:IPSW-SP-PH-indep},~\ref{fig:IPSW-RM-PH-indep} show the improvement of \textsc{ipsw} over naive approach in reducing the estimation bias of survival probability and restrictive mean survival time for mmodels with mild, moderate and strong covariate shift. 

\subsection{Estimation of survival probability under mild and moderate covariate shift}\label{sup_survprob_est}

\begin{figure}[ht!]
	\centering
	\includegraphics[width=0.9\textwidth, height=0.8\textheight]{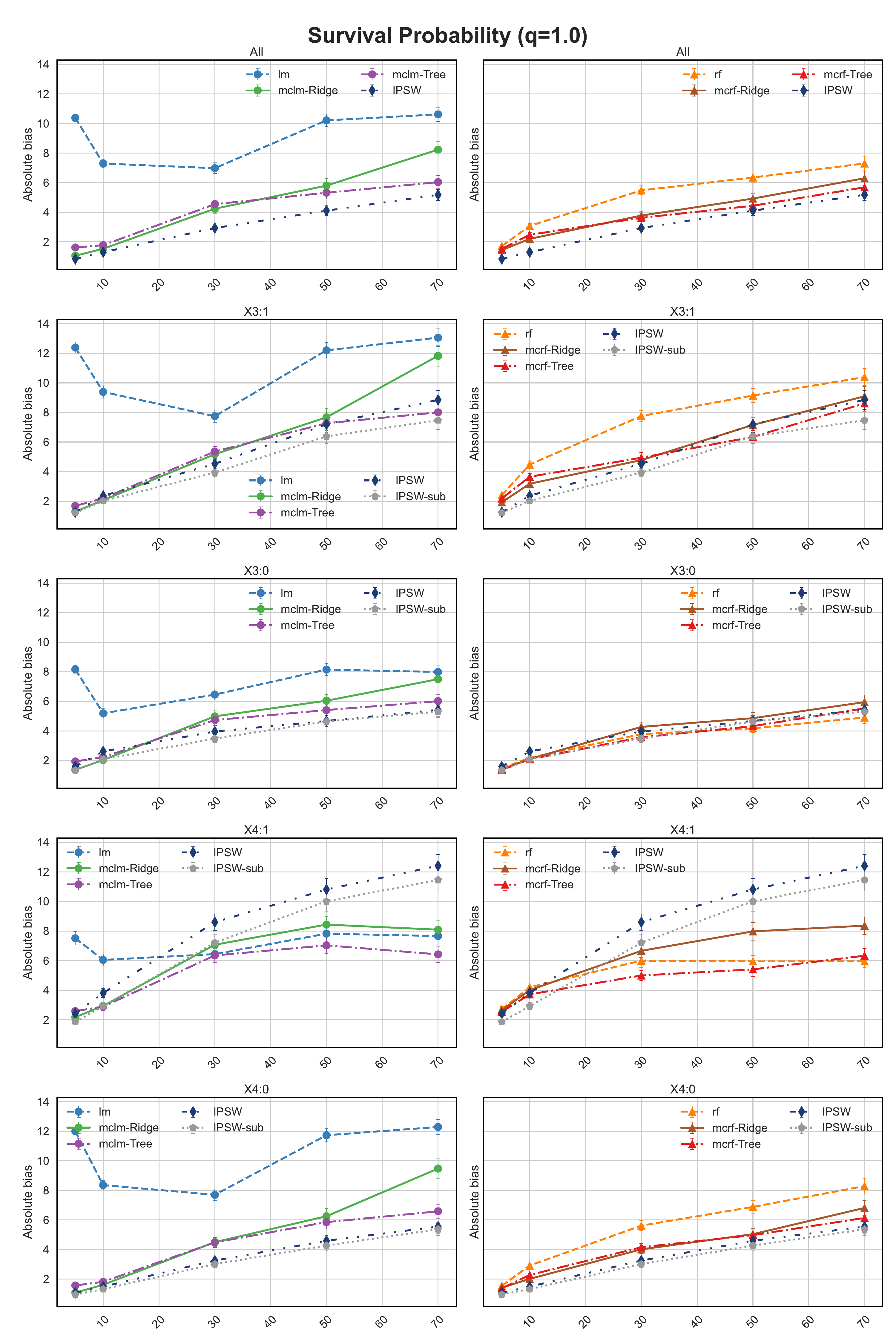}
	\caption{The absolute bias $(\times 10^2)$ of  different  methods in estimating survival probability
		under mild covariate shift ($q = 1$) with a total sample size of $N = 1000$ (600 source and 400 target samples). The survival time is generated from a proportional hazard model, and the censoring is covariate-independent. Results for all subpopulations are presented based on $100$ simulation replications. Left panel: linear model is used to estimate the survival probability using pseudo observations. Right panel: random forest is used to estimate the survival probability using pseudo observations.}
	\label{fig:SP-PH-indep-1}
\end{figure}

\begin{figure}
	\centering
	\includegraphics[width=0.9\textwidth, height=0.8\textheight]{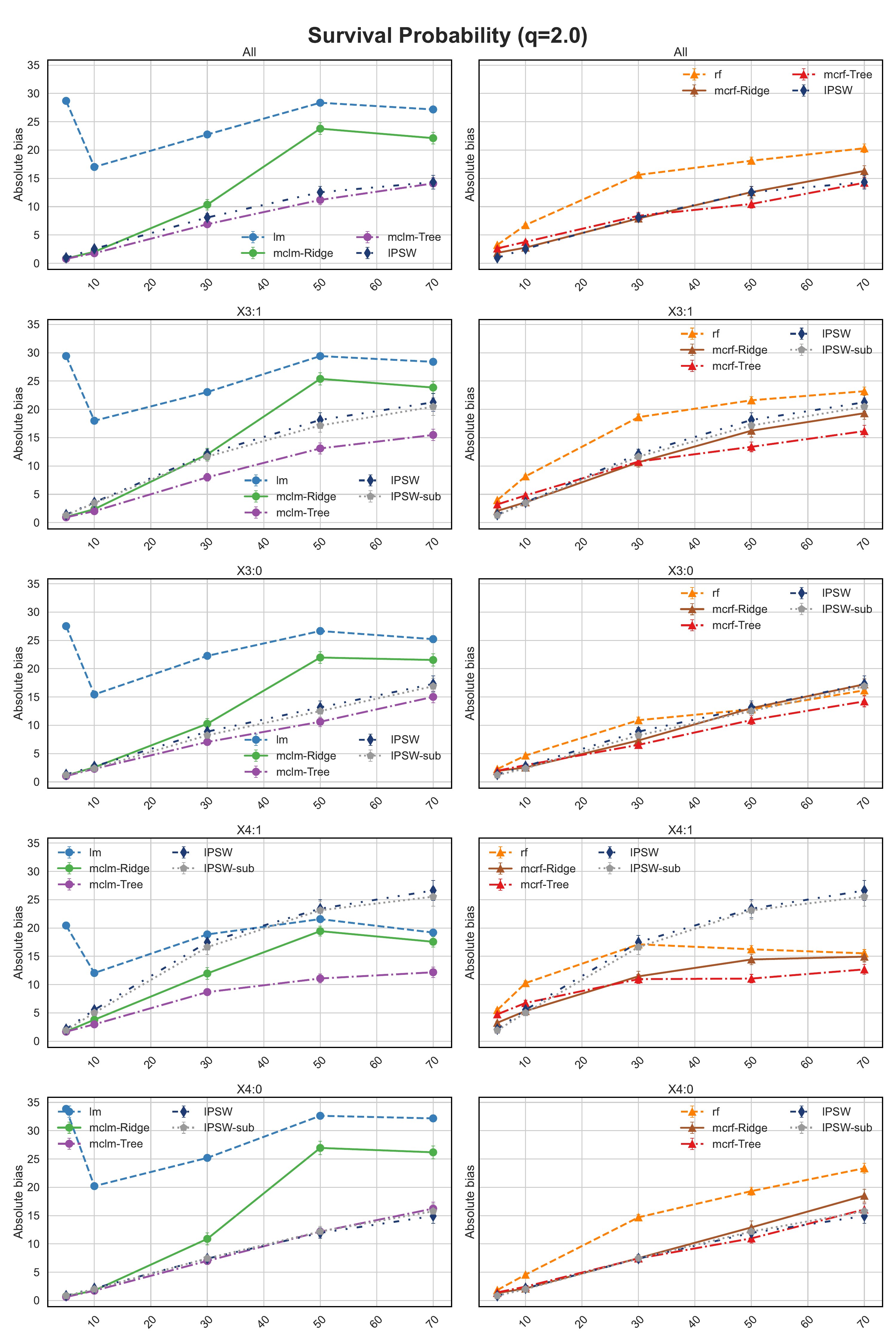}
	\caption{The absolute bias $(\times 10^2)$ for different methods in estimating survival probability
		under mild covariate shift ($q = 2$) with a total sample size of $N = 1000$ (600 source and 400 target samples). The survival time adheres to the proportional hazard model, and the censoring is covariate-independent. Results for all subpopulations are presented based on $100$ simulation replications.}
	\label{fig:SP-PH-indep-2}
\end{figure}

Figures~\ref{fig:SP-PH-indep-1}-\ref{fig:SP-PH-indep-2} present results for moderate shift,

\subsection{Estimation of restricted mean survival time}\label{sup_RM_est}
\begin{figure}
	\centering
	\includegraphics[width=0.9\textwidth, height=0.8\textheight]{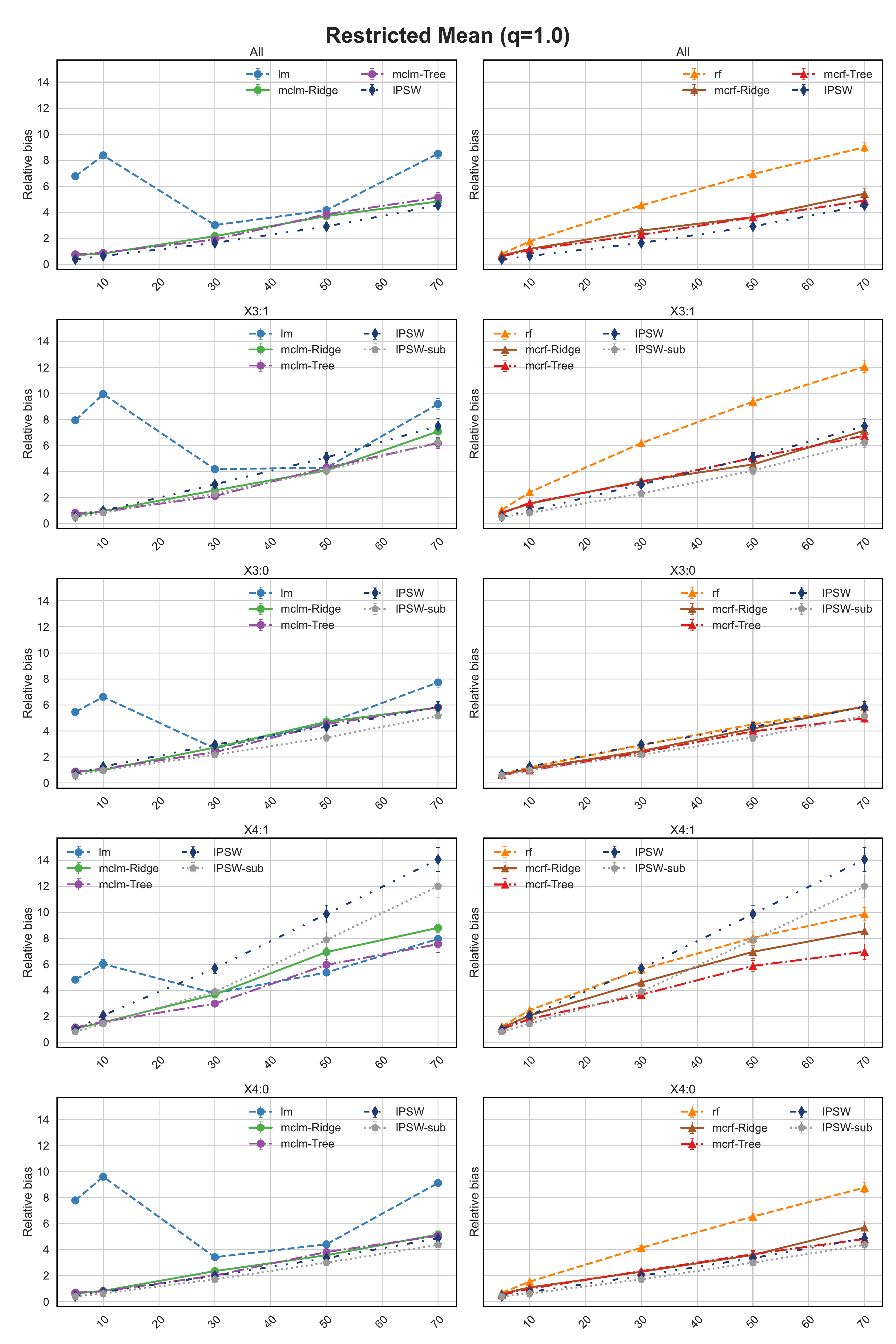}
	\caption{The relative bias $(\times 10^2)$ for all methods in estimating restricted survival mean time
		under mild covariate shift ($q = 1$) with a total sample size of $N = 1000$ (600 source and 400 target samples). The survival time adheres to the proportional hazard model, and the censoring is covariate-independent. Results for all subpopulations are presented based on $100$ simulation replications.}
	\label{fig:RM-PH-indep-1}
\end{figure}

\begin{figure}
	\centering
	\includegraphics[width=0.9\textwidth, height=0.8\textheight]{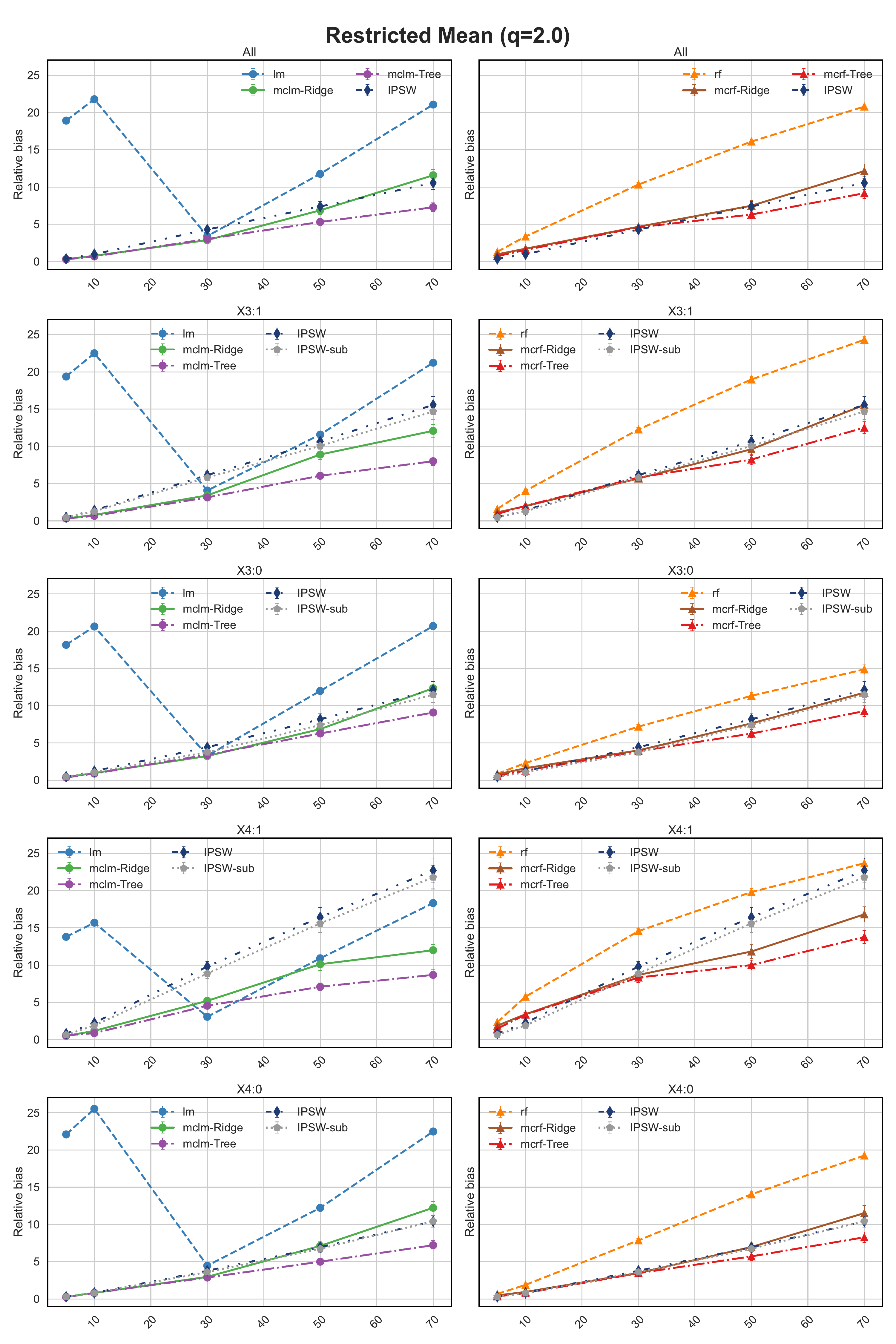}
	\caption{The relative bias $(\times 10^2)$ for all methods in estimating the restricted survival mean time 
		under strong covariate shift ($q = 2$) with a total sample size of $N = 1000$ (600 source and 400 target samples). The survival time adheres to the proportional hazard model, and the censoring is covariate-independent. Results for all subpopulations are presented based on $100$ simulation replications.}
	\label{fig:RM-PH-indep-2}
\end{figure}

\begin{figure}
	\centering
	\includegraphics[width=0.9\textwidth, height=0.8\textheight]{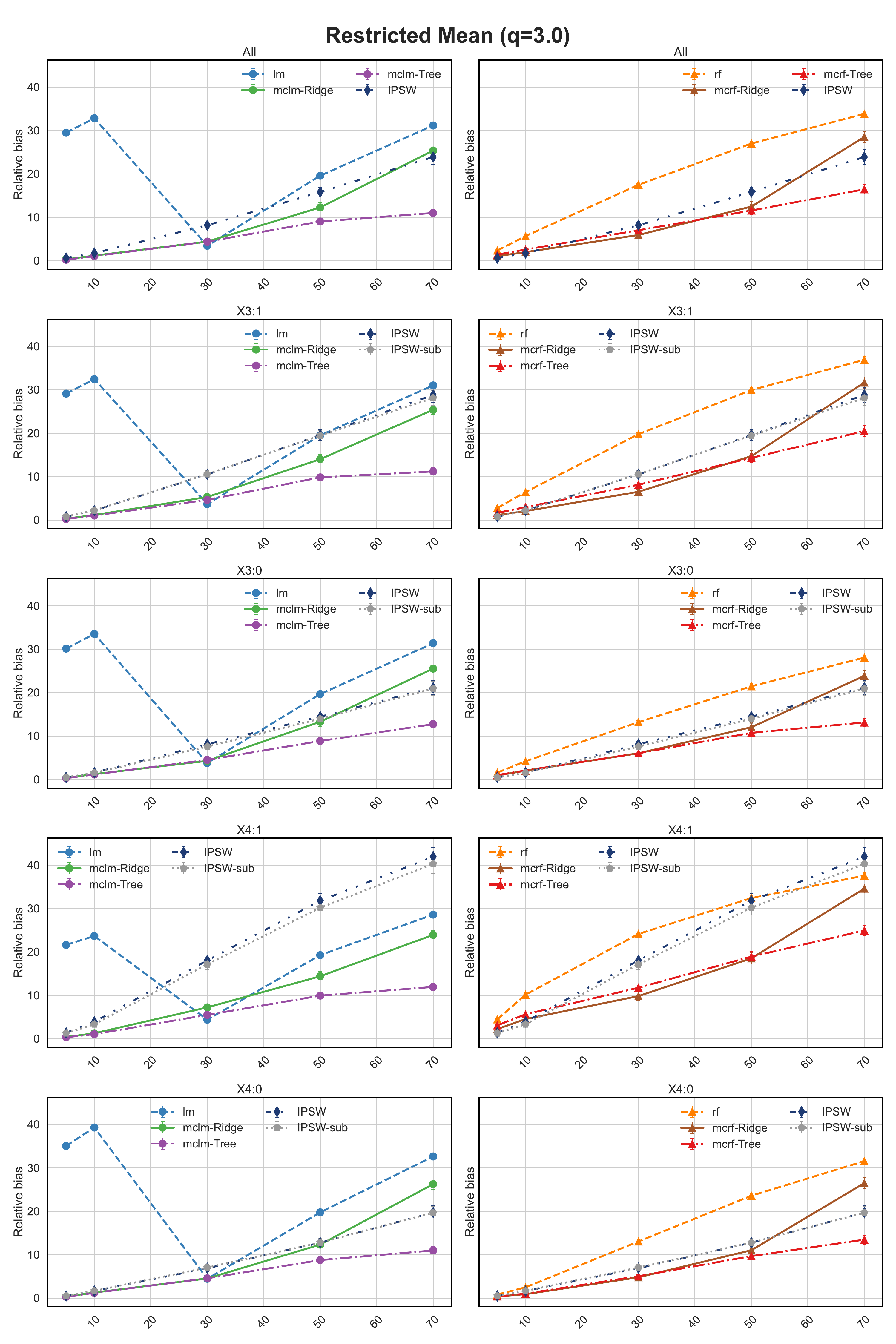}
	\caption{The relative bias $(\times 10^2)$ for all methods in estimating the restricted survival mean time 
		under strong covariate shift ($q = 3$) with a total sample size of $N = 1000$ (600 source and 400 target samples). The survival time adheres to the proportional hazard model, and the censoring is covariate-independent. Results for all subpopulations are presented based on $100$ simulation replications.}
	\label{fig:RM-PH-indep-3}
\end{figure}

Figures~\ref{fig:RM-PH-indep-1} through \ref{fig:RM-PH-indep-3} show the performance in estimating the restrictive mean survival time under mild, moderate and strong covariate shift.  


\subsection{Simulation comparisons under accelerated failure time models}  \label{supp_AFT_simu}
Our simulation also explored different survival and censoring distribution models. For the accelerated failure time model, we consider the survival times where $\log(T_i) \sim \mathcal{N}(\mu, \sigma^2 = 0.64)$ with $\mu = 3.5 - X_i^\top \alpha$. The results are shown in Tables~\ref{simu:aft-1}-\ref{simu:aft-3}.

\begin{table}[ht!]
	\centering
	\resizebox{\textwidth}{!}{
		\begin{tabular}{llccccccccc}
			\hline 
			\hline
			&  Time  &     \textsc{Naive}      &  \textsc{IPSW}     & \textsc{IPSW-sub}   &              \textsc{lm}   &          \textsc{ mclm-Ridge} &   \textsc{mclm-Tree} &    \textsc{ rf} & \textsc{mcrf-Ridge} &   \textsc{ mcrf-Tree} \\
			\hline 
			\multicolumn{11}{l}{Absolute bias $(\times 10^2)$, $N = 1000, q = 1$, accelerate failure hazard, independent censoring, single source }                              \\
			\hline
			\multirow{5}{*}{\textsc{All} } & 5  &   4.91 (0.11) &  \textbf{\textbf{0.55} (0.04)} &                      - &                    3.10 (0.14) &     0.70 (0.06) &                    0.71 (0.06) &                    0.79 (0.06) &     0.72 (0.06) &                    0.86 (0.07) \\
			& 10 &  13.03 (0.19) &  \textbf{\textbf{1.20} (0.08)} &                      - &                    4.81 (0.23) &     1.55 (0.14) &                    1.31 (0.10) &                    2.03 (0.14) &     1.88 (0.14) &                    1.87 (0.13) \\
			& 30 &  25.69 (0.33) &                    3.73 (0.25) &                     - &  \textbf{\textbf{3.17} (0.24)} &     4.11 (0.29) &                    4.12 (0.29) &                    4.38 (0.31) &     4.44 (0.33) &                    4.29 (0.30) \\
			& 50 &  24.87 (0.38) &                    4.69 (0.35) &                      - &  \textbf{\textbf{4.58} (0.33)} &     5.45 (0.39) &                    5.23 (0.41) &                    5.57 (0.37) &     5.62 (0.42) &                    4.95 (0.41) \\
			& 70 &  21.66 (0.41) &                    5.84 (0.44) &                      - &                    5.80 (0.41) &     6.06 (0.50) &                    5.79 (0.44) &                    6.04 (0.40) &     5.82 (0.47) &  \textbf{\textbf{5.16} (0.41)} \\
			\cline{1-11}
			\multirow{5}{*}{ $X_3 = 1$} & 5  &   3.71 (0.18) &  \textbf{\textbf{0.72} (0.06)} &                    0.73 (0.06) &                    3.72 (0.20) &     0.87 (0.07) &                    0.92 (0.07) &                    1.03 (0.07) &     0.96 (0.08) &                    1.14 (0.08) \\
			& 10 &  10.52 (0.32) &                    1.98 (0.15) &                    1.83 (0.15) &                    5.90 (0.33) &     2.09 (0.19) &  \textbf{\textbf{1.75} (0.13)} &                    2.73 (0.19) &     2.58 (0.19) &                    2.61 (0.19) \\
			& 30 &  23.32 (0.55) &                    5.82 (0.40) &                    5.24 (0.39) &  \textbf{\textbf{4.33} (0.36)} &     5.78 (0.40) &                    5.56 (0.39) &                    6.35 (0.40) &     6.53 (0.46) &                    5.91 (0.38) \\
			& 50 &  24.26 (0.60) &                    7.62 (0.56) &                    7.04 (0.53) &  \textbf{\textbf{6.28} (0.44)} &     7.80 (0.54) &                    6.62 (0.50) &                    7.93 (0.54) &     8.28 (0.61) &                    6.89 (0.49) \\
			& 70 &  21.80 (0.62) &                    8.99 (0.66) &                    8.48 (0.60) &                    7.85 (0.53) &     8.83 (0.75) &                    7.28 (0.55) &                    8.43 (0.56) &     8.57 (0.71) &  \textbf{\textbf{6.84} (0.52)} \\
			\cline{1-11}
			\multirow{5}{*}{$X_3 = 0$} & 5  &   5.24 (0.14) &  \textbf{\textbf{0.81} (0.06)} &                    0.82 (0.06) &                    2.53 (0.15) &     0.91 (0.07) &                    0.87 (0.07) &                    0.85 (0.06) &     0.91 (0.07) &                    0.91 (0.07) \\
			& 10 &  13.45 (0.26) &                    1.69 (0.14) &  \textbf{\textbf{1.63} (0.13)} &                    4.01 (0.24) &     1.85 (0.15) &                    1.79 (0.14) &                    1.82 (0.15) &     1.87 (0.14) &                    1.81 (0.14) \\
			& 30 &  24.79 (0.43) &                    3.80 (0.27) &  \textbf{\textbf{3.76} (0.25)} &                    3.97 (0.26) &     3.82 (0.33) &                    3.80 (0.32) &                    3.87 (0.29) &     4.49 (0.35) &                    4.08 (0.33) \\
			& 50 &  22.84 (0.51) &                    5.22 (0.38) &                    4.94 (0.38) &                    5.10 (0.36) &     5.52 (0.42) &                    6.06 (0.47) &  \textbf{\textbf{4.91} (0.36)} &     5.78 (0.44) &                    5.86 (0.44) \\
			& 70 &  19.27 (0.57) &                    5.97 (0.45) &                    5.77 (0.43) &                    5.91 (0.42) &     6.74 (0.57) &                    7.22 (0.57) &  \textbf{\textbf{5.67} (0.42)} &     6.62 (0.50) &                    6.82 (0.52) \\
			\cline{1-11}
			\multirow{5}{*}{$X_4 = 1$} & 5  &   6.35 (0.29) &  \textbf{\textbf{1.09} (0.09)} &                    1.12 (0.09) &                    3.01 (0.19) &     1.49 (0.12) &                    1.42 (0.12) &                    1.37 (0.09) &     1.72 (0.14) &                    1.48 (0.10) \\
			& 10 &  15.31 (0.45) &                    2.71 (0.23) &  \textbf{\textbf{2.60} (0.20)} &                    4.82 (0.30) &     3.02 (0.22) &                    2.82 (0.21) &                    3.10 (0.24) &     3.63 (0.26) &                    3.18 (0.24) \\
			& 30 &  26.41 (0.67) &                    8.78 (0.66) &                    8.21 (0.65) &  \textbf{\textbf{6.00} (0.45)} &     7.15 (0.57) &                    6.76 (0.48) &                    6.20 (0.50) &     8.32 (0.63) &                    6.79 (0.48) \\
			& 50 &  24.74 (0.65) &                    9.82 (0.80) &                    9.37 (0.73) &  \textbf{\textbf{6.32} (0.47)} &     8.24 (0.74) &                    6.78 (0.58) &                    6.65 (0.49) &     8.56 (0.76) &                    7.19 (0.52) \\
			& 70 &  21.15 (0.63) &                   11.92 (0.87) &                   11.54 (0.86) &                    6.94 (0.48) &     9.63 (0.75) &                    7.88 (0.56) &                    6.94 (0.49) &     9.10 (0.69) &  \textbf{\textbf{6.85} (0.50)} \\
			\cline{1-11}
			\multirow{5}{*}{$X_4 = 0$} & 5  &   4.68 (0.12) &                    0.57 (0.05) &  \textbf{\textbf{0.56} (0.05)} &                    3.54 (0.16) &     0.65 (0.07) &                    0.62 (0.05) &                    0.74 (0.06) &     0.64 (0.05) &                    0.79 (0.06) \\
			& 10 &  12.85 (0.21) &                    1.40 (0.10) &                    1.35 (0.10) &                    5.50 (0.25) &     1.62 (0.15) &  \textbf{\textbf{1.32} (0.10)} &                    1.95 (0.14) &     1.80 (0.12) &                    1.75 (0.12) \\
			& 30 &  26.47 (0.41) &  \textbf{\textbf{3.39} (0.31)} &                    3.48 (0.29) &                    3.47 (0.27) &     4.11 (0.33) &                    4.13 (0.33) &                    4.28 (0.31) &     4.66 (0.35) &                    4.59 (0.30) \\
			& 50 &  25.90 (0.45) &  \textbf{\textbf{4.92} (0.39)} &                    4.99 (0.39) &                    5.20 (0.34) &     6.03 (0.47) &                    6.04 (0.47) &                    5.80 (0.39) &     6.49 (0.51) &                    5.77 (0.47) \\
			& 70 &  22.70 (0.50) &                    6.24 (0.46) &                    6.40 (0.49) &                    6.62 (0.44) &     7.29 (0.61) &                    6.82 (0.58) &                    6.75 (0.41) &     7.26 (0.61) &  \textbf{\textbf{5.93} (0.52)} \\
			\bottomrule
		\end{tabular}
	}
	\caption{ 
		The absolute bias $(\times 10^2)$ for all methods 
		under mild covariate shift ($q = 1$) with a total sample size of $N = 1000$. The survival time follows the accelerated failure hazard model with covariate-independent censoring. 
		Results for all subpopulations are presented based on $100$ simulation replications.  
		\label{simu:aft-1}
	}
\end{table}

\begin{table}[ht!]
	\centering
	\resizebox{\textwidth}{!}{
		\begin{tabular}{llccccccccc}
			\hline 
			\hline
			&  Time  &     \textsc{Naive}      &  \textsc{IPSW}     & \textsc{IPSW-sub}   &              \textsc{lm}   &          \textsc{ mclm-Ridge} &   \textsc{mclm-Tree} &    \textsc{ rf} & \textsc{mcrf-Ridge} &   \textsc{ mcrf-Tree} \\
			\hline 
			\multicolumn{11}{l}{Absolute bias $(\times 10^2)$, $N = 1000, q = 2$, accelerate failure hazard, independent censoring, single source }                              \\
			\hline
			\multirow{5}{*}{All} & 5  &   7.59 (0.13) &  \textbf{\textbf{0.45} (0.04)} &                     - &                     8.48 (0.28) &     0.54 (0.07) &                     0.53 (0.05) &                     1.43 (0.09) &     1.05 (0.13) &                     1.30 (0.16) \\
			& 10 &  20.60 (0.19) &                    1.42 (0.16) &                      - &                    13.20 (0.41) &     2.00 (0.29) &   \textbf{\textbf{1.29} (0.08)} &                     4.08 (0.19) &     2.88 (0.23) &                     3.44 (0.20) \\
			& 30 &  42.42 (0.29) &                    7.86 (0.62) &                      - &   \textbf{\textbf{4.64} (0.33)} &     6.98 (0.50) &                     5.99 (0.46) &                    11.51 (0.44) &     8.11 (0.54) &                     7.39 (0.73) \\
			& 50 &  41.27 (0.36) &                   12.71 (1.09) &                      - &                    10.34 (0.60) &    12.00 (0.89) &                     9.79 (0.70) &                    12.92 (0.56) &    11.79 (0.81) &   \textbf{\textbf{8.79} (0.60)} \\
			& 70 &  36.15 (0.36) &                   16.66 (1.37) &                      - &                    13.56 (0.63) &    15.32 (1.15) &  \textbf{\textbf{12.14} (0.95)} &                    12.86 (0.62) &    13.56 (1.01) &                    12.44 (0.82) \\
			\cline{1-11}
			\multirow{5}{*}{$X_3 = 1$} & 5  &   6.36 (0.23) &  \textbf{\textbf{0.58} (0.06)} &                    0.58 (0.06) &                     8.86 (0.35) &     0.62 (0.08) &                     0.63 (0.06) &                     1.60 (0.10) &     1.21 (0.16) &                     1.47 (0.16) \\
			& 10 &  18.15 (0.34) &                    2.11 (0.22) &                    2.14 (0.21) &                    13.91 (0.49) &     2.30 (0.33) &   \textbf{\textbf{1.46} (0.10)} &                     4.71 (0.21) &     3.49 (0.27) &                     4.01 (0.23) \\
			& 30 &  39.91 (0.56) &                   11.45 (0.82) &                   11.46 (0.81) &   \textbf{\textbf{5.91} (0.41)} &     8.51 (0.61) &                     7.04 (0.54) &                    13.11 (0.54) &    10.25 (0.71) &                     8.84 (0.86) \\
			& 50 &  39.97 (0.58) &                   16.02 (1.15) &                   16.17 (1.19) &                    11.25 (0.69) &    15.00 (1.00) &                    11.73 (0.79) &                    15.19 (0.65) &    14.23 (0.98) &  \textbf{\textbf{10.88} (0.73)} \\
			& 70 &  35.32 (0.60) &                   20.57 (1.45) &                   20.51 (1.52) &                    14.13 (0.72) &    18.75 (1.27) &                    14.06 (1.00) &                    14.72 (0.71) &    16.50 (1.21) &  \textbf{\textbf{14.06} (0.90)} \\
			\cline{1-11}
			\multirow{5}{*}{$X_3 = 0$} & 5  &   7.91 (0.15) &                    0.55 (0.04) &  \textbf{\textbf{0.55} (0.04)} &                     7.92 (0.28) &     0.61 (0.07) &                     0.62 (0.06) &                     1.25 (0.09) &     1.06 (0.12) &                     1.19 (0.15) \\
			& 10 &  21.13 (0.24) &                    1.97 (0.19) &                    2.02 (0.19) &                    12.14 (0.41) &     2.09 (0.26) &   \textbf{\textbf{1.59} (0.11)} &                     3.39 (0.22) &     2.83 (0.22) &                     2.94 (0.22) \\
			& 30 &  42.40 (0.44) &                    8.05 (0.63) &                    8.19 (0.62) &   \textbf{\textbf{4.72} (0.38)} &     6.74 (0.57) &                     6.56 (0.52) &                     9.34 (0.49) &     7.67 (0.68) &                     7.27 (0.63) \\
			& 50 &  40.28 (0.51) &                   13.61 (1.20) &                   13.82 (1.17) &   \textbf{\textbf{9.59} (0.66)} &    12.37 (0.86) &                    10.92 (0.79) &                     9.70 (0.64) &    12.80 (0.89) &                     9.75 (0.73) \\
			& 70 &  35.09 (0.54) &                   17.78 (1.70) &                   18.04 (1.70) &                    13.01 (0.76) &    17.08 (1.28) &                    14.18 (1.18) &  \textbf{\textbf{10.68} (0.74)} &    15.39 (1.27) &                    13.94 (1.06) \\
			\cline{1-11}
			\multirow{5}{*}{$X_4 = 1$} & 5  &  10.87 (0.35) &                    0.93 (0.09) &  \textbf{\textbf{0.90} (0.08)} &                     6.49 (0.40) &     0.94 (0.10) &                     1.17 (0.12) &                     2.39 (0.17) &     1.88 (0.23) &                     2.19 (0.20) \\
			& 10 &  25.88 (0.49) &                    3.16 (0.33) &                    3.21 (0.36) &                    10.23 (0.59) &     2.60 (0.28) &   \textbf{\textbf{2.14} (0.16)} &                     6.26 (0.32) &     5.82 (0.39) &                     6.04 (0.33) \\
			& 30 &  44.57 (0.68) &                   14.68 (1.12) &                   13.73 (1.10) &   \textbf{\textbf{7.13} (0.50)} &    10.52 (0.88) &                     7.66 (0.68) &                    13.59 (0.65) &    12.24 (0.89) &                     9.95 (0.89) \\
			& 50 &  40.13 (0.64) &                   20.37 (1.51) &                   18.93 (1.45) &   \textbf{\textbf{9.89} (0.69)} &    17.04 (1.19) &                    12.24 (0.79) &                    13.11 (0.67) &    16.89 (1.17) &                    10.20 (0.81) \\
			& 70 &  33.27 (0.64) &                   24.76 (1.93) &                   22.78 (1.77) &  \textbf{\textbf{10.88} (0.72)} &    18.96 (1.39) &                    13.13 (1.16) &                    11.71 (0.69) &    17.30 (1.25) &                    12.54 (0.95) \\
			\cline{1-11}
			\multirow{5}{*}{$X_4 = 0$} & 5  &   7.16 (0.14) &                    0.41 (0.03) &                    0.41 (0.03) &                    10.04 (0.27) &     0.49 (0.07) &   \textbf{\textbf{0.38} (0.03)} &                     0.96 (0.07) &     0.76 (0.09) &                     0.96 (0.15) \\
			& 10 &  20.14 (0.21) &                    1.52 (0.13) &                    1.53 (0.14) &                    15.50 (0.35) &     2.21 (0.33) &   \textbf{\textbf{1.49} (0.10)} &                     2.80 (0.18) &     2.11 (0.18) &                     2.25 (0.17) \\
			& 30 &  44.05 (0.37) &                    7.50 (0.55) &                    7.64 (0.57) &   \textbf{\textbf{4.61} (0.34)} &     7.01 (0.53) &                     6.77 (0.46) &                    10.27 (0.46) &     7.95 (0.62) &                     7.41 (0.70) \\
			& 50 &  44.18 (0.45) &                   12.55 (1.09) &                   12.91 (1.12) &                    11.49 (0.60) &    12.40 (0.95) &                    10.23 (0.80) &                    12.97 (0.60) &    12.34 (0.92) &  \textbf{\textbf{10.15} (0.72)} \\
			& 70 &  39.39 (0.48) &                   15.90 (1.29) &                   16.35 (1.29) &                    15.75 (0.68) &    15.92 (1.36) &  \textbf{\textbf{13.48} (0.97)} &                    13.90 (0.68) &    15.22 (1.09) &                    14.52 (0.95) \\
			\bottomrule
		\end{tabular}
		
	}
	\caption{ 
		The absolute bias $(\times 10^2)$ for all methods 
		under moderate covariate shift ($q = 2$) with a total sample size of $N = 1000$. The survival time follows the accelerated failure hazard model with covariate-independent censoring. 
		Results for all subpopulations are presented based on $100$ simulation replications. 
	}
\end{table}
\begin{table}
	\centering
	\resizebox{\textwidth}{!}{
		\begin{tabular}{llccccccccc}
			\hline 
			\hline
			&  Time  &     \textsc{Naive}      &  \textsc{IPSW}     & \textsc{IPSW-sub}   &              \textsc{lm}   &          \textsc{ mclm-Ridge} &   \textsc{mclm-Tree} &    \textsc{ rf} & \textsc{mcrf-Ridge} &   \textsc{ mcrf-Tree} \\
			\hline  
			\multicolumn{11}{l}{Absolute bias $(\times 10^2)$, $N = 1000, q = 3$, accelerate failure hazard, independent censoring, single source }                              \\
			\hline 
			\multirow{5}{*}{\textsc{All}} & 5  &   8.93 (0.12) &   0.46 (0.07) &    - &                    13.79 (0.42) &  \textbf{\textbf{0.43} (0.03)} &                     0.59 (0.09) &                     2.39 (0.11) &     2.05 (0.27) &                     2.49 (0.20) \\
			& 10 &  24.79 (0.17) &   2.63 (0.34) &     - &                    19.88 (0.59) &                    2.20 (0.21) &   \textbf{\textbf{1.77} (0.10)} &                     7.03 (0.27) &     5.68 (0.37) &                     6.35 (0.35) \\
			& 30 &  51.02 (0.28) &  17.52 (1.41) &    - &   \textbf{\textbf{6.76} (0.42)} &                   11.79 (0.92) &                     9.86 (0.67) &                    18.36 (0.61) &    13.03 (0.89) &                    12.56 (0.98) \\
			& 50 &  49.77 (0.30) &  25.69 (1.76) &    - &                    14.24 (0.74) &                   21.07 (1.41) &                    15.53 (1.12) &                    20.35 (0.71) &    20.49 (1.25) &  \textbf{\textbf{13.44} (0.91)} \\
			& 70 &  43.37 (0.34) &  30.63 (2.31) &     - &                    18.33 (0.84) &                   23.92 (1.40) &                    19.32 (1.45) &                    19.23 (0.84) &    23.95 (1.43) &  \textbf{\textbf{17.96} (1.37)} \\
			\cline{1-11}
			\multirow{5}{*}{$X_3 = 1$} & 5  &   8.75 (0.26) &   0.56 (0.08) &   0.56 (0.08) &                    13.60 (0.48) &  \textbf{\textbf{0.43} (0.04)} &                     0.63 (0.09) &                     2.64 (0.12) &     2.19 (0.31) &                     2.69 (0.21) \\
			& 10 &  24.24 (0.40) &   3.57 (0.45) &   3.53 (0.50) &                    19.66 (0.72) &                    2.29 (0.20) &   \textbf{\textbf{1.93} (0.12)} &                     7.72 (0.31) &     6.15 (0.41) &                     7.08 (0.39) \\
			& 30 &  50.08 (0.52) &  22.80 (1.72) &  22.21 (1.75) &   \textbf{\textbf{7.41} (0.47)} &                   13.68 (1.06) &                    10.95 (0.76) &                    20.09 (0.66) &    15.35 (0.95) &                    14.36 (1.20) \\
			& 50 &  49.33 (0.55) &  30.67 (1.85) &  30.04 (1.81) &  \textbf{\textbf{14.58} (0.82)} &                   23.54 (1.53) &                    16.65 (1.19) &                    22.39 (0.76) &    24.07 (1.33) &                    15.13 (0.97) \\
			& 70 &  42.95 (0.55) &  36.53 (2.41) &  35.93 (2.34) &                    18.56 (0.92) &                   25.44 (1.68) &                    19.27 (1.39) &                    21.16 (0.88) &    25.53 (1.60) &  \textbf{\textbf{18.21} (1.30)} \\
			\cline{1-11}
			\multirow{5}{*}{$X_3 = 0$} & 5  &   8.92 (0.15) &   0.57 (0.06) &   0.58 (0.07) &                    14.14 (0.40) &  \textbf{\textbf{0.50} (0.05)} &                     0.70 (0.09) &                     1.98 (0.13) &     1.98 (0.24) &                     2.19 (0.19) \\
			& 10 &  24.83 (0.23) &   1.96 (0.16) &   2.02 (0.19) &                    20.37 (0.50) &                    2.33 (0.26) &   \textbf{\textbf{1.82} (0.12)} &                     5.71 (0.26) &     5.17 (0.35) &                     5.08 (0.33) \\
			& 30 &  51.44 (0.37) &  13.33 (1.11) &  13.88 (1.11) &   \textbf{\textbf{6.55} (0.42)} &                   11.20 (0.86) &                    10.03 (0.60) &                    15.09 (0.69) &    12.25 (0.93) &                    11.44 (0.85) \\
			& 50 &  49.81 (0.45) &  22.66 (1.68) &  22.88 (1.74) &  \textbf{\textbf{13.91} (0.76)} &                   23.09 (1.49) &                    17.75 (1.26) &                    16.58 (0.83) &    20.61 (1.35) &                    15.56 (1.02) \\
			& 70 &  43.43 (0.56) &  24.83 (1.93) &  25.59 (2.05) &                    18.26 (0.87) &                   24.98 (1.72) &                    23.36 (1.72) &  \textbf{\textbf{16.25} (0.93)} &    26.11 (1.77) &                    22.87 (1.69) \\
			\cline{1-11}
			\multirow{5}{*}{$X_4 = 1$} & 5  &  14.18 (0.40) &   1.04 (0.15) &   0.99 (0.14) &                    10.49 (0.53) &  \textbf{\textbf{0.67} (0.07)} &                     1.07 (0.20) &                     3.69 (0.21) &     3.20 (0.40) &                     4.04 (0.32) \\
			& 10 &  33.74 (0.55) &   5.06 (0.65) &   4.82 (0.59) &                    14.55 (0.77) &                    2.59 (0.19) &   \textbf{\textbf{2.46} (0.18)} &                     9.82 (0.43) &     8.64 (0.63) &                     9.30 (0.57) \\
			& 30 &  53.99 (0.69) &  25.50 (1.94) &  25.26 (1.96) &   \textbf{\textbf{8.25} (0.54)} &                   18.14 (1.31) &                    13.33 (0.91) &                    19.58 (0.70) &    20.43 (1.28) &                    14.61 (1.15) \\
			& 50 &  47.69 (0.65) &  35.93 (1.93) &  34.44 (1.96) &  \textbf{\textbf{12.05} (0.79)} &                   22.05 (1.63) &                    17.23 (1.31) &                    18.72 (0.80) &    23.37 (1.32) &                    14.91 (1.06) \\
			& 70 &  38.92 (0.65) &  34.52 (2.15) &  33.96 (2.04) &  \textbf{\textbf{13.46} (0.87)} &                   24.14 (1.65) &                    18.91 (1.59) &                    16.41 (0.81) &    24.48 (1.50) &                    16.66 (1.19) \\
			\cline{1-11}
			\multirow{5}{*}{$X_4 = 0$} & 5  &   8.27 (0.13) &   0.36 (0.04) &   0.35 (0.04) &                    16.34 (0.42) &  \textbf{\textbf{0.35} (0.03)} &                     0.37 (0.04) &                     1.48 (0.09) &     1.34 (0.21) &                     1.49 (0.14) \\
			& 10 &  24.05 (0.20) &   1.96 (0.22) &   1.94 (0.21) &                    23.99 (0.56) &                    2.29 (0.28) &   \textbf{\textbf{1.84} (0.09)} &                     5.07 (0.24) &     3.93 (0.31) &                     4.43 (0.30) \\
			& 30 &  53.49 (0.31) &  12.64 (1.10) &  13.01 (1.06) &   \textbf{\textbf{6.64} (0.42)} &                   11.18 (0.88) &                     9.33 (0.55) &                    17.50 (0.65) &    11.50 (0.85) &                    12.29 (0.98) \\
			& 50 &  54.16 (0.35) &  20.63 (1.55) &  20.89 (1.56) &                    16.83 (0.72) &                   24.03 (1.46) &                    16.05 (1.08) &                    21.50 (0.76) &    23.12 (1.36) &  \textbf{\textbf{15.12} (1.01)} \\
			& 70 &  48.32 (0.47) &  28.43 (2.32) &  28.51 (2.35) &                    22.37 (0.90) &                   27.90 (1.60) &  \textbf{\textbf{21.35} (1.52)} &                    21.51 (0.96) &    28.10 (1.59) &                    21.43 (1.51) \\
			\bottomrule
		\end{tabular}
	}
	\caption{
		The absolute bias $(\times 10^2)$ for all methods 
		under strong covariate shift ($q = 3$) with a total sample size of $N = 1000$. The survival time follows the accelerated failure hazard model with covariate-independent censoring. 
		Results for all subpopulations are presented based on $100$ simulation replications.
		\label{simu:aft-3}
	}
\end{table} 

\subsection{Simulations under the dependent censoring assumption}\label{sup_dep_simu}
Additionally, Tables~\ref{simu:prop-dep-sp-1}-\ref{simu:prop-dep-sp-3} show the results for the scenario where the survival time adheres to the $\lambda(t|X_i) = \eta \nu t^{\nu - 1}\exp( X_i^\top \alpha)$,  $C_i$ following a Weibull model $\lambda^c(t | X_i) = \eta_c \nu_c t^{\nu_c - 1}\exp(X_i^\top \alpha_c),$ $\eta = 0.0001, \nu = 3, \alpha = (0, 2, 1, -1.2, 0.8)^\top$, and $\eta_c = 0.0001, \nu_c = 2.7, \alpha_c = (1, 0.5, -0.5, -0.5)^\top$.

\begin{table}[ht!]
	\centering
	\resizebox{\textwidth}{!}{
		\begin{tabular}{llccccccccc}
			\hline 
			\hline
			&  Time  &     \textsc{Naive}      &  \textsc{IPSW}     & \textsc{IPSW-sub}   &              \textsc{lm}   &          \textsc{ mclm-Ridge} &   \textsc{mclm-Tree} &    \textsc{ rf} & \textsc{mcrf-Ridge} &   \textsc{ mcrf-Tree} \\
			\hline 
			\multicolumn{11}{l}{SP, Absolute bias $(\times 10^2)$, $N = 1000, q = 1$, proportional hazard, Weibull censoring, single source }                              \\
			\hline
			\multirow{5}{*}{All} & 5  &  19.09 (0.21) &  \textbf{\textbf{0.81} (0.06)} &                      - &  10.33 (0.25) &   1.14 (0.10) &  1.84 (0.14) &                    1.58 (0.10) &                    1.28 (0.09) &                    1.47 (0.10) \\
			& 10 &  30.81 (0.27) &  \textbf{\textbf{1.28} (0.11)} &                      - &   7.53 (0.28) &   1.49 (0.13) &  1.83 (0.13) &                    2.65 (0.16) &                    1.75 (0.13) &                    1.98 (0.13) \\
			& 30 &  36.42 (0.32) &                    4.19 (0.31) &                      - &   5.40 (0.35) &   4.28 (0.30) &  4.71 (0.32) &                    3.93 (0.27) &                    3.06 (0.25) &  \textbf{\textbf{3.00} (0.22)} \\
			& 50 &  32.28 (0.37) &                    7.45 (0.51) &                     - &   8.30 (0.44) &   7.86 (0.52) &  6.65 (0.49) &  \textbf{\textbf{4.65} (0.32)} &                    5.84 (0.43) &                    4.78 (0.35) \\
			& 70 &  28.88 (0.36) &                    9.24 (0.73) &                     - &   9.40 (0.51) &   8.78 (0.75) &  5.53 (0.56) &                    5.48 (0.41) &                    5.83 (0.61) &  \textbf{\textbf{4.67} (0.40)} \\
			\cline{1-11}
			\multirow{5}{*}{$X_3 = 1$} & 5  &  14.53 (0.32) &                    1.18 (0.10) &  \textbf{\textbf{1.10} (0.09)} &  12.27 (0.37) &   1.38 (0.11) &  1.90 (0.13) &                    2.27 (0.12) &                    1.70 (0.12) &                    2.12 (0.12) \\
			& 10 &  25.50 (0.46) &                    2.32 (0.19) &  \textbf{\textbf{1.91} (0.16)} &   9.59 (0.42) &   2.17 (0.18) &  2.20 (0.17) &                    3.99 (0.22) &                    2.57 (0.20) &                    3.18 (0.21) \\
			& 30 &  34.01 (0.51) &                    6.33 (0.48) &                    5.66 (0.39) &   6.11 (0.45) &   5.06 (0.39) &  5.40 (0.36) &                    6.28 (0.41) &                    4.31 (0.36) &  \textbf{\textbf{3.98} (0.30)} \\
			& 50 &  31.50 (0.53) &                   10.71 (0.83) &                    9.86 (0.69) &  10.01 (0.51) &  10.62 (0.69) &  7.61 (0.52) &                    7.33 (0.45) &                    7.90 (0.53) &  \textbf{\textbf{6.22} (0.48)} \\
			& 70 &  29.48 (0.57) &                   13.95 (1.06) &                   13.01 (0.96) &  11.87 (0.63) &  11.45 (1.00) &  7.06 (0.62) &                    8.52 (0.55) &                    8.05 (0.82) &  \textbf{\textbf{5.83} (0.53)} \\
			\cline{1-11}
			\multirow{5}{*}{$X_3 = 0$} & 5  &  20.43 (0.29) &                    1.51 (0.11) &  \textbf{\textbf{1.29} (0.10)} &   8.17 (0.27) &   1.50 (0.12) &  2.21 (0.17) &                    1.35 (0.11) &                    1.31 (0.10) &                    1.33 (0.11) \\
			& 10 &  31.63 (0.38) &                    2.45 (0.17) &                    1.98 (0.15) &   5.42 (0.32) &   2.20 (0.14) &  2.37 (0.16) &                    1.98 (0.15) &                    2.05 (0.15) &  \textbf{\textbf{1.85} (0.14)} \\
			& 30 &  34.75 (0.43) &                    4.38 (0.33) &                    4.20 (0.30) &   5.24 (0.36) &   5.05 (0.34) &  5.17 (0.34) &  \textbf{\textbf{2.92} (0.23)} &                    4.00 (0.30) &                    4.19 (0.28) \\
			& 50 &  29.70 (0.45) &                    6.81 (0.45) &                    6.97 (0.46) &   6.80 (0.45) &   6.74 (0.52) &  6.93 (0.56) &  \textbf{\textbf{3.87} (0.30)} &                    5.85 (0.45) &                    5.42 (0.34) \\
			& 70 &  25.62 (0.42) &                    8.75 (0.57) &                    8.78 (0.63) &   7.09 (0.50) &   7.80 (0.58) &  6.75 (0.56) &  \textbf{\textbf{5.21} (0.36)} &                    6.67 (0.52) &                    7.14 (0.49) \\
			\cline{1-11}
			\multirow{5}{*}{$X_4 = 1$} & 5  &  23.16 (0.50) &                    2.17 (0.16) &  \textbf{\textbf{1.65} (0.13)} &   7.24 (0.44) &   2.12 (0.17) &  2.74 (0.21) &                    2.54 (0.18) &                    2.44 (0.18) &                    2.45 (0.17) \\
			& 10 &  34.51 (0.60) &                    3.95 (0.30) &  \textbf{\textbf{2.88} (0.24)} &   5.75 (0.41) &   3.07 (0.26) &  2.92 (0.25) &                    3.87 (0.26) &                    3.35 (0.26) &                    3.29 (0.24) \\
			& 30 &  35.74 (0.68) &                   10.03 (0.65) &                    8.46 (0.59) &   6.04 (0.46) &   6.88 (0.54) &  6.13 (0.43) &                    5.08 (0.40) &                    6.53 (0.48) &  \textbf{\textbf{4.84} (0.37)} \\
			& 50 &  29.03 (0.64) &                   13.97 (1.03) &                   12.52 (0.95) &   6.28 (0.47) &  12.30 (0.90) &  9.81 (0.67) &  \textbf{\textbf{5.19} (0.37)} &                    9.95 (0.78) &                    7.53 (0.63) \\
			& 70 &  23.69 (0.61) &                   19.00 (1.49) &                   17.51 (1.55) &   5.86 (0.49) &  12.23 (1.38) &  9.08 (0.80) &  \textbf{\textbf{5.36} (0.39)} &                    9.75 (1.05) &                    7.22 (0.64) \\
			\cline{1-11}
			\multirow{5}{*}{$X_4 = 0$} & 5  &  18.53 (0.23) &                    1.02 (0.07) &  \textbf{\textbf{0.91} (0.07)} &  12.05 (0.24) &   1.20 (0.09) &  1.74 (0.14) &                    1.43 (0.09) &                    1.26 (0.09) &                    1.38 (0.09) \\
			& 10 &  30.84 (0.29) &                    1.40 (0.11) &  \textbf{\textbf{1.27} (0.10)} &   8.87 (0.31) &   1.59 (0.13) &  1.84 (0.14) &                    2.45 (0.16) &                    1.77 (0.12) &                    1.87 (0.13) \\
			& 30 &  38.53 (0.36) &                    3.94 (0.26) &                    3.88 (0.26) &   5.96 (0.37) &   4.20 (0.29) &  4.81 (0.34) &                    4.22 (0.28) &  \textbf{\textbf{3.08} (0.23)} &                    3.10 (0.25) \\
			& 50 &  35.50 (0.47) &                    7.29 (0.52) &                    7.14 (0.55) &  10.21 (0.52) &   7.17 (0.52) &  6.32 (0.50) &                    5.67 (0.38) &                    5.60 (0.42) &  \textbf{\textbf{4.80} (0.34)} \\
			& 70 &  32.94 (0.45) &                    8.83 (0.69) &                    8.97 (0.71) &  12.26 (0.62) &   8.93 (0.63) &  6.35 (0.53) &                    7.00 (0.52) &                    5.81 (0.58) &  \textbf{\textbf{5.03} (0.44)} \\
			\bottomrule
		\end{tabular}
		
	}
	\caption{ The absolute bias $(\times 10^2)$ for all methods 
		under mild covariate shift ($q = 1$) with a total sample size of $N = 1000$. The survival time follows the proportional 
		hazard model, and censoring follows the covariate-dependent Weibull distribution. Results for all subpopulations are presented based on $100$ simulation replications. 
		\label{simu:prop-dep-sp-1}
	}
\end{table}

\begin{table}[ht!]
	\centering
	\resizebox{\textwidth}{!}{
		\begin{tabular}{llccccccccc}
			\hline 
			\hline
			&  Time  &     \textsc{Naive}      &  \textsc{IPSW}     & \textsc{IPSW-sub}   &              \textsc{lm}   &          \textsc{ mclm-Ridge} &   \textsc{mclm-Tree} &    \textsc{ rf} & \textsc{mcrf-Ridge} &   \textsc{ mcrf-Tree} \\
			\hline 
			\multicolumn{11}{l}{SP, Absolute bias $(\times 10^2)$, $N = 1000, q = 2$, proportional hazard, Weibull censoring, single source }                              \\
			\hline
			\midrule
			\multirow{5}{*}{All} & 5  &  30.67 (0.20) &   0.74 (0.08) &                      - &  28.59 (0.36) &  \textbf{\textbf{0.70} (0.05)} &                    0.89 (0.06) &                     2.83 (0.11) &   1.49 (0.15) &                     2.36 (0.12) \\
			& 10 &  49.45 (0.24) &   2.44 (0.17) &                      - &  18.24 (0.39) &                    1.89 (0.13) &  \textbf{\textbf{1.54} (0.11)} &                     5.11 (0.19) &   1.94 (0.17) &                     2.75 (0.23) \\
			& 30 &  59.16 (0.28) &  12.99 (1.03) &                      - &  17.78 (0.46) &                   10.13 (0.63) &                    9.62 (0.54) &                    10.03 (0.41) &   9.48 (0.58) &   \textbf{\textbf{6.38} (0.43)} \\
			& 50 &  52.49 (0.31) &  24.87 (1.99) &                      - &  24.62 (0.69) &                   17.72 (1.19) &                   14.25 (1.00) &                    12.52 (0.86) &  14.70 (0.96) &  \textbf{\textbf{12.34} (0.81)} \\
			& 70 &  46.14 (0.33) &  36.80 (3.13) &                      - &  25.59 (0.79) &                   24.69 (1.38) &                   20.06 (1.61) &  \textbf{\textbf{16.12} (1.01)} &  19.60 (1.41) &                    20.11 (1.84) \\
			\cline{1-11}
			\multirow{5}{*}{$X_3 = 1$} & 5  &  26.80 (0.37) &   1.04 (0.15) &  \textbf{\textbf{0.88} (0.16)} &  29.28 (0.45) &                    0.89 (0.06) &                    1.02 (0.08) &                     3.57 (0.15) &   1.76 (0.18) &                     3.03 (0.15) \\
			& 10 &  44.54 (0.46) &   3.28 (0.23) &                    2.87 (0.23) &  19.30 (0.48) &                    2.16 (0.17) &  \textbf{\textbf{1.74} (0.12)} &                     6.60 (0.24) &   2.46 (0.21) &                     3.75 (0.29) \\
			& 30 &  56.66 (0.49) &  15.42 (1.15) &                   14.11 (1.02) &  17.96 (0.57) &                   11.92 (0.74) &                    9.64 (0.64) &                    13.47 (0.53) &  10.50 (0.70) &   \textbf{\textbf{6.93} (0.55)} \\
			& 50 &  51.07 (0.45) &  31.47 (2.12) &                   29.16 (2.01) &  25.23 (0.73) &                   19.28 (1.26) &                   14.83 (1.02) &                    16.26 (0.90) &  16.97 (1.12) &  \textbf{\textbf{13.65} (0.94)} \\
			& 70 &  45.20 (0.51) &  46.94 (3.71) &                   43.67 (3.68) &  26.26 (0.87) &                   25.62 (1.55) &                   20.17 (1.53) &  \textbf{\textbf{19.24} (1.06)} &  21.27 (1.71) &                    20.80 (1.86) \\
			\cline{1-11}
			\multirow{5}{*}{$X_3 = 0$} & 5  &  31.68 (0.25) &   1.07 (0.09) &                    0.97 (0.08) &  27.50 (0.47) &  \textbf{\textbf{0.92} (0.06)} &                    1.04 (0.08) &                     1.82 (0.11) &   1.55 (0.14) &                     1.54 (0.10) \\
			& 10 &  50.34 (0.34) &   2.26 (0.20) &                    2.27 (0.17) &  16.54 (0.47) &                    2.19 (0.15) &  \textbf{\textbf{2.04} (0.15)} &                     3.01 (0.22) &   2.30 (0.18) &                     2.35 (0.23) \\
			& 30 &  58.23 (0.42) &  10.36 (0.77) &                   10.76 (0.79) &  17.53 (0.49) &                   10.94 (0.68) &                   11.03 (0.57) &   \textbf{\textbf{5.27} (0.39)} &  10.05 (0.68) &                     8.22 (0.50) \\
			& 50 &  51.02 (0.49) &  20.61 (1.52) &                   21.36 (1.56) &  23.65 (0.74) &                   18.37 (1.31) &                   15.99 (1.12) &   \textbf{\textbf{8.50} (0.77)} &  16.96 (1.18) &                    14.67 (0.91) \\
			& 70 &  44.47 (0.47) &  30.92 (2.21) &                   32.02 (2.30) &  24.45 (0.79) &                   24.21 (1.61) &                   22.56 (1.80) &  \textbf{\textbf{12.69} (0.98)} &  20.58 (1.47) &                    22.97 (2.11) \\
			\cline{1-11}
			\multirow{5}{*}{$X_4 = 1$} & 5  &  41.61 (0.52) &   1.87 (0.21) &  \textbf{\textbf{1.48} (0.19)} &  20.26 (0.55) &                    1.57 (0.12) &                    1.70 (0.14) &                     5.13 (0.23) &   2.82 (0.26) &                     4.55 (0.24) \\
			& 10 &  57.67 (0.64) &   5.05 (0.42) &                    4.43 (0.34) &  12.74 (0.53) &                    3.51 (0.27) &  \textbf{\textbf{2.81} (0.19)} &                     8.66 (0.38) &   4.38 (0.39) &                     5.55 (0.44) \\
			& 30 &  57.73 (0.54) &  22.02 (1.42) &                   20.29 (1.43) &  14.61 (0.58) &                   14.20 (0.98) &                   11.15 (0.78) &                    12.73 (0.56) &  12.95 (0.84) &   \textbf{\textbf{8.28} (0.60)} \\
			& 50 &  46.31 (0.59) &  38.50 (2.45) &                   36.40 (2.44) &  17.18 (0.82) &                   20.98 (1.41) &                   17.14 (1.18) &  \textbf{\textbf{12.12} (0.80)} &  18.84 (1.28) &                    12.60 (0.95) \\
			& 70 &  36.87 (0.59) &  52.35 (3.92) &                   50.30 (3.83) &  15.47 (0.86) &                   22.84 (1.97) &                   21.79 (1.58) &  \textbf{\textbf{12.40} (0.79)} &  22.70 (2.27) &                    20.90 (1.90) \\
			\cline{1-11}
			\multirow{5}{*}{$X_4 = 0$} & 5  &  29.25 (0.22) &   0.60 (0.05) &                    0.57 (0.05) &  33.81 (0.35) &  \textbf{\textbf{0.54} (0.05)} &                    0.66 (0.05) &                     1.43 (0.09) &   0.95 (0.10) &                     1.09 (0.08) \\
			& 10 &  49.45 (0.25) &   1.88 (0.14) &                    1.88 (0.13) &  21.75 (0.43) &                    1.62 (0.12) &  \textbf{\textbf{1.43} (0.11)} &                     2.96 (0.17) &   1.46 (0.11) &                     1.70 (0.15) \\
			& 30 &  63.22 (0.30) &  11.69 (0.74) &                   12.13 (0.74) &  19.83 (0.50) &                   10.71 (0.66) &                    9.63 (0.64) &                     8.49 (0.46) &   9.53 (0.58) &   \textbf{\textbf{7.28} (0.48)} \\
			& 50 &  58.74 (0.34) &  23.33 (1.61) &                   24.20 (1.65) &  29.35 (0.69) &                   19.51 (1.28) &                   14.57 (1.07) &  \textbf{\textbf{13.36} (0.94)} &  15.02 (0.98) &                    13.46 (0.94) \\
			& 70 &  53.96 (0.38) &  36.10 (2.71) &                   37.08 (2.75) &  32.14 (0.81) &                   27.23 (1.37) &                   21.08 (1.73) &  \textbf{\textbf{19.30} (1.16)} &  20.51 (1.30) &                    21.34 (2.01) \\
			\bottomrule
		\end{tabular}
		
	}
	\caption{ 
		The absolute bias $(\times 10^2)$ for all methods 
		under moderate covariate shift ($q = 2$) with a total sample size of $N = 1000$. The survival time follows the proportional 
		hazard model, and censoring follows the covariate-dependent Weibull distribution. Results for all subpopulations are presented based on $100$ simulation replications.  
	}
\end{table}

\begin{table}[ht!]
	\centering
	\resizebox{\textwidth}{!}{
		\begin{tabular}{llccccccccc}
			\hline 
			\hline
			&  Time  &     \textsc{Naive}      &  \textsc{IPSW}     & \textsc{IPSW-sub}   &              \textsc{lm}   &          \textsc{ mclm-Ridge} &   \textsc{mclm-Tree} &    \textsc{ rf} & \textsc{mcrf-Ridge} &   \textsc{ mcrf-Tree} \\
			\hline 
			\multicolumn{11}{l}{SP, Absolute bias $(\times 10^2)$, $N = 1000, q = 3$, proportional hazard, Weibull censoring, single source }                              \\
			\hline
			\multirow{5}{*}{All} & 5  &  37.04 (0.20) &    0.79 (0.19) &      - &  44.36 (0.48) &   0.57 (0.05) &  \textbf{\textbf{0.50} (0.04)} &                     5.19 (0.16) &   2.08 (0.25) &                     4.22 (0.25) \\
			& 10 &  60.46 (0.23) &    3.57 (0.23) &      - &  24.38 (0.53) &   2.61 (0.16) &  \textbf{\textbf{2.41} (0.14)} &                     9.94 (0.27) &   3.20 (0.26) &                     5.46 (0.38) \\
			& 30 &  71.47 (0.27) &   24.10 (1.46) &      - &  29.83 (0.86) &  19.56 (1.59) &                   15.74 (0.83) &                    18.07 (0.80) &  15.60 (1.11) &   \textbf{\textbf{9.54} (0.79)} \\
			& 50 &  63.47 (0.28) &   51.91 (3.95) &     - &  38.50 (0.87) &  38.03 (2.03) &                   24.23 (1.72) &                    27.02 (1.31) &  29.48 (1.95) &  \textbf{\textbf{19.66} (1.53)} \\
			& 70 &  55.88 (0.32) &  90.64 (11.24) &      - &  37.48 (1.12) &  39.96 (2.11) &                   38.98 (3.17) &  \textbf{\textbf{30.18} (1.67)} &  31.63 (2.05) &                    32.87 (2.72) \\
			\cline{1-11}
			\multirow{5}{*}{$X_3 = 1$} & 5  &  36.05 (0.44) &    1.04 (0.27) &    1.01 (0.29) &  43.75 (0.54) &   0.63 (0.05) &  \textbf{\textbf{0.54} (0.04)} &                     6.13 (0.19) &   2.11 (0.26) &                     5.01 (0.27) \\
			& 10 &  58.84 (0.45) &    4.30 (0.34) &    4.41 (0.39) &  24.20 (0.64) &   2.80 (0.17) &  \textbf{\textbf{2.39} (0.16)} &                    11.64 (0.31) &   3.57 (0.33) &                     6.66 (0.42) \\
			& 30 &  69.87 (0.40) &   27.37 (1.72) &   27.14 (1.67) &  29.68 (0.88) &  21.84 (1.69) &                   15.83 (0.93) &                    22.02 (0.82) &  17.03 (1.26) &  \textbf{\textbf{10.57} (0.83)} \\
			& 50 &  61.85 (0.41) &   58.41 (4.11) &   58.13 (3.98) &  38.00 (0.89) &  39.32 (2.32) &                   23.70 (1.84) &                    30.23 (1.21) &  30.08 (1.99) &  \textbf{\textbf{19.66} (1.51)} \\
			& 70 &  54.61 (0.47) &  97.23 (12.14) &  94.44 (10.98) &  37.21 (1.14) &  41.77 (2.24) &                   36.41 (2.97) &  \textbf{\textbf{32.54} (1.54)} &  34.00 (2.10) &                    33.74 (2.71) \\
			\cline{1-11}
			\multirow{5}{*}{$X_3 = 0$} & 5  &  37.23 (0.23) &    0.83 (0.07) &    0.79 (0.07) &  45.48 (0.52) &   0.74 (0.06) &  \textbf{\textbf{0.67} (0.05)} &                     3.45 (0.16) &   2.18 (0.28) &                     2.84 (0.24) \\
			& 10 &  60.73 (0.27) &    3.09 (0.20) &    3.05 (0.18) &  24.77 (0.53) &   2.76 (0.18) &  \textbf{\textbf{2.63} (0.17)} &                     6.76 (0.31) &   3.40 (0.28) &                     3.78 (0.36) \\
			& 30 &  71.71 (0.37) &   20.78 (1.16) &   20.78 (1.20) &  30.14 (0.90) &  18.33 (1.53) &                   16.44 (0.89) &  \textbf{\textbf{11.39} (0.83)} &  17.83 (1.08) &                    13.47 (0.87) \\
			& 50 &  64.08 (0.44) &   43.50 (2.97) &   43.68 (3.17) &  39.46 (0.93) &  39.11 (1.67) &                   30.50 (1.84) &  \textbf{\textbf{22.22} (1.48)} &  34.40 (2.02) &                    25.48 (1.84) \\
			& 70 &  56.19 (0.49) &   76.16 (7.15) &   77.58 (7.47) &  38.01 (1.20) &  39.93 (1.86) &                   47.23 (3.72) &  \textbf{\textbf{27.92} (1.80)} &  30.44 (2.37) &                    38.71 (3.00) \\
			\cline{1-11}
			\multirow{5}{*}{$X_4 = 1$} & 5  &  55.27 (0.62) &    2.86 (0.55) &    2.14 (0.54) &  31.65 (0.63) &   1.14 (0.10) &  \textbf{\textbf{1.13} (0.09)} &                     9.91 (0.34) &   4.34 (0.60) &                     8.72 (0.57) \\
			& 10 &  73.85 (0.60) &    8.54 (0.90) &    7.16 (0.87) &  15.67 (0.68) &   3.65 (0.22) &  \textbf{\textbf{2.71} (0.18)} &                    17.53 (0.55) &   6.01 (0.62) &                    11.95 (0.74) \\
			& 30 &  68.74 (0.45) &   36.82 (2.24) &   36.72 (2.40) &  25.11 (0.88) &  21.81 (1.47) &                   17.01 (1.00) &                    21.54 (0.74) &  15.85 (1.23) &  \textbf{\textbf{10.50} (0.84)} \\
			& 50 &  54.12 (0.47) &   65.85 (4.30) &   63.63 (4.18) &  27.93 (0.95) &  32.42 (2.17) &                   22.39 (1.66) &                    23.38 (1.07) &  27.13 (1.89) &  \textbf{\textbf{17.48} (1.41)} \\
			& 70 &  43.13 (0.51) &   80.95 (7.58) &   77.84 (7.17) &  24.18 (1.14) &  30.30 (2.12) &                   32.75 (2.66) &  \textbf{\textbf{21.98} (1.29)} &  25.43 (1.91) &                    27.48 (2.30) \\
			\cline{1-11}
			\multirow{5}{*}{$X_4 = 0$} & 5  &  34.62 (0.21) &    0.54 (0.04) &    0.52 (0.04) &  53.46 (0.49) &   0.53 (0.04) &  \textbf{\textbf{0.43} (0.04)} &                     1.83 (0.11) &   0.85 (0.13) &                     1.19 (0.12) \\
			& 10 &  59.68 (0.25) &    2.88 (0.15) &    2.90 (0.14) &  30.62 (0.57) &   2.76 (0.13) &                    2.77 (0.13) &                     4.56 (0.24) &   2.59 (0.21) &   \textbf{\textbf{2.36} (0.22)} \\
			& 30 &  76.76 (0.30) &   22.39 (1.19) &   23.10 (1.09) &  33.19 (0.89) &  20.85 (1.65) &                   16.06 (0.77) &                    15.70 (0.89) &  17.32 (1.18) &  \textbf{\textbf{12.08} (0.84)} \\
			& 50 &  72.09 (0.38) &   47.43 (3.09) &   48.21 (3.07) &  46.04 (0.92) &  43.38 (2.13) &                   26.20 (1.97) &                    29.71 (1.54) &  33.19 (2.13) &  \textbf{\textbf{23.11} (1.76)} \\
			& 70 &  66.56 (0.41) &  90.07 (11.25) &  91.13 (11.15) &  47.14 (1.16) &  47.68 (2.35) &                   44.16 (3.69) &  \textbf{\textbf{36.30} (1.97)} &  36.64 (2.25) &                    37.43 (3.06) \\
			\bottomrule
		\end{tabular}
	}
	\caption{ The absolute bias $(\times 10^2)$ for all methods 
		under strong covariate shift ($q = 3$) with a total sample size of $N = 1000$. The survival time follows the proportional 
		hazard model and censoring follows the covariate-dependent Weibull distribution. Results for all subpopulations are presented based on $100$ simulation replications. 
		\label{simu:prop-dep-sp-3}
	}
\end{table}

\subsection{Performance under multiple source domains} \label{simu.multi}
We assessed the performance of our proposed methods in scenarios involving double source domains and a single target domain. Tables~\ref{simu:double-sp-1}-\ref{simu:double-rm-3} showcase the absolute bias in predicted survival probabilities and the relative bias in restricted mean survival times. For \textsc{ipsw} approaches~(\textsc{ipsw} and \textsc{ipsw-sub}), propensity scores were estimated for each source-target pair, following which a weighted average is  calculated. Whereas for multicalibration, we aggregated both source samples, learned an initial function and refined it through post-processing.

\begin{table}[ht!]
	\centering
	\resizebox{\textwidth}{!}{
		\begin{tabular}{llccccccccc}
			\hline
			\hline
			&  Time  &     \textsc{Naive}      &  \textsc{IPSW}     & \textsc{IPSW-sub}   &              \textsc{lm}   &          \textsc{ mclm-Ridge} &   \textsc{mclm-Tree} &    \textsc{ rf} & \textsc{mcrf-Ridge} &   \textsc{ mcrf-Tree} \\
			\hline 
			\multicolumn{11}{l}{SP, Absolute bias $(\times 10^2)$, $N = 1000, q = 1$, proportional hazard, independent censoring, double sources }                              \\
			\hline 
			\multirow{5}{*}{\textsc{All} } & 5  &  20.67 (0.18) &  \textbf{\textbf{0.77} (0.05)} &                      - &  11.11 (0.26) &                    1.07 (0.07) &                    1.38 (0.09) &                    1.80 (0.09) &     1.39 (0.08) &                    1.27 (0.09) \\
			& 10 &  33.05 (0.25) &  \textbf{\textbf{1.47} (0.11)} &                     - &   7.32 (0.32) &                    1.74 (0.13) &                    1.76 (0.13) &                    3.17 (0.17) &     1.97 (0.15) &                    1.95 (0.14) \\
			& 30 &  37.12 (0.32) &  \textbf{\textbf{3.98} (0.31)} &                      - &   8.10 (0.39) &                    4.54 (0.34) &                    4.89 (0.33) &                    6.31 (0.33) &     4.26 (0.34) &                    4.04 (0.32) \\
			& 50 &  30.11 (0.32) &                    5.10 (0.42) &                      - &  10.89 (0.42) &                    5.76 (0.46) &                    5.95 (0.38) &                    6.60 (0.37) &     4.76 (0.31) &  \textbf{\textbf{4.60} (0.33)} \\
			& 70 &  24.19 (0.29) &                    5.93 (0.51) &                      - &  11.13 (0.45) &                    8.81 (0.58) &                    6.78 (0.52) &                    7.54 (0.40) &     6.19 (0.43) &  \textbf{\textbf{5.49} (0.43)} \\
			\cline{1-11}
			\multirow{5}{*}{$X_3 = 1$} & 5  &  16.22 (0.30) &                    1.26 (0.10) &  \textbf{\textbf{1.09} (0.08)} &  13.23 (0.36) &                    1.36 (0.11) &                    1.42 (0.11) &                    2.39 (0.12) &     1.81 (0.12) &                    1.72 (0.12) \\
			& 10 &  27.76 (0.40) &                    2.53 (0.19) &                    2.06 (0.16) &   9.50 (0.42) &  \textbf{\textbf{1.95} (0.15)} &                    2.13 (0.16) &                    4.35 (0.23) &     2.48 (0.19) &                    2.61 (0.18) \\
			& 30 &  35.97 (0.51) &                    6.41 (0.46) &                    5.47 (0.40) &   9.38 (0.49) &                    6.18 (0.46) &                    5.94 (0.41) &                    8.73 (0.43) &     5.74 (0.44) &  \textbf{\textbf{4.97} (0.39)} \\
			& 50 &  31.03 (0.52) &                    9.02 (0.67) &                    8.12 (0.58) &  13.31 (0.53) &                    8.34 (0.60) &                    7.17 (0.52) &                    9.71 (0.49) &     7.15 (0.56) &  \textbf{\textbf{6.52} (0.51)} \\
			& 70 &  25.82 (0.49) &                    9.74 (0.77) &                    9.53 (0.75) &  13.62 (0.57) &                   11.63 (0.77) &                    8.34 (0.61) &                   10.47 (0.52) &     8.56 (0.56) &  \textbf{\textbf{7.41} (0.50)} \\
			\cline{1-11}
			\multirow{5}{*}{$X_3 = 0$} & 5  &  22.10 (0.26) &                    1.47 (0.11) &  \textbf{\textbf{1.23} (0.09)} &   8.83 (0.32) &                    1.30 (0.10) &                    1.61 (0.13) &                    1.56 (0.10) &     1.54 (0.11) &                    1.37 (0.10) \\
			& 10 &  34.05 (0.34) &                    2.68 (0.19) &                    2.33 (0.15) &   5.42 (0.32) &                    2.39 (0.19) &  \textbf{\textbf{2.20} (0.16)} &                    2.49 (0.18) &     2.47 (0.18) &                    2.37 (0.16) \\
			& 30 &  35.05 (0.41) &                    5.47 (0.38) &                    4.89 (0.35) &   7.29 (0.37) &                    4.88 (0.34) &                    5.06 (0.34) &  \textbf{\textbf{4.55} (0.30)} &     5.03 (0.38) &                    4.65 (0.36) \\
			& 50 &  27.04 (0.47) &                    6.88 (0.48) &                    6.37 (0.47) &   8.33 (0.49) &                    6.35 (0.51) &                    6.75 (0.45) &  \textbf{\textbf{4.41} (0.35)} &     5.23 (0.42) &                    5.35 (0.37) \\
			& 70 &  21.00 (0.41) &                    7.13 (0.50) &                    6.93 (0.48) &   8.61 (0.48) &                    8.26 (0.55) &                    7.38 (0.60) &  \textbf{\textbf{5.62} (0.37)} &     7.10 (0.56) &                    6.57 (0.55) \\
			\cline{1-11}
			\multirow{5}{*}{$X_4 = 1$} & 5  &  24.41 (0.47) &                    2.07 (0.16) &  \textbf{\textbf{1.92} (0.13)} &   7.43 (0.43) &                    2.23 (0.17) &                    2.32 (0.18) &                    2.42 (0.18) &     2.58 (0.19) &                    2.07 (0.17) \\
			& 10 &  36.15 (0.59) &                    4.10 (0.30) &                    3.16 (0.22) &   5.60 (0.37) &                    3.41 (0.26) &  \textbf{\textbf{2.54} (0.19)} &                    4.04 (0.28) &     3.68 (0.35) &                    3.01 (0.22) \\
			& 30 &  35.37 (0.67) &                    9.80 (0.73) &                    8.45 (0.69) &   7.23 (0.49) &                    7.89 (0.56) &                    6.30 (0.42) &                    6.81 (0.47) &     7.82 (0.58) &  \textbf{\textbf{5.70} (0.46)} \\
			& 50 &  26.36 (0.62) &                   12.57 (0.92) &                   11.42 (0.94) &   8.08 (0.53) &                    8.40 (0.61) &                    7.97 (0.58) &  \textbf{\textbf{5.97} (0.46)} &     8.30 (0.57) &                    6.39 (0.47) \\
			& 70 &  20.66 (0.57) &                   13.14 (1.09) &                   12.39 (1.06) &   8.84 (0.46) &                    9.42 (0.82) &                    7.57 (0.71) &                    6.70 (0.39) &     9.92 (0.72) &  \textbf{\textbf{6.69} (0.59)} \\
			\cline{1-11}
			\multirow{5}{*}{$X_4 = 0$} & 5  &  20.35 (0.20) &                    0.96 (0.06) &  \textbf{\textbf{0.84} (0.06)} &  12.97 (0.28) &                    1.00 (0.07) &                    1.28 (0.09) &                    1.66 (0.09) &     1.19 (0.08) &                    1.23 (0.09) \\
			& 10 &  33.34 (0.29) &                    1.70 (0.14) &  \textbf{\textbf{1.68} (0.12)} &   8.73 (0.35) &                    1.86 (0.14) &                    1.92 (0.15) &                    2.89 (0.18) &     1.98 (0.13) &                    2.01 (0.14) \\
			& 30 &  39.06 (0.39) &                    4.01 (0.34) &  \textbf{\textbf{3.74} (0.31)} &   9.00 (0.41) &                    4.78 (0.37) &                    5.10 (0.37) &                    6.52 (0.31) &     4.70 (0.32) &                    4.24 (0.34) \\
			& 50 &  32.53 (0.37) &                    5.22 (0.43) &                    5.06 (0.44) &  12.70 (0.43) &                    6.73 (0.50) &                    6.31 (0.39) &                    7.41 (0.39) &     5.22 (0.39) &  \textbf{\textbf{4.82} (0.35)} \\
			& 70 &  26.35 (0.31) &                    6.00 (0.50) &                    6.12 (0.50) &  12.83 (0.44) &                    9.54 (0.64) &                    7.34 (0.56) &                    8.46 (0.42) &     6.30 (0.49) &  \textbf{\textbf{5.77} (0.45)} \\
			\bottomrule
		\end{tabular}
	}
	\caption{ 
		The absolute bias $(\times 10^2)$ for all methods 
		under mild covariate shift ($q = 1$) with a total sample size of $N = 1000$. We consider the scenario with double source domains, wherein the survival time follows the proportional hazard model and the censoring is covariate-independent.   
		Results for all subpopulations are presented based on $100$ simulation replications. \label{simu:double-sp-1}
	}
\end{table}
\begin{table}[ht!]
	\centering
	\resizebox{\textwidth}{!}{
		\begin{tabular}{llccccccccc}
			\hline
			\hline
			&  Time  &     \textsc{Naive}      &  \textsc{IPSW}     & \textsc{IPSW-sub}   &              \textsc{lm}   &          \textsc{ mclm-Ridge} &   \textsc{mclm-Tree} &    \textsc{ rf} & \textsc{mcrf-Ridge} &   \textsc{ mcrf-Tree} \\
			\hline 
			\multicolumn{11}{l}{SP, Absolute bias $(\times 10^2)$, $N = 1000, q = 2$, proportional hazard, independent censoring, double sources }                              \\
			\hline 
			\multirow{5}{*}{\textsc{All} } & 5  &  32.66 (0.20) &                     1.35 (0.24) &     - &  30.37 (0.43) &                    0.91 (0.05) &   \textbf{\textbf{0.75} (0.05)} &   3.51 (0.15) &     1.32 (0.11) &                     2.13 (0.23) \\
			& 10 &  52.74 (0.25) &                     3.44 (0.37) &     - &  16.45 (0.39) &                    2.22 (0.13) &   \textbf{\textbf{2.22} (0.13)} &   7.52 (0.25) &     2.85 (0.20) &                     3.54 (0.31) \\
			& 30 &  60.25 (0.30) &                    11.51 (0.86) &     - &  24.55 (0.54) &                   12.25 (1.07) &   \textbf{\textbf{6.68} (0.53)} &  17.63 (0.54) &     8.51 (0.73) &                     7.68 (0.69) \\
			& 50 &  49.53 (0.34) &                    16.50 (1.33) &    - &  30.36 (0.61) &                   26.02 (1.03) &                    11.36 (0.92) &  20.65 (0.77) &    11.94 (0.97) &  \textbf{\textbf{10.59} (0.88)} \\
			& 70 &  39.79 (0.33) &                    20.52 (1.76) &     - &  28.60 (0.58) &                   24.84 (1.07) &                    17.19 (1.17) &  23.35 (0.78) &    19.20 (1.09) &  \textbf{\textbf{16.84} (1.24)} \\
			\cline{1-11}
			\multirow{5}{*}{$X_3 = 1$} & 5  &  28.07 (0.38) &                     1.65 (0.31) &   1.59 (0.26) &  31.66 (0.51) &                    0.95 (0.07) &   \textbf{\textbf{0.83} (0.06)} &   4.04 (0.18) &     1.36 (0.11) &                     2.45 (0.25) \\
			& 10 &  48.17 (0.44) &                     4.12 (0.47) &   3.92 (0.42) &  17.37 (0.48) &                    2.59 (0.16) &   \textbf{\textbf{2.41} (0.15)} &   8.81 (0.29) &     3.14 (0.21) &                     3.99 (0.30) \\
			& 30 &  59.33 (0.44) &                    15.08 (1.02) &  13.69 (1.07) &  25.57 (0.62) &                   14.18 (1.25) &   \textbf{\textbf{7.82} (0.58)} &  20.63 (0.62) &    10.48 (0.90) &                     9.81 (0.79) \\
			& 50 &  50.03 (0.43) &                    21.90 (1.57) &  21.14 (1.58) &  31.94 (0.66) &                   27.30 (1.08) &                    12.82 (1.08) &  23.97 (0.81) &    14.40 (1.08) &  \textbf{\textbf{11.73} (1.05)} \\
			& 70 &  40.73 (0.41) &                    26.64 (2.11) &  25.96 (2.17) &  30.29 (0.63) &                   26.54 (1.10) &                    18.33 (1.32) &  26.15 (0.81) &    20.98 (1.17) &  \textbf{\textbf{17.57} (1.32)} \\
			\cline{1-11}
			\multirow{5}{*}{$X_3 = 0$} & 5  &  34.31 (0.23) &                     1.55 (0.15) &   1.51 (0.15) &  28.47 (0.46) &                    0.99 (0.07) &   \textbf{\textbf{0.88} (0.08)} &   2.76 (0.17) &     1.57 (0.14) &                     1.98 (0.22) \\
			& 10 &  53.68 (0.36) &                     3.66 (0.27) &   3.29 (0.27) &  15.11 (0.43) &                    2.75 (0.20) &   \textbf{\textbf{2.74} (0.18)} &   5.72 (0.30) &     3.88 (0.34) &                     4.27 (0.38) \\
			& 30 &  58.32 (0.44) &                    10.79 (0.85) &  10.84 (0.83) &  23.09 (0.55) &                   10.93 (1.12) &   \textbf{\textbf{7.67} (0.57)} &  13.28 (0.56) &     8.18 (0.83) &                     7.89 (0.72) \\
			& 50 &  46.91 (0.50) &                    15.39 (1.24) &  15.83 (1.29) &  28.03 (0.68) &                   24.38 (1.03) &  \textbf{\textbf{12.01} (0.90)} &  15.76 (0.86) &    13.26 (0.99) &                    12.73 (0.91) \\
			& 70 &  37.07 (0.50) &  \textbf{\textbf{18.48} (1.53)} &  18.78 (1.60) &  26.10 (0.68) &                   23.47 (1.00) &                    18.48 (1.16) &  19.33 (0.86) &    19.21 (1.10) &                    19.23 (1.28) \\
			\cline{1-11}
			\multirow{5}{*}{$X_4 = 1$} & 5  &  44.78 (0.58) &                     2.62 (0.30) &   2.27 (0.26) &  21.86 (0.62) &                    1.32 (0.11) &   \textbf{\textbf{1.18} (0.11)} &   6.13 (0.31) &     2.53 (0.28) &                     4.01 (0.32) \\
			& 10 &  61.80 (0.57) &                     6.73 (0.56) &   5.86 (0.46) &  10.99 (0.50) &                    3.85 (0.29) &   \textbf{\textbf{2.95} (0.19)} &  11.73 (0.44) &     5.47 (0.45) &                     6.66 (0.53) \\
			& 30 &  57.66 (0.46) &                    20.80 (1.44) &  18.90 (1.41) &  20.36 (0.63) &                   13.90 (1.11) &   \textbf{\textbf{9.33} (0.72)} &  18.91 (0.58) &    11.12 (0.93) &                    11.07 (0.92) \\
			& 50 &  42.81 (0.52) &                    30.45 (1.94) &  29.03 (1.85) &  22.76 (0.72) &                   20.59 (0.83) &                    11.42 (0.77) &  17.70 (0.71) &    12.89 (0.91) &  \textbf{\textbf{11.21} (0.84)} \\
			& 70 &  31.97 (0.47) &                    32.68 (2.10) &  32.00 (1.90) &  20.28 (0.61) &                   19.36 (0.90) &                    15.44 (1.02) &  17.71 (0.67) &    16.61 (0.95) &  \textbf{\textbf{14.40} (1.04)} \\
			\cline{1-11}
			\multirow{5}{*}{$X_4 = 0$} & 5  &  31.08 (0.23) &                     1.06 (0.22) &   1.07 (0.21) &  35.72 (0.41) &                    0.85 (0.06) &   \textbf{\textbf{0.84} (0.05)} &   1.91 (0.11) &     0.95 (0.07) &                     1.39 (0.19) \\
			& 10 &  52.56 (0.26) &                     2.72 (0.36) &   2.66 (0.35) &  20.03 (0.43) &  \textbf{\textbf{2.00} (0.14)} &                     2.37 (0.13) &   4.89 (0.21) &     2.31 (0.18) &                     2.52 (0.22) \\
			& 30 &  64.07 (0.34) &                     9.92 (0.80) &   9.63 (0.79) &  27.18 (0.56) &                   12.22 (1.18) &   \textbf{\textbf{6.25} (0.54)} &  16.82 (0.61) &     8.60 (0.75) &                     7.40 (0.69) \\
			& 50 &  54.69 (0.41) &                    15.09 (1.16) &  15.20 (1.10) &  35.13 (0.65) &                   29.94 (1.17) &                    12.64 (1.11) &  22.51 (0.88) &    12.95 (1.12) &  \textbf{\textbf{11.50} (1.00)} \\
			& 70 &  45.26 (0.43) &                    20.19 (1.61) &  20.16 (1.57) &  33.83 (0.67) &                   28.51 (1.34) &  \textbf{\textbf{19.10} (1.38)} &  26.93 (0.91) &    22.05 (1.26) &                    19.60 (1.41) \\
			\bottomrule
		\end{tabular}
	}
	\caption{ The absolute bias $(\times 10^2)$ for all methods 
		under moderate covariate shift ($q = 2$) with a total sample size of $N = 1000$. We consider the scenario with double source domains, wherein the survival time follows the proportional hazard model and the censoring is covariate-independent.   
		Results for all subpopulations are presented based on $100$ simulation replications. 
	}
\end{table}
\begin{table}[ht!]
	\centering
	\resizebox{\textwidth}{!}{
		\begin{tabular}{llccccccccc}
			\hline
			\hline
			&  Time  &     \textsc{Naive}      &  \textsc{IPSW}     & \textsc{IPSW-sub}   &              \textsc{lm}   &          \textsc{ mclm-Ridge} &   \textsc{mclm-Tree} &    \textsc{ rf} & \textsc{mcrf-Ridge} &   \textsc{ mcrf-Tree} \\
			\hline 
			\multicolumn{11}{l}{SP, Absolute bias $(\times 10^2)$, $N = 1000, q = 3$, proportional hazard, independent censoring, double sources }                              \\
			\hline 
			\multirow{5}{*}{\textsc{All} } & 5  &  37.95 (0.22) &   2.55 (0.36) &    - &  46.19 (0.57) &                    1.01 (0.06) &   \textbf{\textbf{1.01} (0.06)} &   6.23 (0.21) &     2.86 (0.40) &                     4.20 (0.27) \\
			& 10 &  62.14 (0.23) &   6.97 (0.78) &     - &  22.44 (0.56) &  \textbf{\textbf{2.96} (0.18)} &                     3.02 (0.16) &  12.78 (0.32) &     4.61 (0.40) &                     6.76 (0.54) \\
			& 30 &  71.80 (0.23) &  28.14 (1.53) &     - &  38.80 (0.73) &                   28.55 (1.71) &  \textbf{\textbf{10.27} (0.84)} &  29.64 (0.82) &    17.01 (1.50) &                    11.71 (0.98) \\
			& 50 &  59.75 (0.26) &  38.83 (2.31) &    - &  45.42 (0.62) &                   43.97 (0.90) &                    23.48 (1.66) &  39.25 (0.95) &    35.03 (1.47) &  \textbf{\textbf{23.00} (1.66)} \\
			& 70 &  48.56 (0.30) &  39.70 (2.82) &    - &  41.30 (0.62) &                   39.60 (0.96) &                    30.89 (2.00) &  38.91 (0.84) &    35.13 (1.20) &  \textbf{\textbf{29.02} (1.96)} \\
			\cline{1-11}
			\multirow{5}{*}{$X_3 = 1$} & 5  &  33.90 (0.38) &   2.99 (0.48) &   2.45 (0.40) &  45.85 (0.65) &                    1.10 (0.06) &   \textbf{\textbf{1.05} (0.06)} &   6.78 (0.23) &     2.81 (0.37) &                     4.61 (0.30) \\
			& 10 &  58.12 (0.40) &   8.49 (1.04) &   7.18 (0.91) &  22.31 (0.62) &  \textbf{\textbf{3.13} (0.22)} &                     3.14 (0.16) &  14.00 (0.35) &     4.59 (0.44) &                     7.54 (0.55) \\
			& 30 &  70.52 (0.37) &  31.47 (1.85) &  27.33 (1.76) &  39.04 (0.79) &                   29.04 (1.71) &  \textbf{\textbf{11.73} (0.90)} &  32.33 (0.86) &    18.03 (1.59) &                    13.98 (1.07) \\
			& 50 &  59.29 (0.32) &  44.25 (2.15) &  40.61 (2.10) &  45.65 (0.66) &                   44.16 (0.91) &  \textbf{\textbf{23.77} (1.68)} &  41.16 (0.94) &    37.26 (1.37) &                    25.30 (1.64) \\
			& 70 &  48.33 (0.38) &  44.13 (2.41) &  43.25 (2.34) &  41.38 (0.65) &                   40.43 (0.81) &                    31.17 (1.87) &  39.78 (0.83) &    35.62 (1.22) &  \textbf{\textbf{29.65} (1.84)} \\
			\cline{1-11}
			\multirow{5}{*}{$X_3 = 0$} & 5  &  39.44 (0.28) &   2.01 (0.37) &   1.97 (0.35) &  46.76 (0.59) &  \textbf{\textbf{1.03} (0.08)} &                     1.03 (0.08) &   5.24 (0.25) &     3.35 (0.47) &                     3.67 (0.27) \\
			& 10 &  63.32 (0.28) &   5.83 (0.75) &   5.68 (0.71) &  22.65 (0.61) &  \textbf{\textbf{3.00} (0.17)} &                     3.14 (0.19) &  10.59 (0.39) &     5.27 (0.47) &                     5.97 (0.60) \\
			& 30 &  71.63 (0.35) &  23.77 (1.60) &  24.24 (1.55) &  38.37 (0.73) &                   28.12 (1.78) &  \textbf{\textbf{10.38} (0.82)} &  24.85 (0.87) &    17.36 (1.57) &                    11.67 (0.89) \\
			& 50 &  59.49 (0.47) &  34.03 (2.30) &  34.56 (2.40) &  45.00 (0.71) &                   43.80 (0.92) &                    25.16 (1.81) &  35.83 (1.11) &    32.63 (1.62) &  \textbf{\textbf{22.59} (1.69)} \\
			& 70 &  48.38 (0.52) &  39.83 (2.84) &  41.09 (3.05) &  41.12 (0.78) &                   40.00 (0.98) &                    32.79 (2.17) &  37.31 (1.04) &    35.65 (1.14) &  \textbf{\textbf{31.19} (2.16)} \\
			\cline{1-11}
			\multirow{5}{*}{$X_4= 1$} & 5  &  57.02 (0.63) &   4.64 (0.60) &   3.85 (0.52) &  34.16 (0.75) &                    1.38 (0.10) &   \textbf{\textbf{1.20} (0.12)} &  11.18 (0.43) &     5.30 (0.72) &                     8.33 (0.46) \\
			& 10 &  75.85 (0.50) &  15.65 (1.51) &  13.83 (1.38) &  14.49 (0.69) &                    4.55 (0.31) &   \textbf{\textbf{3.56} (0.26)} &  20.75 (0.57) &     7.70 (0.92) &                    13.44 (0.87) \\
			& 30 &  67.77 (0.40) &  43.12 (2.24) &  41.91 (2.29) &  33.14 (0.80) &                   25.64 (1.48) &  \textbf{\textbf{10.93} (0.88)} &  30.40 (0.80) &    19.94 (1.45) &                    15.70 (1.21) \\
			& 50 &  49.62 (0.41) &  49.00 (1.70) &  47.87 (1.76) &  34.72 (0.66) &                   33.55 (0.82) &  \textbf{\textbf{19.40} (1.37)} &  32.13 (0.80) &    29.04 (1.13) &                    21.46 (1.36) \\
			& 70 &  36.87 (0.41) &  38.14 (1.71) &  37.35 (1.78) &  29.43 (0.68) &                   29.02 (0.85) &                    24.06 (1.61) &  28.63 (0.76) &    26.38 (0.91) &  \textbf{\textbf{22.88} (1.58)} \\
			\cline{1-11}
			\multirow{5}{*}{$X_4 = 0$} & 5  &  35.56 (0.22) &   1.84 (0.36) &   1.90 (0.37) &  54.69 (0.57) &  \textbf{\textbf{1.00} (0.05)} &                     1.02 (0.05) &   2.75 (0.16) &     1.43 (0.24) &                     1.57 (0.19) \\
			& 10 &  61.39 (0.23) &   4.86 (0.78) &   5.00 (0.79) &  28.15 (0.60) &  \textbf{\textbf{2.74} (0.16)} &                     3.10 (0.13) &   7.11 (0.29) &     3.42 (0.30) &                     3.40 (0.38) \\
			& 30 &  77.04 (0.27) &  21.27 (1.52) &  22.10 (1.55) &  42.78 (0.74) &                   30.90 (1.89) &                    10.58 (0.89) &  29.09 (0.89) &    16.98 (1.61) &  \textbf{\textbf{10.38} (0.96)} \\
			& 50 &  67.74 (0.35) &  33.83 (2.10) &  34.78 (2.09) &  53.00 (0.66) &                   51.36 (1.03) &                    26.85 (1.93) &  44.26 (1.10) &    39.78 (1.71) &  \textbf{\textbf{25.13} (1.87)} \\
			& 70 &  57.20 (0.41) &  41.65 (2.74) &  41.74 (2.78) &  49.72 (0.66) &                   47.24 (1.17) &                    36.33 (2.34) &  46.19 (0.95) &    41.49 (1.47) &  \textbf{\textbf{33.79} (2.28)} \\
			\bottomrule
		\end{tabular}
	}
	\caption{ The absolute bias $(\times 10^2)$ for all methods 
		under strong covariate shift ($q = 3$) with a total sample size of $N = 1000$. We consider the scenario with double source domains, wherein the survival time follows the proportional hazard model and the censoring is covariate-independent.   
		Results for all subpopulations are presented based on $100$ simulation replications.  
	}
\end{table}

\begin{table}[ht!]
	\centering
	\resizebox{\textwidth}{!}{
		\begin{tabular}{llccccccccc}
			\hline
			\hline
			&  Time  &     \textsc{Naive}      &  \textsc{IPSW}     & \textsc{IPSW-sub}   &              \textsc{lm}   &          \textsc{ mclm-Ridge} &   \textsc{mclm-Tree} &    \textsc{ rf} & \textsc{mcrf-Ridge} &   \textsc{ mcrf-Tree} \\
			\hline 
			\multicolumn{11}{l}{RM, Relative bias $(\times 10^2)$, $N = 1000, q = 1$, proportional hazard, independent censoring, double source }                              \\
			\hline
			\multirow{5}{*}{\textsc{All} } & 5  &   9.83 (0.09) &  \textbf{\textbf{0.33} (0.02)} &                      - &                    7.36 (0.16) &     0.39 (0.03) &                    0.55 (0.04) &   0.81 (0.04) &                    0.69 (0.04) &                    0.51 (0.04) \\
			& 10 &  19.78 (0.14) &  \textbf{\textbf{0.61} (0.05)} &                      - &                    8.83 (0.22) &     0.82 (0.06) &                    0.65 (0.05) &   1.80 (0.08) &                    1.17 (0.08) &                    0.80 (0.06) \\
			& 30 &  39.61 (0.19) &  \textbf{\textbf{2.12} (0.16)} &                      - &                    2.61 (0.18) &     2.39 (0.16) &                    2.58 (0.17) &   4.80 (0.18) &                    2.73 (0.18) &                    2.41 (0.19) \\
			& 50 &  48.16 (0.22) &                    3.74 (0.31) &                      - &                    4.54 (0.26) &     4.03 (0.29) &                    4.38 (0.32) &   7.31 (0.27) &                    4.17 (0.32) &  \textbf{\textbf{3.58} (0.30)} \\
			& 70 &  53.04 (0.24) &                    5.32 (0.48) &                      - &                    9.02 (0.37) &     5.45 (0.39) &                    5.40 (0.36) &   9.42 (0.36) &                    4.83 (0.37) &  \textbf{\textbf{4.78} (0.32)} \\
			\cline{1-11}
			\multirow{5}{*}{$X_3 = 1$} & 5  &   7.17 (0.16) &                    0.51 (0.04) &  \textbf{\textbf{0.43} (0.03)} &                    8.56 (0.22) &     0.46 (0.03) &                    0.57 (0.04) &   1.05 (0.05) &                    0.88 (0.05) &                    0.63 (0.05) \\
			& 10 &  15.49 (0.22) &                    1.10 (0.08) &                    0.88 (0.06) &                   10.61 (0.28) &     0.96 (0.08) &  \textbf{\textbf{0.74} (0.06)} &   2.34 (0.10) &                    1.47 (0.10) &                    1.09 (0.07) \\
			& 30 &  34.12 (0.34) &                    3.68 (0.28) &                    2.70 (0.22) &                    3.59 (0.25) &     2.70 (0.19) &  \textbf{\textbf{2.64} (0.19)} &   6.27 (0.23) &                    3.58 (0.23) &                    2.98 (0.22) \\
			& 50 &  42.88 (0.43) &                    6.50 (0.53) &                    5.18 (0.42) &  \textbf{\textbf{4.96} (0.31)} &     5.43 (0.43) &                    5.05 (0.37) &   9.58 (0.37) &                    5.65 (0.49) &                    4.99 (0.38) \\
			& 70 &  48.11 (0.47) &                    9.23 (0.73) &                    7.94 (0.59) &                    9.98 (0.44) &     7.44 (0.58) &  \textbf{\textbf{6.32} (0.49)} &  12.36 (0.45) &                    7.49 (0.60) &                    6.79 (0.48) \\
			\cline{1-11}
			\multirow{5}{*}{$X_3 = 0$} & 5  &  10.86 (0.14) &                    0.60 (0.04) &  \textbf{\textbf{0.53} (0.04)} &                    6.05 (0.17) &     0.56 (0.04) &                    0.69 (0.05) &   0.70 (0.04) &                    0.66 (0.05) &                    0.59 (0.04) \\
			& 10 &  21.28 (0.20) &                    1.31 (0.09) &                    1.03 (0.08) &                    6.89 (0.24) &     1.06 (0.08) &                    0.98 (0.07) &   1.40 (0.09) &                    1.24 (0.08) &  \textbf{\textbf{0.97} (0.08)} \\
			& 30 &  41.04 (0.25) &                    3.41 (0.24) &                    2.75 (0.20) &  \textbf{\textbf{2.41} (0.17)} &     2.86 (0.21) &                    3.03 (0.20) &   3.31 (0.21) &                    2.81 (0.21) &                    2.83 (0.21) \\
			& 50 &  49.21 (0.30) &                    5.64 (0.43) &                    4.77 (0.39) &                    4.59 (0.29) &     4.65 (0.29) &                    4.92 (0.32) &   4.93 (0.31) &                    4.78 (0.29) &  \textbf{\textbf{4.39} (0.30)} \\
			& 70 &  53.71 (0.34) &                    8.12 (0.57) &                    6.96 (0.55) &                    7.96 (0.42) &     6.58 (0.51) &                    6.37 (0.43) &   6.39 (0.40) &  \textbf{\textbf{5.83} (0.46)} &                    6.24 (0.44) \\
			\cline{1-11}
			\multirow{5}{*}{$X_4 = 1$} & 5  &  12.51 (0.26) &                    0.84 (0.07) &  \textbf{\textbf{0.75} (0.07)} &                    4.81 (0.24) &     0.84 (0.07) &                    0.93 (0.07) &   1.23 (0.08) &                    1.24 (0.09) &                    0.98 (0.07) \\
			& 10 &  23.57 (0.35) &                    1.88 (0.14) &                    1.40 (0.10) &                    5.70 (0.31) &     1.35 (0.10) &  \textbf{\textbf{1.05} (0.08)} &   2.44 (0.13) &                    2.04 (0.15) &                    1.38 (0.11) \\
			& 30 &  43.12 (0.54) &                    6.69 (0.50) &                    5.16 (0.39) &  \textbf{\textbf{3.19} (0.25)} &     3.91 (0.29) &                    3.68 (0.28) &   5.72 (0.31) &                    5.02 (0.36) &                    3.92 (0.29) \\
			& 50 &  51.06 (0.63) &                   11.54 (0.87) &                    9.51 (0.74) &  \textbf{\textbf{5.10} (0.37)} &     7.21 (0.58) &                    6.47 (0.45) &   8.13 (0.46) &                    7.98 (0.59) &                    6.08 (0.50) \\
			& 70 &  55.39 (0.68) &                   15.87 (1.21) &                   13.69 (1.06) &                    8.14 (0.48) &    10.32 (0.71) &                    8.37 (0.66) &   9.80 (0.54) &                   10.14 (0.64) &  \textbf{\textbf{7.91} (0.59)} \\
			\cline{1-11}
			\multirow{5}{*}{$X_4 = 0$} & 5  &   9.42 (0.11) &                    0.41 (0.03) &  \textbf{\textbf{0.37} (0.03)} &                    8.59 (0.17) &     0.38 (0.03) &                    0.51 (0.04) &   0.66 (0.04) &                    0.62 (0.04) &                    0.44 (0.04) \\
			& 10 &  19.37 (0.15) &                    0.78 (0.06) &  \textbf{\textbf{0.69} (0.06)} &                   10.30 (0.23) &     0.88 (0.07) &                    0.79 (0.06) &   1.56 (0.08) &                    1.07 (0.07) &                    0.84 (0.06) \\
			& 30 &  39.70 (0.22) &                    2.18 (0.19) &  \textbf{\textbf{1.87} (0.16)} &                    3.02 (0.21) &     2.32 (0.19) &                    2.43 (0.18) &   4.44 (0.19) &                    2.39 (0.17) &                    2.29 (0.19) \\
			& 50 &  48.53 (0.25) &                    3.75 (0.33) &  \textbf{\textbf{3.48} (0.30)} &                    4.97 (0.25) &     3.90 (0.32) &                    4.31 (0.33) &   7.07 (0.26) &                    4.18 (0.30) &                    3.58 (0.29) \\
			& 70 &  53.58 (0.26) &                    5.49 (0.46) &                    5.22 (0.45) &                    9.92 (0.37) &     5.77 (0.41) &                    5.14 (0.36) &   9.47 (0.34) &                    5.03 (0.42) &  \textbf{\textbf{4.41} (0.32)} \\
			\bottomrule
		\end{tabular}
		
	}
	\caption{ 
		The relative bias $(\times 10^2)$ for all methods 
		under mild covariate shift ($q = 1$) with a total sample size of $N = 1000$. We consider the scenario of double source domains, wherein the survival time follows the proportional hazard model, and the censoring is covariate-independent. Results for all subpopulations are presented based on $100$ simulation replications. 
	}
\end{table}

\begin{table}[ht!]
	\centering
	\resizebox{\textwidth}{!}{
		\begin{tabular}{llcccccccccc}
			\hline
			\hline
			&  Time  &     \textsc{Naive}      &  \textsc{IPSW}     & \textsc{IPSW-sub}   &              \textsc{lm}   &          \textsc{ mclm-Ridge} &   \textsc{mclm-Tree} &    \textsc{ rf} & \textsc{mcrf-Ridge} &   \textsc{ mcrf-Tree} \\
			\hline 
			\multicolumn{11}{l}{RM, Relative bias $(\times 10^2)$, $N = 1000, q = 2$, proportional hazard, independent censoring, double source }                              \\
			\hline
			\multirow{5}{*}{\textsc{All} } & 5  &  15.13 (0.10) &   0.51 (0.11) &    - &                   20.07 (0.27) &                    0.31 (0.02) &  \textbf{\textbf{0.27} (0.02)} &   1.47 (0.06) &     0.82 (0.07) &                    0.77 (0.06) \\
			& 10 &  30.32 (0.14) &   1.39 (0.21) &     - &                   22.72 (0.28) &                    0.83 (0.05) &  \textbf{\textbf{0.74} (0.05)} &   3.67 (0.11) &     1.73 (0.14) &                    1.92 (0.14) \\
			& 30 &  57.46 (0.16) &   6.14 (0.49) &    - &  \textbf{\textbf{3.08} (0.20)} &                    3.55 (0.25) &                    3.62 (0.24) &  11.62 (0.27) &     4.39 (0.36) &                    5.47 (0.50) \\
			& 50 &  67.31 (0.16) &  10.81 (0.85) &     - &                   13.49 (0.40) &                    8.97 (0.67) &  \textbf{\textbf{6.92} (0.45)} &  17.94 (0.39) &     8.23 (0.71) &                    7.28 (0.69) \\
			& 70 &  72.32 (0.16) &  15.24 (1.26) &     - &                   23.10 (0.46) &                   12.27 (0.93) &  \textbf{\textbf{8.06} (0.66)} &  23.16 (0.53) &    13.24 (0.98) &                    9.91 (0.76) \\
			\cline{1-11}
			\multirow{5}{*}{$X_3 = 1$} & 5  &  12.32 (0.19) &   0.63 (0.14) &   0.60 (0.12) &                   20.93 (0.31) &                    0.34 (0.02) &  \textbf{\textbf{0.31} (0.02)} &   1.69 (0.07) &     0.88 (0.08) &                    0.94 (0.07) \\
			& 10 &  26.36 (0.26) &   1.70 (0.27) &   1.60 (0.23) &                   23.72 (0.33) &                    0.90 (0.06) &  \textbf{\textbf{0.78} (0.05)} &   4.21 (0.14) &     1.96 (0.15) &                    2.21 (0.15) \\
			& 30 &  53.33 (0.30) &   7.83 (0.62) &   7.18 (0.60) &  \textbf{\textbf{3.46} (0.26)} &                    4.12 (0.29) &                    3.64 (0.24) &  13.38 (0.33) &     5.12 (0.40) &                    6.32 (0.54) \\
			& 50 &  63.59 (0.28) &  14.18 (1.00) &  13.00 (1.02) &                   13.64 (0.44) &                   10.01 (0.79) &  \textbf{\textbf{7.32} (0.51)} &  20.47 (0.46) &     9.50 (0.83) &                    9.17 (0.75) \\
			& 70 &  68.95 (0.27) &  19.45 (1.47) &  18.32 (1.50) &                   23.53 (0.49) &                   12.69 (0.97) &  \textbf{\textbf{8.22} (0.66)} &  26.11 (0.58) &    16.32 (1.05) &                   12.13 (0.89) \\
			\cline{1-11}
			\multirow{5}{*}{$X_3 = 0$} & 5  &  16.21 (0.12) &   0.52 (0.05) &   0.50 (0.05) &                   18.80 (0.29) &                    0.37 (0.03) &  \textbf{\textbf{0.36} (0.02)} &   1.16 (0.07) &     0.87 (0.07) &                    0.63 (0.06) \\
			& 10 &  31.74 (0.17) &   1.51 (0.13) &   1.38 (0.13) &                   21.23 (0.32) &                    0.89 (0.06) &  \textbf{\textbf{0.88} (0.06)} &   2.88 (0.14) &     1.81 (0.14) &                    1.89 (0.18) \\
			& 30 &  58.55 (0.20) &   5.69 (0.42) &   5.43 (0.43) &  \textbf{\textbf{3.19} (0.19)} &                    3.56 (0.27) &                    4.03 (0.29) &   8.93 (0.29) &     4.40 (0.37) &                    5.09 (0.50) \\
			& 50 &  68.13 (0.19) &  10.26 (0.78) &  10.26 (0.76) &                   13.22 (0.41) &                    9.69 (0.77) &                    8.17 (0.51) &  14.02 (0.44) &     8.86 (0.72) &  \textbf{\textbf{7.48} (0.68)} \\
			& 70 &  72.95 (0.19) &  14.83 (1.17) &  14.89 (1.20) &                   22.35 (0.47) &                   12.26 (0.96) &  \textbf{\textbf{9.46} (0.73)} &  18.48 (0.61) &    12.59 (0.98) &                   10.72 (0.78) \\
			\cline{1-11}
			\multirow{5}{*}{$X_4 = 1$} & 5  &  22.81 (0.34) &   0.94 (0.13) &   0.84 (0.10) &                   14.62 (0.37) &  \textbf{\textbf{0.45} (0.03)} &                    0.52 (0.04) &   2.71 (0.12) &     1.64 (0.14) &                    1.64 (0.13) \\
			& 10 &  40.37 (0.42) &   2.83 (0.27) &   2.40 (0.22) &                   16.36 (0.37) &                    1.11 (0.08) &  \textbf{\textbf{0.99} (0.06)} &   6.45 (0.22) &     3.73 (0.29) &                    4.03 (0.24) \\
			& 30 &  65.52 (0.40) &  11.60 (0.90) &  10.67 (0.84) &  \textbf{\textbf{3.43} (0.26)} &                    6.28 (0.50) &                    5.05 (0.45) &  16.28 (0.42) &     8.85 (0.82) &                    9.42 (0.76) \\
			& 50 &  73.57 (0.35) &  21.42 (1.53) &  19.99 (1.46) &                   12.58 (0.51) &                   12.70 (0.87) &  \textbf{\textbf{8.65} (0.55)} &  21.87 (0.48) &    12.09 (0.92) &                   11.00 (0.84) \\
			& 70 &  77.35 (0.33) &  29.81 (2.06) &  28.27 (1.90) &                   20.20 (0.59) &                   12.42 (0.88) &  \textbf{\textbf{9.34} (0.71)} &  25.82 (0.60) &    19.37 (1.06) &                   14.92 (0.93) \\
			\cline{1-11}
			\multirow{5}{*}{$X_4 = 0$} & 5  &  13.97 (0.11) &   0.38 (0.10) &   0.39 (0.09) &                   23.47 (0.26) &                    0.31 (0.02) &  \textbf{\textbf{0.24} (0.02)} &   0.71 (0.04) &     0.43 (0.04) &                    0.37 (0.03) \\
			& 10 &  29.03 (0.15) &   1.09 (0.20) &   1.09 (0.19) &                   26.64 (0.30) &                    0.93 (0.05) &  \textbf{\textbf{0.83} (0.05)} &   1.97 (0.09) &     0.96 (0.07) &                    1.02 (0.10) \\
			& 30 &  57.10 (0.17) &   4.98 (0.46) &   4.87 (0.45) &                    3.68 (0.27) &  \textbf{\textbf{3.17} (0.21)} &                    3.53 (0.20) &   8.91 (0.26) &     3.35 (0.23) &                    4.11 (0.39) \\
			& 50 &  67.45 (0.17) &   8.95 (0.76) &   8.77 (0.74) &                   13.95 (0.39) &                    8.55 (0.68) &                    7.11 (0.45) &  15.76 (0.41) &     8.10 (0.66) &  \textbf{\textbf{6.60} (0.65)} \\
			& 70 &  72.76 (0.16) &  13.36 (1.07) &  13.34 (1.01) &                   24.59 (0.44) &                   12.67 (0.99) &  \textbf{\textbf{8.43} (0.66)} &  21.71 (0.56) &    12.14 (0.96) &                    9.62 (0.69) \\
			\bottomrule
		\end{tabular}
	}
	\caption{
		The relative bias $(\times 10^2)$ for all methods 
		under moderate covariate shift ($q = 2$) with a total sample size of $N = 1000$. We consider the scenario of double source domains, wherein the survival time follows the proportional hazard model and the censoring is covariate-independent. Results for all subpopulations are presented based on $100$ simulation replications.     
	}
\end{table}

\begin{table}[ht!]
	\centering
	\resizebox{\textwidth}{!}{
		\begin{tabular}{llcccccccccc}
			\hline
			\hline
			&  Time  &     \textsc{Naive}      &  \textsc{IPSW}     & \textsc{IPSW-sub}   &              \textsc{lm}   &          \textsc{ mclm-Ridge} &   \textsc{mclm-Tree} &    \textsc{ rf} & \textsc{mcrf-Ridge} &   \textsc{ mcrf-Tree} \\
			\hline 
			\multicolumn{11}{l}{RM, Relative bias $(\times 10^2)$, $N = 1000, q = 3$, proportional hazard, independent censoring, double source }                              \\
			\hline
			\multirow{5}{*}{\textsc{All} } & 5  &  17.18 (0.12) &   0.85 (0.14) &     - &                   30.63 (0.34) &     0.37 (0.02) &   \textbf{\textbf{0.27} (0.02)} &   2.63 (0.11) &                    1.11 (0.10) &    1.38 (0.10) \\
			& 10 &  34.88 (0.15) &   2.79 (0.35) &    - &                   33.65 (0.37) &     1.20 (0.05) &   \textbf{\textbf{1.16} (0.06)} &   6.27 (0.15) &                    2.78 (0.22) &    3.40 (0.22) \\
			& 30 &  64.95 (0.14) &  13.51 (0.85) &     - &  \textbf{\textbf{3.35} (0.24)} &     5.07 (0.33) &                     4.18 (0.33) &  18.62 (0.36) &                    6.93 (0.63) &    8.39 (0.74) \\
			& 50 &  74.77 (0.12) &  24.03 (1.30) &    - &                   21.10 (0.47) &    15.08 (1.34) &   \textbf{\textbf{8.72} (0.60)} &  29.02 (0.54) &                   15.78 (1.37) &   12.81 (0.87) \\
			& 70 &  79.49 (0.10) &  32.45 (1.93) &     - &                   33.23 (0.52) &    28.15 (0.88) &  \textbf{\textbf{13.69} (1.04)} &  37.32 (0.68) &                   31.82 (1.11) &   20.44 (1.22) \\
			\cline{1-11}
			\multirow{5}{*}{$X_3 = 1$} & 5  &  14.78 (0.20) &   0.93 (0.18) &   0.78 (0.15) &                   30.48 (0.38) &     0.39 (0.02) &   \textbf{\textbf{0.29} (0.02)} &   2.90 (0.12) &                    1.11 (0.10) &    1.47 (0.11) \\
			& 10 &  31.52 (0.25) &   3.21 (0.45) &   2.70 (0.38) &                   33.48 (0.42) &     1.22 (0.06) &   \textbf{\textbf{1.20} (0.06)} &   6.82 (0.16) &                    2.80 (0.24) &    3.53 (0.22) \\
			& 30 &  61.80 (0.26) &  15.45 (1.16) &  12.95 (1.04) &  \textbf{\textbf{3.64} (0.24)} &     5.72 (0.40) &                     4.44 (0.35) &  20.46 (0.39) &                    7.73 (0.70) &    9.38 (0.77) \\
			& 50 &  72.15 (0.23) &  27.64 (1.44) &  23.66 (1.40) &                   21.25 (0.50) &    15.51 (1.30) &   \textbf{\textbf{9.39} (0.66)} &  31.39 (0.56) &                   17.65 (1.31) &   15.08 (0.97) \\
			& 70 &  77.22 (0.20) &  36.58 (1.88) &  32.79 (1.80) &                   33.34 (0.54) &    28.37 (0.88) &  \textbf{\textbf{14.29} (1.04)} &  39.72 (0.68) &                   35.09 (0.98) &   23.83 (1.27) \\
			\cline{1-11}
			\multirow{5}{*}{$X_3 = 0$} & 5  &  18.07 (0.14) &   0.69 (0.15) &   0.68 (0.14) &                   30.88 (0.36) &     0.38 (0.03) &   \textbf{\textbf{0.31} (0.02)} &   2.15 (0.11) &                    1.18 (0.11) &    1.28 (0.10) \\
			& 10 &  36.08 (0.18) &   2.28 (0.36) &   2.20 (0.34) &                   33.94 (0.40) &     1.20 (0.06) &   \textbf{\textbf{1.12} (0.06)} &   5.27 (0.18) &                    2.86 (0.23) &    3.28 (0.27) \\
			& 30 &  65.96 (0.15) &  11.37 (0.92) &  11.24 (0.89) &  \textbf{\textbf{3.48} (0.30)} &     4.78 (0.32) &                     4.58 (0.31) &  15.28 (0.40) &                    6.76 (0.61) &    7.60 (0.73) \\
			& 50 &  75.61 (0.13) &  20.14 (1.47) &  20.45 (1.45) &                   20.81 (0.48) &    16.24 (1.51) &   \textbf{\textbf{9.20} (0.63)} &  24.72 (0.61) &                   15.16 (1.51) &   11.15 (0.73) \\
			& 70 &  80.21 (0.13) &  28.64 (2.05) &  29.11 (2.08) &                   32.99 (0.54) &    27.95 (0.88) &  \textbf{\textbf{15.37} (1.02)} &  32.93 (0.79) &                   27.01 (1.37) &   17.12 (1.10) \\
			\cline{1-11}
			\multirow{5}{*}{$X_4 = 1$} & 5  &  29.46 (0.39) &   1.56 (0.23) &   1.32 (0.21) &                   22.68 (0.45) &     0.42 (0.03) &   \textbf{\textbf{0.34} (0.03)} &   5.05 (0.23) &                    2.28 (0.20) &    3.04 (0.21) \\
			& 10 &  49.99 (0.41) &   5.55 (0.61) &   4.88 (0.56) &                   24.57 (0.46) &     1.30 (0.08) &   \textbf{\textbf{1.22} (0.10)} &  11.37 (0.30) &                    6.10 (0.47) &    7.40 (0.38) \\
			& 30 &  75.67 (0.30) &  25.40 (1.82) &  24.19 (1.70) &  \textbf{\textbf{4.56} (0.34)} &     7.76 (0.62) &                     5.22 (0.47) &  25.46 (0.48) &                   11.84 (0.97) &   13.82 (0.94) \\
			& 50 &  82.44 (0.25) &  39.48 (2.10) &  37.94 (2.02) &                   20.64 (0.55) &    15.39 (1.24) &  \textbf{\textbf{10.19} (0.67)} &  34.05 (0.60) &                   22.74 (1.50) &   20.52 (1.20) \\
			& 70 &  85.40 (0.23) &  47.96 (2.40) &  46.71 (2.30) &                   30.51 (0.60) &    26.40 (0.78) &  \textbf{\textbf{13.69} (0.95)} &  40.43 (0.69) &                   37.36 (0.80) &   26.90 (1.22) \\
			\cline{1-11}
			\multirow{5}{*}{$X_4 = 0$} & 5  &  15.50 (0.12) &   0.62 (0.15) &   0.64 (0.15) &                   36.23 (0.33) &     0.36 (0.02) &   \textbf{\textbf{0.31} (0.02)} &   0.94 (0.06) &                    0.44 (0.04) &    0.39 (0.04) \\
			& 10 &  33.02 (0.15) &   1.97 (0.36) &   2.05 (0.37) &                   39.99 (0.39) &     1.25 (0.05) &                     1.25 (0.05) &   2.71 (0.11) &  \textbf{\textbf{1.06} (0.08)} &    1.11 (0.13) \\
			& 30 &  64.34 (0.14) &   9.42 (0.90) &   9.85 (0.92) &  \textbf{\textbf{3.78} (0.31)} &     4.60 (0.29) &                     4.34 (0.29) &  14.11 (0.35) &                    5.38 (0.57) &    6.01 (0.64) \\
			& 50 &  74.85 (0.11) &  17.92 (1.29) &  18.48 (1.31) &                   21.36 (0.46) &    15.48 (1.49) &   \textbf{\textbf{8.26} (0.61)} &  25.89 (0.55) &                   13.42 (1.39) &   10.14 (0.77) \\
			& 70 &  79.93 (0.10) &  25.60 (1.78) &  26.17 (1.80) &                   34.81 (0.50) &    29.35 (0.93) &  \textbf{\textbf{14.02} (1.10)} &  35.47 (0.70) &                   28.96 (1.36) &   18.30 (1.17) \\
			\bottomrule
		\end{tabular}
	}
	\caption{ 
		The relative bias $(\times 10^2)$ for all methods 
		under strong covariate shift ($q = 3$) with a total sample size of $N = 1000$. We consider the scenario of double source domains, wherein the survival time follows the proportional hazard model and the censoring is covariate-independent. Results for all subpopulations are presented based on $100$ simulation replications.  \label{simu:double-rm-3}       
	}
\end{table}

\subsection{Comparison of C-index}

\begin{figure}[ht!]
	\centering
	\includegraphics[width = 0.9\textwidth, height=0.38\textheight]{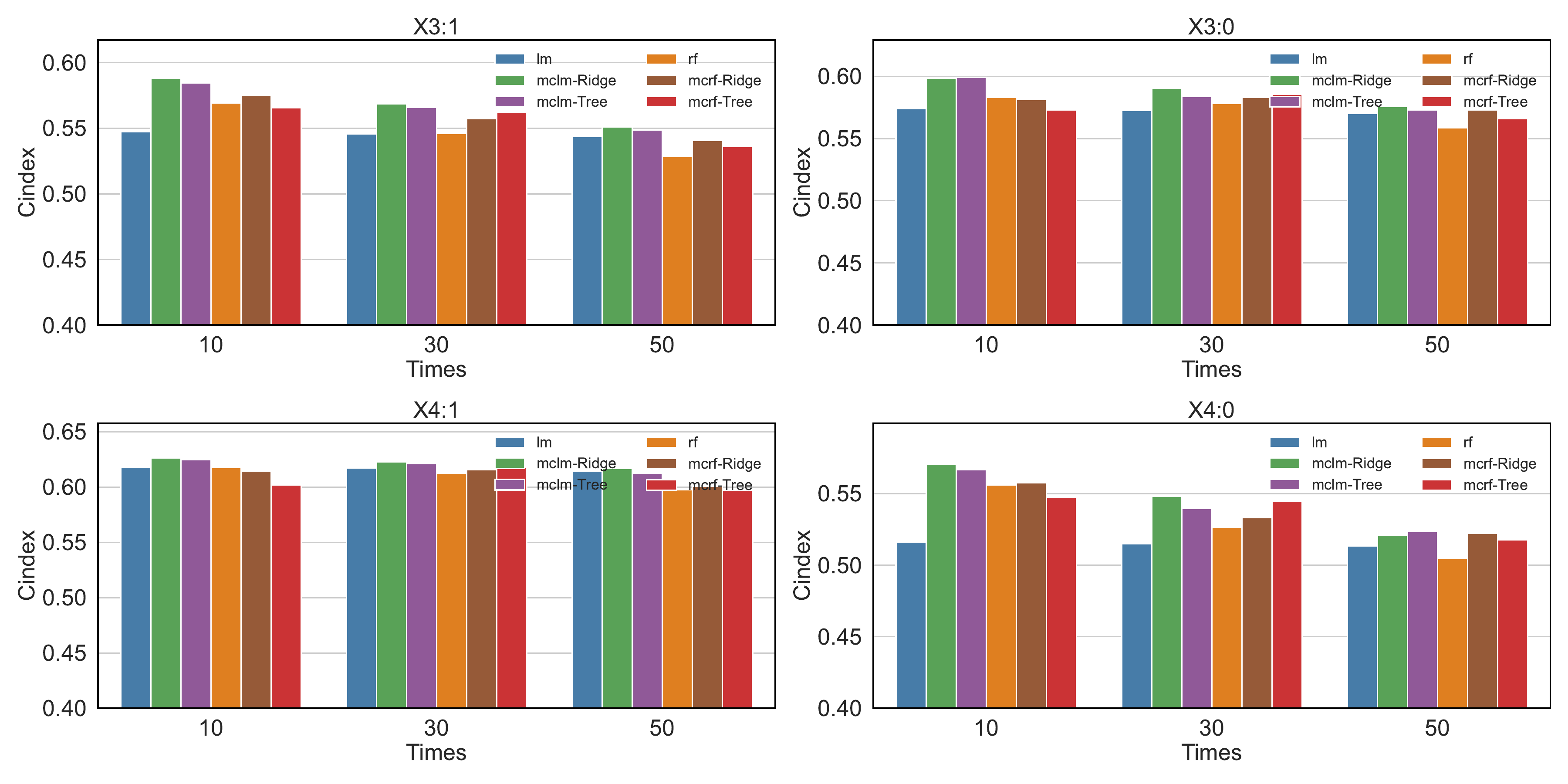}
	\caption{C-index for all methods under moderate covariate shift ($q = 2$) for various subpopulations. The proportional hazard model with independent censoring following a uniform distribution is considered. Results are based on an average of $100$ simulation replications.  }
	\label{fig:Cindex-prop-indep-single-1}
\end{figure}

\begin{figure}[ht!]
	\centering
	\includegraphics[width = 0.9\textwidth, height=0.38\textheight]{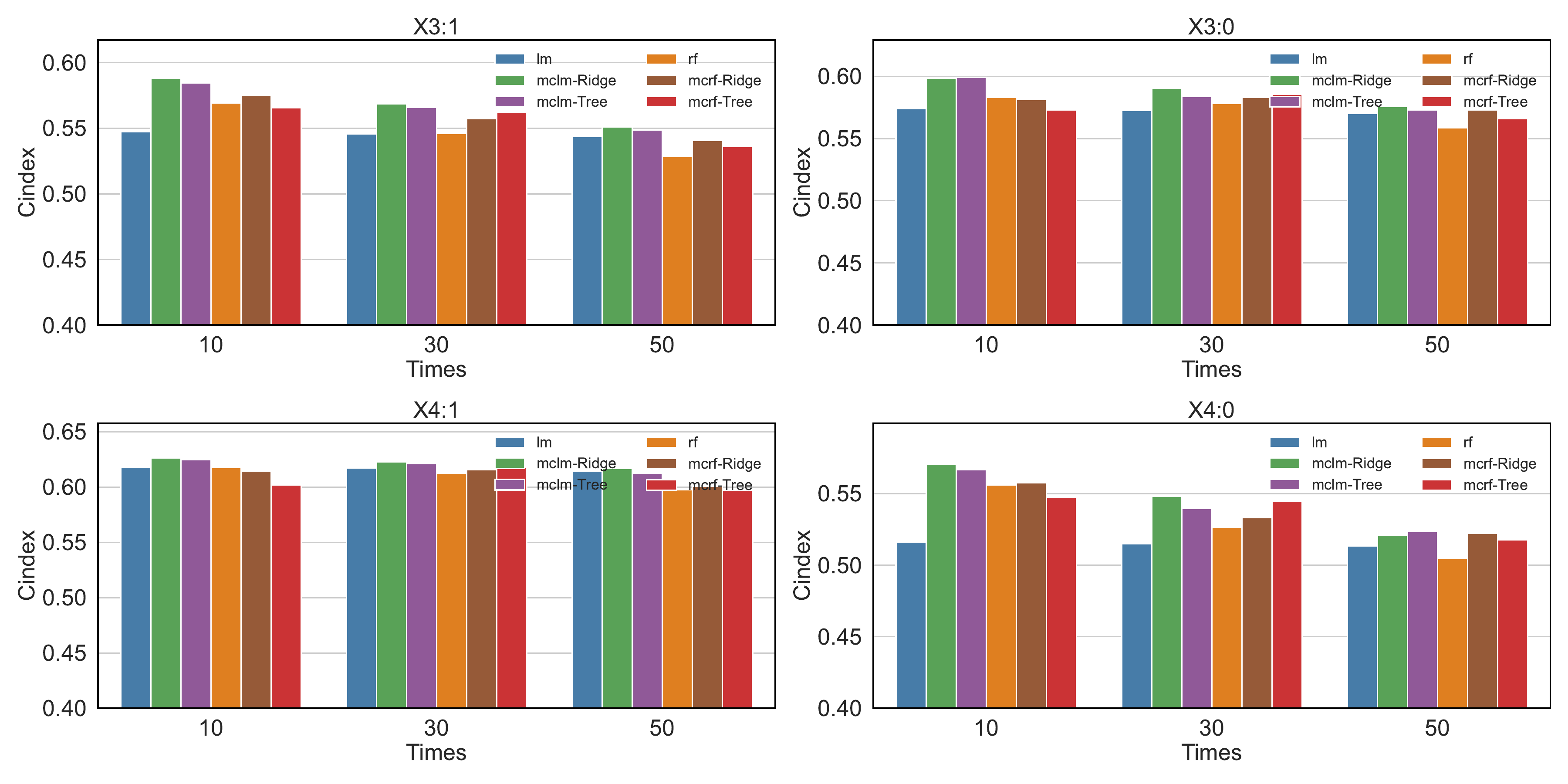}
	\caption{C-index for all methods under moderate covariate shift ($q = 2$) for various subpopulations. The proportional hazard model with independent censoring following a uniform distribution is considered. Results are based on an average of $100$ simulation replications.  }
	\label{fig:Cindex-prop-indep-single-2}
\end{figure}

\section{Extended simulations to mimic real daata}

\subsection{Adjusting parameter of membership probability odds}

We conduct additional numerical experiments with different parameter configurations per one reviewer's request to gain more insights. We explore scenarios with $1500$ source and $300$ target samples and vary $\omega_0 \in \{0.5, 1, 2\}$, the first parameter of $\omega = (0, 0.5, 0.45, -0.9, -0.7)^\top$, to mimic the real-world studies. We maintained a total of $5$ covariates, including $2$ categorical variables that mimic gender and race, as in previous simulations. 

\begin{figure}
	\centering
	\includegraphics[width = \textwidth]{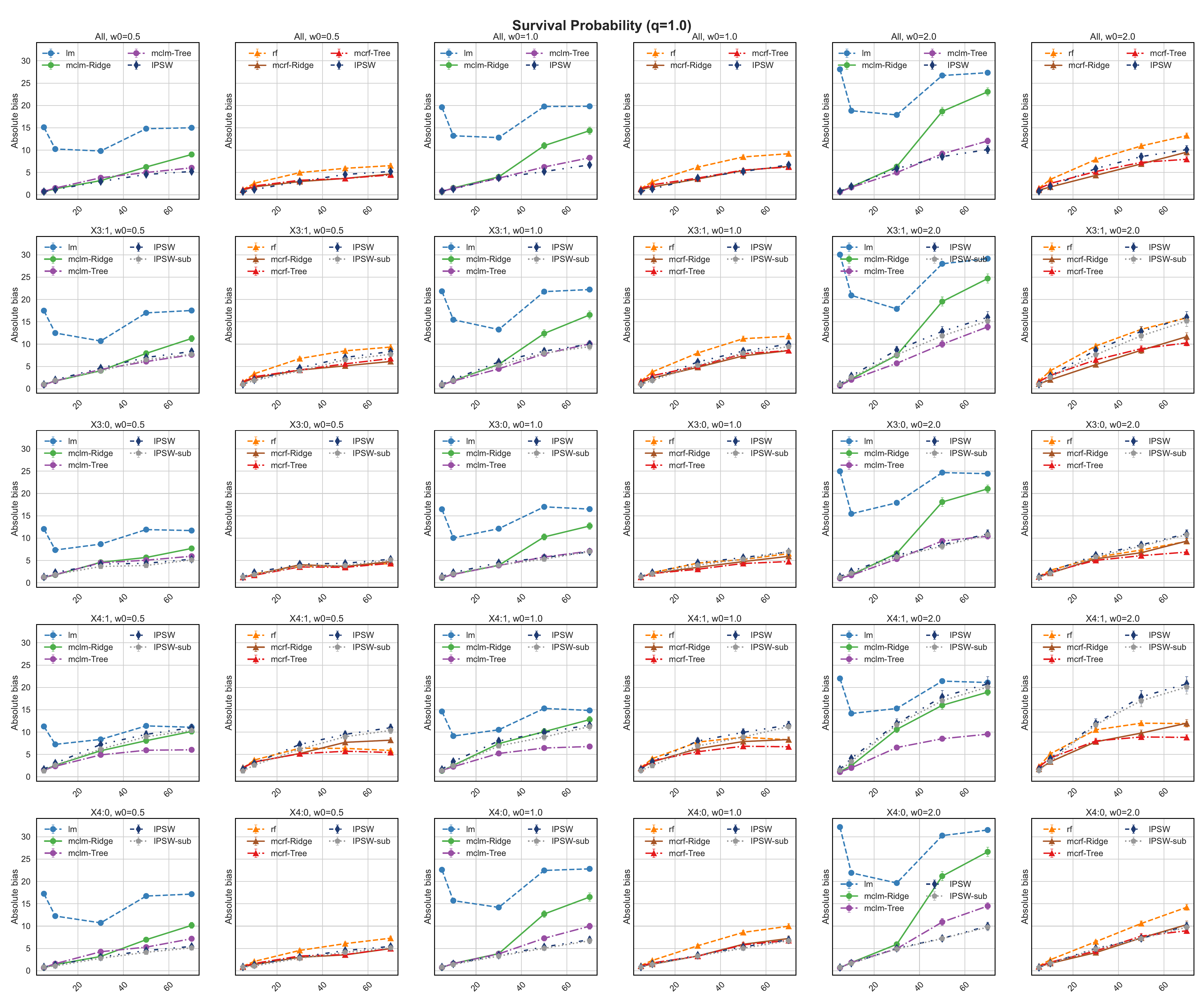}
	\caption{ The absolute bias $(\times 10^2)$ for all methods in estimating survival probability
		under mild covariate shift ($q = 1$) with $1500$ source and $300$ target samples. Survival time adheres to the proportional hazard model and censoring is covariate independent. The results for all subpopulations are presented based on simulation replications $100$.}
	\label{fig:SP-PH-indep-1500-300-1-comb}
\end{figure}

\begin{figure}
	\centering
	\includegraphics[width = \textwidth]{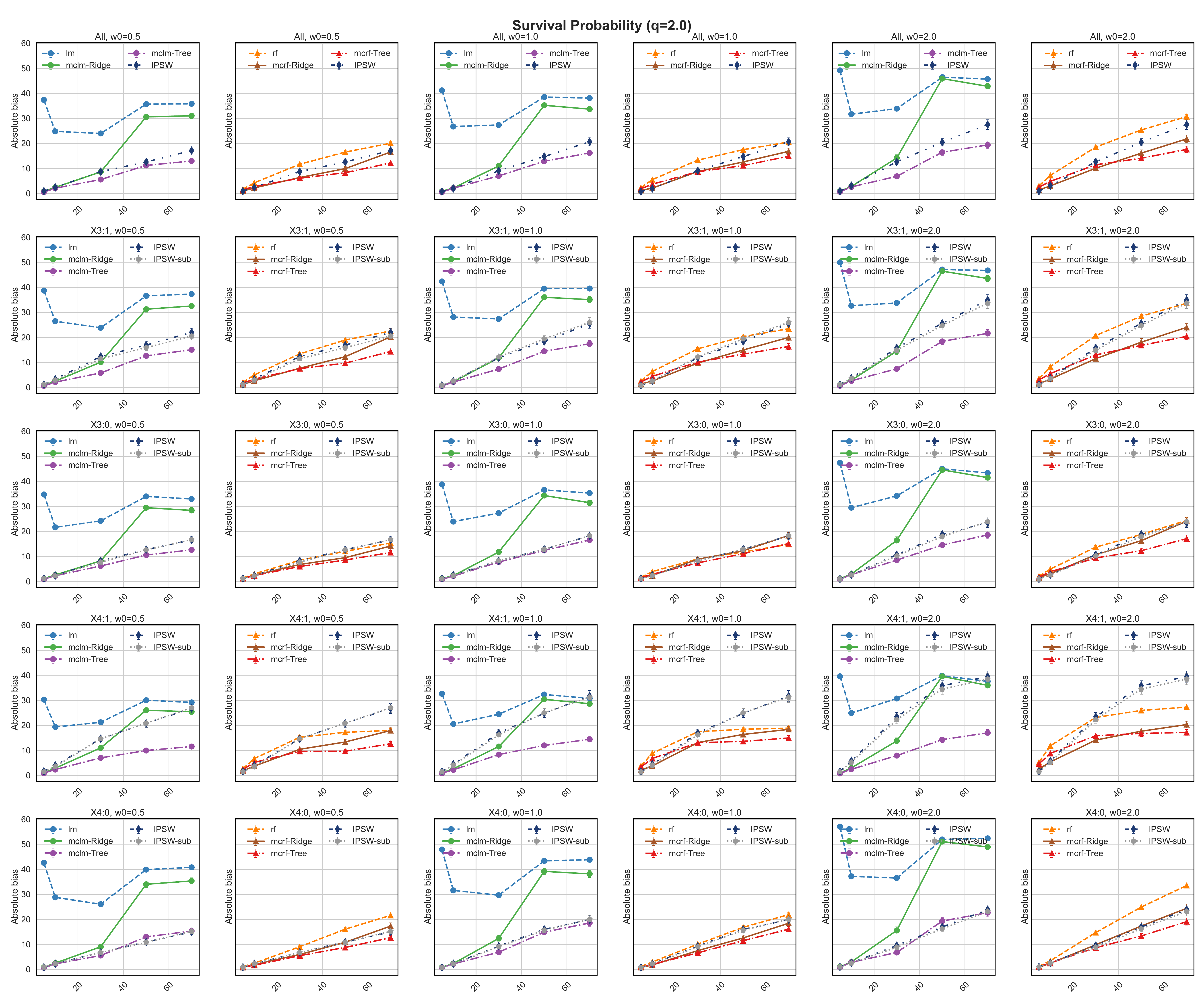}
	\caption{ The absolute bias $(\times 10^2)$ for all methods in estimating survival probability
		under mild covariate shift ($q = 2$) with $1500$ source and $300$ target samples. Survival time adheres to the proportional hazard model and censoring is covariate independent. The results for all subpopulations are presented based on simulation replications $100$.}
	\label{fig:SP-PH-indep-1500-300-2-comb}
\end{figure}

The results are consistent with previous simulations, as depicted in Fig~\ref{fig:SP-PH-indep-1500-300-1-comb} and~\ref{fig:SP-PH-indep-1500-300-2-comb}. Traditional machine learning and IPSW improve with larger sample sizes but fall short under strong covariate shift, especially for minorities with significant source-target imbalances. Our methods, however, show less bias and often outperform imputation and IPSW, even under mild shifts. Adjusting $\omega_0$ or parameter $q$ affects the performance of all methods, and our multicalibrated predictors consistently demonstrate superior results. 

\subsection{Extending the number of covariates}

To make our simulation analogous to a more complicated real-world study, we extended our setting to include $13$ covariates with $8$ continuous and $5$ categorical variables, which model clinical variables such as sex, race, smoking status, hypertension, and diabetes. To be more specific, we generated $(X_{i1}, X_{i2})$ from a mean $0$ bivariate normal distribution with variance $2$ and correlation $1/4$, and the remaining continuous variables from independent normal distributions with variance $1.5$. The discrete variables were sampled as follows: $X_{i8} \sim \text{Binomial}(0.4)$, $X_{i9} \sim \text{Binomial}(0.2 + 0.1X_{i8} )$, and $X_{i10} \sim \text{Binomial}(0.4-0.1X_{i8} + 0.1X_{i9})$, representing sex, race, and smoke status, respectively. Meanwhile, $X_{11} \sim \text{Binomial}(0.35)$ and $X_{12} \sim \text{Binomial}(0.25)$ were used to signify hypertension and diabetes.

We introduced correlations among certain covariates to better mimic the practical scenario where interactions exist. The covariates $(1, X_{i1}, X_{i2}, \ldots, X_{i2})$ influence survival time and domain membership. Again, we consider Weibull hazard function and Weibull censoring. 
Some covariates are ``dummies'' in the sense that they have little impact on either and can better test the efficacy of our black-box algorithm. In our experiment, we evaluated all methods at prefixed times $\{5, 10, 30, 50\}$ due to shorter overall survival times in the current scenario and adjusted parameter $q$ to modify the extent of covariate shift. 

\begin{figure}
	\centering
	\includegraphics[width=0.7\linewidth, height=0.66\textheight] 
	{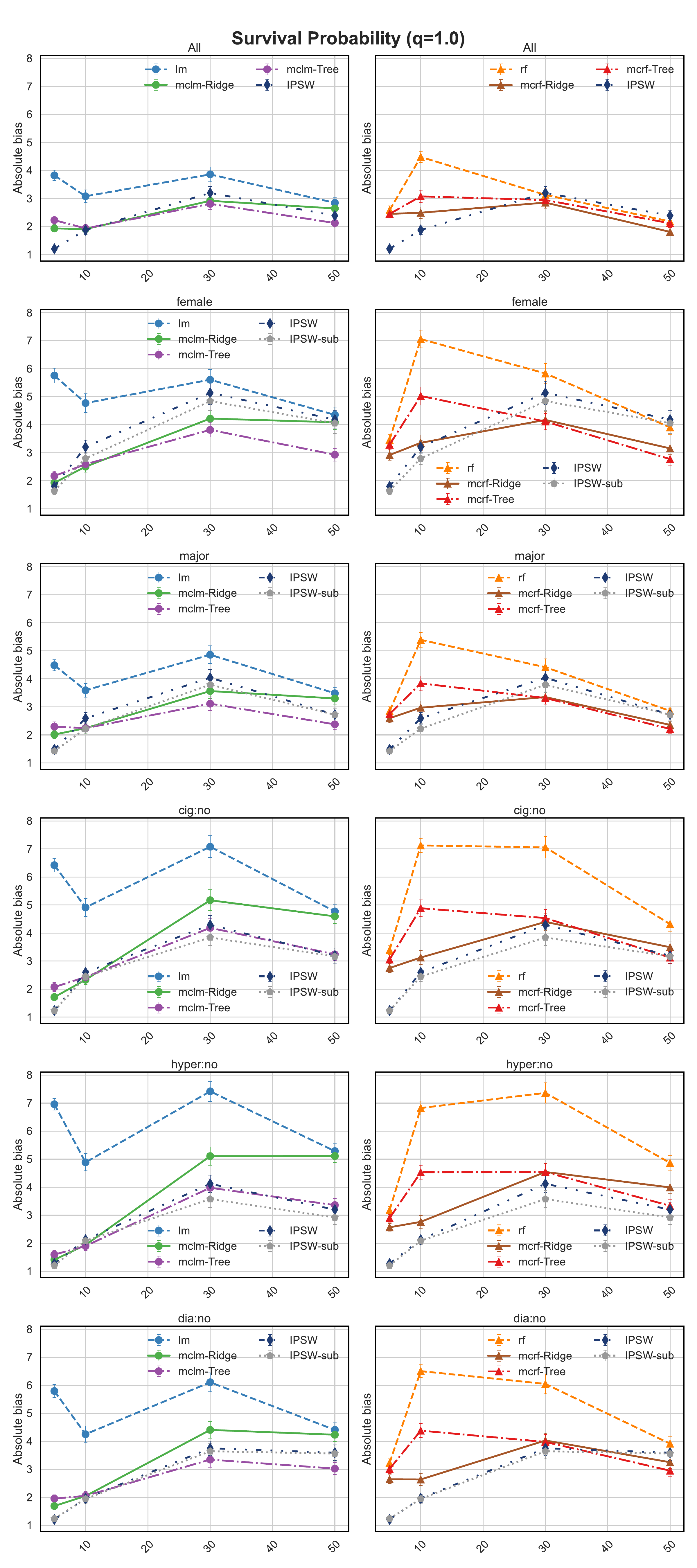}
	\caption{ Results for all subpopulations in the setting with $13$ covariates based on $100$ replications. The absolute bias $(\times 10^2)$ for all methods in estimating survival probability are reported under mild covariate shift ($q = 1$) with $1500$ source and $300$ target samples.
		Survival time adheres to the proportional hazard model, and censoring is covariate-independent.   }
	\label{fig:SP-highdim-1}
\end{figure}

Figures~\ref{fig:SP-highdim-1} and~\ref{fig:SP-highdim-2} show the detailed performance of all methods. As the shift increased, the bias for the linear model, random forest, and \textsc{ipsw} in predicting target survival probabilities grew significantly, whereas our methods demonstrated notable improvement. These findings align with results from other simulations, suggesting their efficiency in simple and complicated scenarios.

\section{Additional analysis of CRIC/MESA data set}
\subsection{Covariate shift between CRIC and MESA cohort}
\begin{figure}[ht!]
	\centering
	\includegraphics[width=0.95\textwidth, height=0.35\textheight]{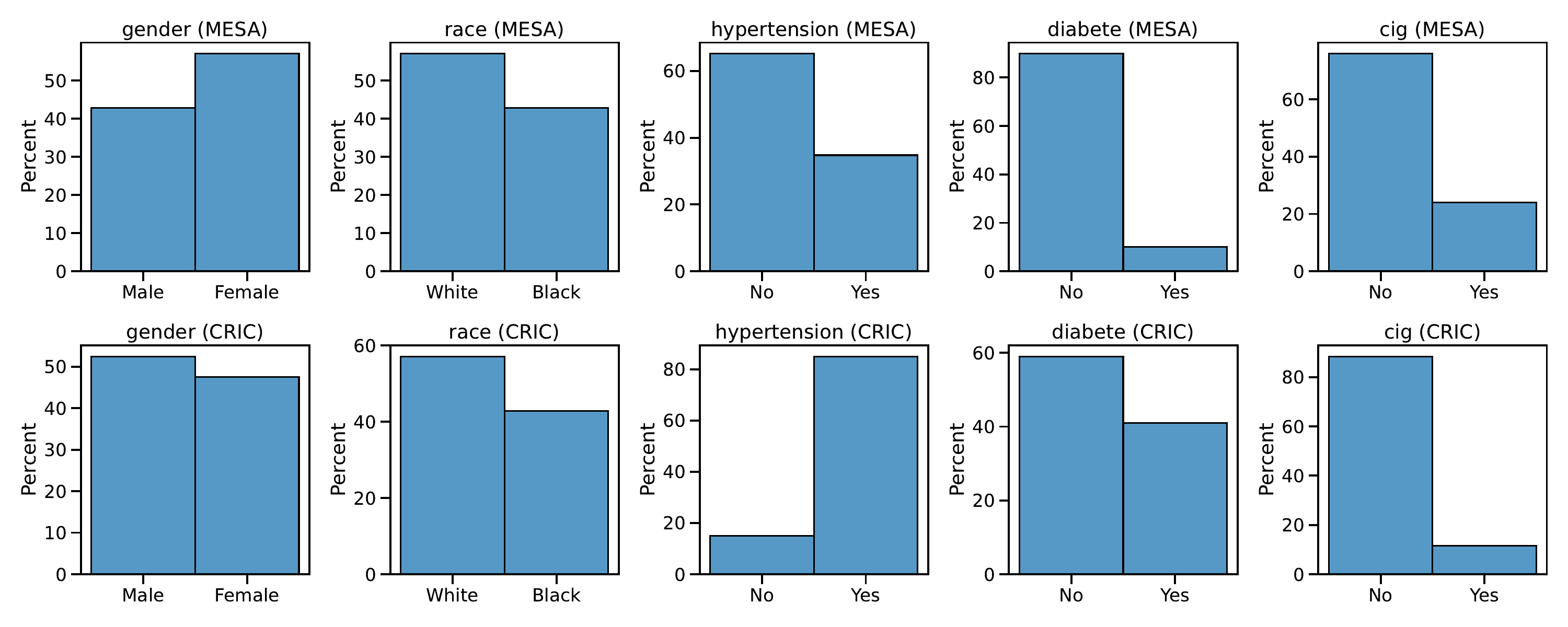} 
	\caption{The boxplots of $5$ categorical variables for \textsc{mesa} (top panel) and \textsc{cric} (bottom panel)  cohorts, highlighting the presence of covariate shift between the two cohorts.
	}
	\label{fig:hist_CRIC_MESA}
\end{figure}

Figure \ref{fig:hist_CRIC_MESA} shows the boxplots of $5$ categorical variables for \textsc{mesa} (top panel) and \textsc{cric} (bottom panel)  cohorts, highlighting the presence of covariate shift between the two cohorts.

\subsection{Results from competing risk analysis}
\begin{figure}[h]
	\centering
	\includegraphics[width=0.9\linewidth]{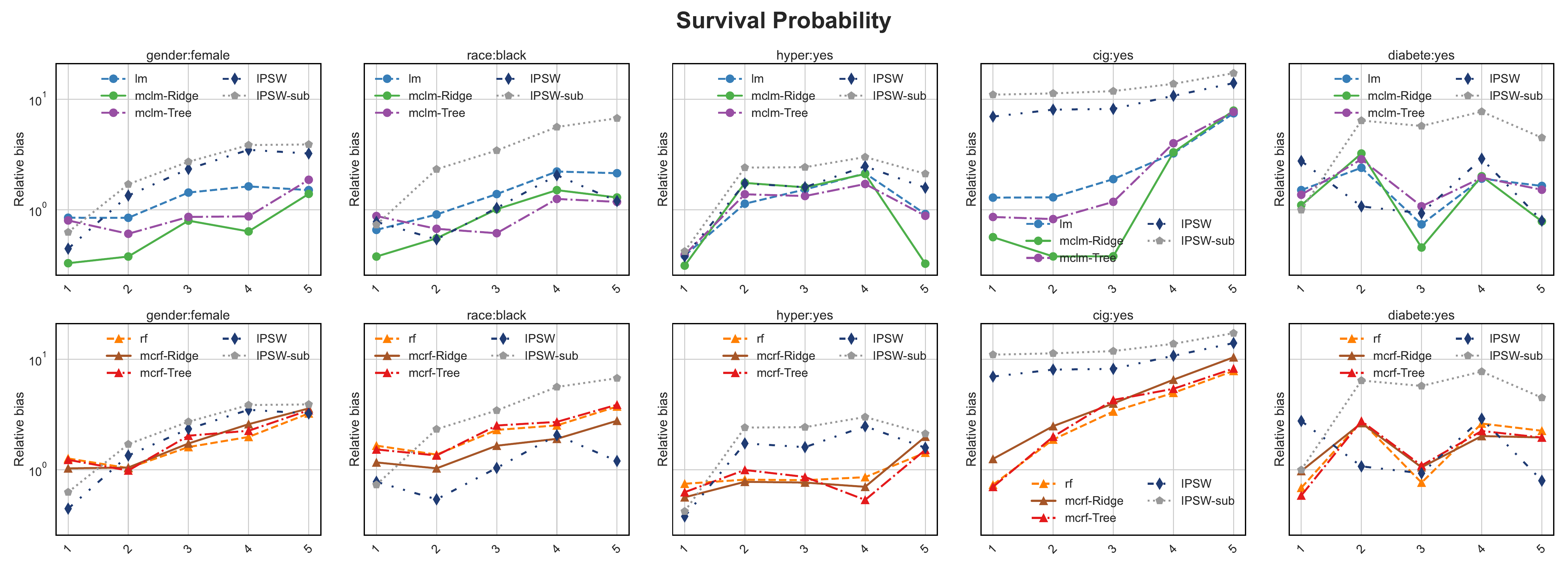}
	\caption{Relative error for survival probability when new pseudo observations are used based on competing risk model. }
	\label{fig:Competing-real}
\end{figure}

As a sensitivity analysis, we also perform the analysis treating none-\textsc{cvd} death as competing risk by estimating the cumulative incidence using the Aalen-Johansen estimatior.  As Figure~\ref{fig:Competing-real} suggests, the results are almost identical to what we observed  before. The non-CVD deaths are not too many   in our cohorts and have no large  influence on the main estimates.

\subsection{C-index for CRIC/MESA data analysis}
\begin{figure}[htb!]
	\centering
	\includegraphics[width=0.95\textwidth,height=0.55\textheight]{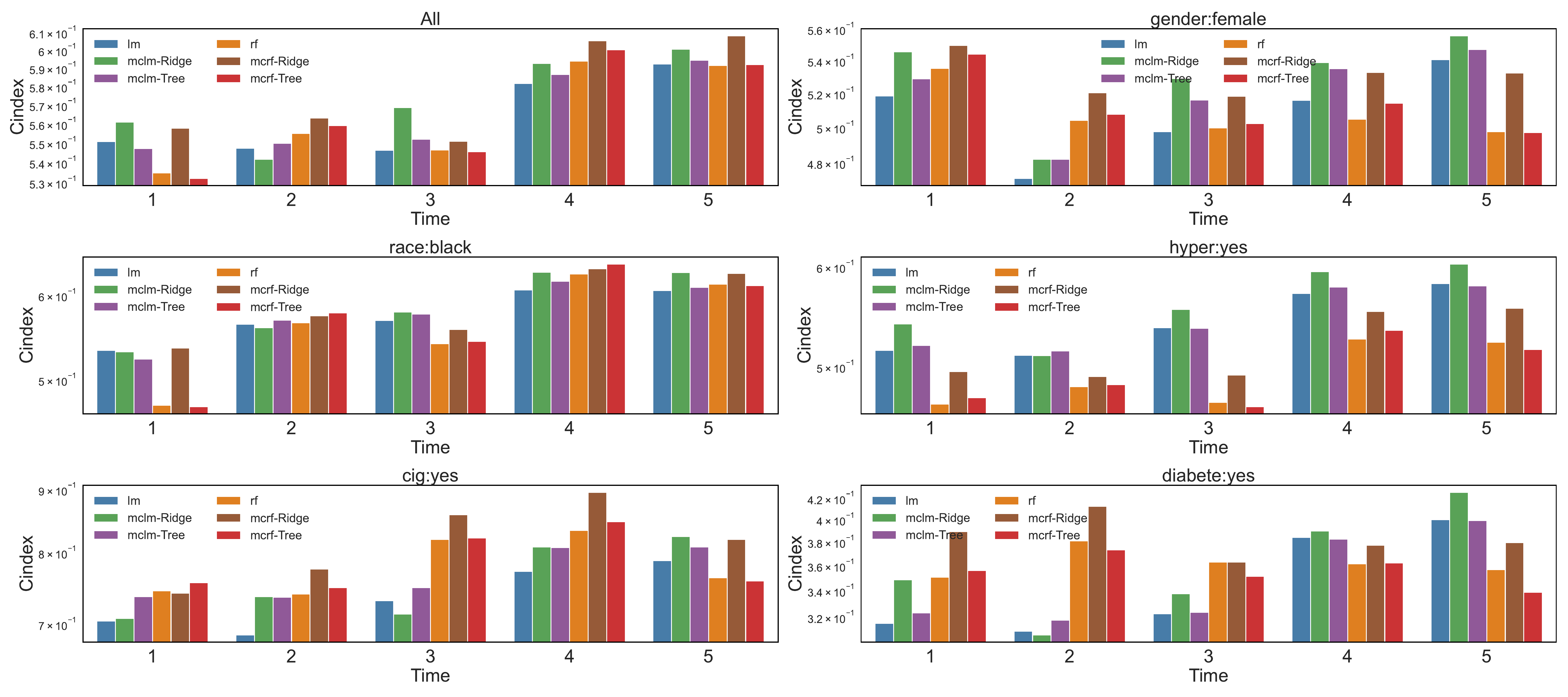}
	\caption{C-index of predicting survival probability from different methods for all \textsc{mesa} cohort and various subpopulations at different follow-up time points. 
	}
	\label{fig:Cindex_MESA_CIRC}
\end{figure}

Figure \ref{fig:Cindex_MESA_CIRC} shows C-index of predicting survival probability from different methods for all \textsc{mesa} cohort and various subpopulations at different follow-up time points. 


\end{document}

%% file: notation.tex
\newcommand{\bbR}{\mathbb{R}}

\newcommand{\sH}{\mathcal{H}}
\newcommand{\sN}{\mathcal{N}}

\newcommand{\sS}{\mathcal{S}}
\newcommand{\sT}{\mathcal{T}}

\newcommand{\sU}{\mathcal{U}}

\newcommand{\sC}{\mathcal{C}}

\newcommand{\vw}{\mathbf{w}}

\newcommand{\Expect}{\mathbb{E}}
\newcommand{\intd}{\,\mathrm{d}}

\usepackage{amsmath}
\DeclareMathOperator*{\argmax}{arg\,max}
\DeclareMathOperator*{\argmin}{arg\,min}

\usepackage{accents}